\documentclass{aa}  

\usepackage{graphicx,txfonts}

\usepackage{mathtools,amsmath,amssymb,CJK,lineno,hyperref,epigraph,textcomp,gensymb,color,graphicx,adjustbox,txfonts}
\include{journal_abbrv.tex}

\modulolinenumbers[2]
 \newcommand{\Afrho}{Af$\rho$\mbox{ }}

\usepackage[colorinlistoftodos]{todonotes}
\let\Oldtodo\todo
\renewcommand{\todo}[1]{\Oldtodo[inline]{#1}}

\begin{document} 


\title{Modeling the past and future activity of the Halleyids meteor showers}

\author{A. Egal
    \inst{1, 2, 3}\fnmsep\thanks{\email{aegal@uwo.ca}}
    \and
    P. Wiegert \inst{1,2}
    \and
    P. G. Brown \inst{1,2}
    \and
    M. Campbell-Brown \inst{1,2}
    \and
    D. Vida \inst{1,2}
    }

\institute{
    Department of Physics and Astronomy, The University of Western Ontario, London, Ontario N6A 3K7, Canada
    \and
    Institute for Earth and Space Exploration (IESX), The University of Western Ontario, London, Ontario N6A 3K7, Canada
    \and
    IMCCE, Observatoire de Paris, PSL Research University, CNRS, Sorbonne Universit\'{e}s, UPMC Univ. Paris 06, Univ. Lille, France
    }

\date{Received XYZ; accepted XYZ}

 
  \abstract
   {We present a new numerical model of the $\eta$-Aquariid and Orionid meteor showers. }
   {The model investigates the origin, variability and age of the $\eta$-Aquariid and Orionid apparitions from 1985 to the present day, in order to forecast their activity over the next several decades. }
   {Through the numerical integration of millions of simulated meteoroids and a custom-made particle weighting scheme, we model the characteristics of every $\eta$-Aquariid and Orionid apparition between 1985 and 2050. The modeled showers are calibrated using 35 years of meteor observations including the showers activity profiles and interannual variability.}
   {Our model reproduces the general characteristics of the present-day $\eta$-Aquariids, and part of the Orionid activity. Simulations suggest that the age of the $\eta$-Aquariids somewhat exceeds 5000 years, while a greater fraction of the Orionids are composed of older material. The 1:6 mean-motion resonance with Jupiter plays a major role in generating some (but not all) Halleyid stream outbursts. We find consistent evidence for a periodicity of 11.8 years in both the observations and modeled maximum meteor rates for the Orionids.
   A weaker evidence of a 10.7 year period in the peak activity for the $\eta$-Aquariids needs to be investigated with future meteor observations. The extension of our model to future years predicts no significant Orionid outburst through 2050 and four significant $\eta$-Aquariid outbursts in 2023, 2024, 2045 and 2046.}
   {}


   \maketitle
%
\section{Introduction}

In recent years, thanks to modern numerical simulations of meteoroid streams, the timing of meteor outbursts have reached unprecedented accuracy \citep[e.g.,][]{Asher1999}. In contrast, reliable estimates of meteor showers duration and intensity are less robust \citep{Vaubaillon2017}. This is due to both a lack of measurement  of the parent body’s past activity and lack of long-term, consistent records of the resulting meteor showers. 

The modeling of comet 1P/Halley and Halleyids' meteor showers, i.e. the $\eta$-Aquariids and Orionids, suffers from these observational limitations. Simulations of the meteoroids released by the comet a few millennia ago are limited by the increasing uncertainty of 1P/Halley's past activity and dynamics. In addition, the lack of consistent observations of the Halleyids (and especially of the $\eta$-Aquariids) has hampered development and validation of new models of the shower \citep[e.g., see][]{Sekhar2014}. 

In this work, we present the second part of our analysis of the Halleyids meteor showers. In \cite{Egal2020b}, we have documented the long-term ($\sim$several decades) activity of the $\eta$-Aquariids and Orionids. The showers' activity profiles, shapes, and year to year variations were defined through a compilation of visual meteor observations since 1985, more recent measurements of the Canadian Meteor Orbit Radar \citep[CMOR, cf.][]{Brown2008,Brown2010} and the IMO Video Meteor Network (VMN)\citep{Molau2009}. That paper reported results of multi-instrumental measurements of the Halleyids over several decades using consistent analysis techniques. This long-term characterization of the Halleyids provides the needed observations to validate modeling of 1P/Halley's meteoroid complex. 

The purpose of the present work is to apply a new numerical model of meteoroid stream evolution developed in \citet{Egal2019} to the Halleyids. In particular, the long term evolution of these streams can now be constrained by the 35 years of observations reported in \cite{Egal2020b}. Through extensive numerical simulations and the application of an adapted weighting solution to the ejected particles \citep[see for e.g.,][]{Egal2019}, we constrain the model parameters using the $\eta$-Aquariids and Orionids intensity profiles and interannual variability since 1985. The simulation parameters are  calibrated on these tens of annual apparitions of each shower to increase the model's robustness. The dynamical evolution of 1P/Halley's meteoroid streams are characterized and linked to the main  characteristics of the present-day observations of the $\eta$-Aquariid and Orionid meteors including origin, age and variability. After model calibration of the observed showers before 2020, we provide a forecast of the Halleyids activity until 2050. 

The paper is divided to four main parts as follows:
\begin{itemize}
    \item[A)] Section \ref{sec:Halley} reviews the characteristics of 1P/Halley's nucleus, dust production activity and dynamics that will serve as starting conditions for our numerical model. 
    \item[B)] Section \ref{section:halleyids} presents a general description of the $\eta$-Aquariid and Orionid meteor showers, summarizing the main conclusions of \cite{Egal2020b}. A short historical review of prior Halleyids models is provided in Section \ref{sec:models}. 
    \item[C)] Section \ref{sec:model} details our new numerical model of 1P/Halley's meteoroid stream. This description is followed with a reconnaissance analysis of the meteoroid trails' evolution (Section \ref{sec:stream_evolution}) and the predicted associated meteor activity on Earth (Section \ref{section:impactors}). 
    \item[D)] Sections \ref{sec:postdiction} and \ref{sec:forecast} present the core of our analysis. The simulated activity profiles are compared and calibrated with meteor observations, and the activity of the $\eta$-Aquariid and Orionid meteor showers is forecast through 2050. The successes and limitations of our model are discussed in Section \ref{section:discussion}. \vspace*{-2\baselineskip}
\end{itemize}

\section{1P/Halley} \label{sec:Halley} 

1P/Halley is among the most famous of all comets. It is named after Sir Edmond Halley, who recognized the periodic nature of the body for the first time in comet science and successfully predicted its return close to the Sun in 1758. 1P/Halley moves on a high-inclination retrograde orbit, with a period of revolution of about 70-80 years. It is the only Halley-type comet in recorded history that can be observed twice in a human lifetime. The comet often appears bright in the sky, as seen from the Earth and depending on the viewing geometry, and is therefore commonly recorded in historical annals. 

The first observation of the comet has been suggested to date to more than 2 000 years ago based on Chinese records \citep{Kiang1972}. In 1986, the return of 1P/Halley to perihelion generated unprecedented enthusiasm for cometary observations. The nucleus was approached by an international fleet of five spacecraft (Giotto, Vega 1, Vega 2, Suisei and Sakigake), collectively termed the "Halley Armada". In-situ space missions were supported by additional space observations (performed by the ISEE-3, Pioneer 7 and Pioneer 12 spacecraft) as well as by ground-based telescopic observations. The observations of the comet and the associated meteoroid streams were coordinated, collected and archived by the International Halley Watch (IHW) organization.

Most of our knowledge of 1P/Halley's composition, morphology and activity were obtained during the extensive observation campaigns organized during the 1986 apparition. In this section, we review the results from the IHW and contemporary studies relating to 1P/Halley pertinent for meteoroid stream modeling. 

 \subsection{Nucleus} \label{sec:1P_nucleus}
 
The spacecraft flybys of 1P/Halley revealed that the comet's nucleus was asymmetric with an avocado-like shape having dimensions of 15x8x8 km \citep{Whipple1987} and a mean effective cross-sectional area of about 80-100 km$^2$ \citep{Sagdeev1988,Keller1986}. Estimates of the nucleus albedo were 0.04-0.05 \citep{Whipple1987,Keller1987,Hughes1987a}.

 The density of the nucleus is not accurately known. By analyzing the influence of nongravitational forces on the motion of the comet, \cite{Rickman1986} estimated the mass and the density of the nucleus. Using the non-gravitational acceleration coefficients found by \cite{Yeomans1977} and \cite{Kiang1972}, \cite{Rickman1986} estimated a most probable nucleus bulk density of about 0.1 to 0.2 g~cm$^{-3}$, with an upper limit of order 0.3 g~cm$^{-3}$ when the asymmetric sublimation rate around perihelion was used. Applying the same technique, but adding data closer to perihelion, \cite{Sagdeev1988} estimated a density of 0.6 g~cm$^{-3}$. Work by \cite{Whipple1987} suggested upper limits of 0.4 to 0.5 g~cm$^{-3}$.  
 
   \begin{figure*}[!ht]
     \centering
     \includegraphics[width=.49\textwidth]{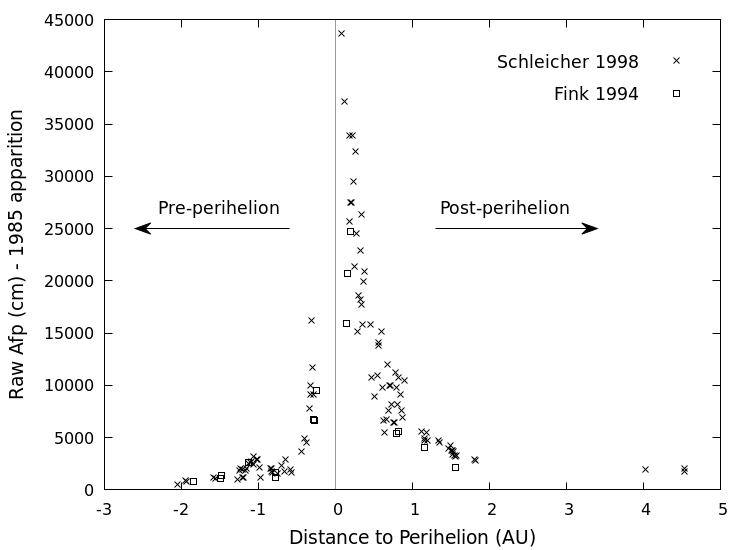}
     \includegraphics[width=.49\textwidth]{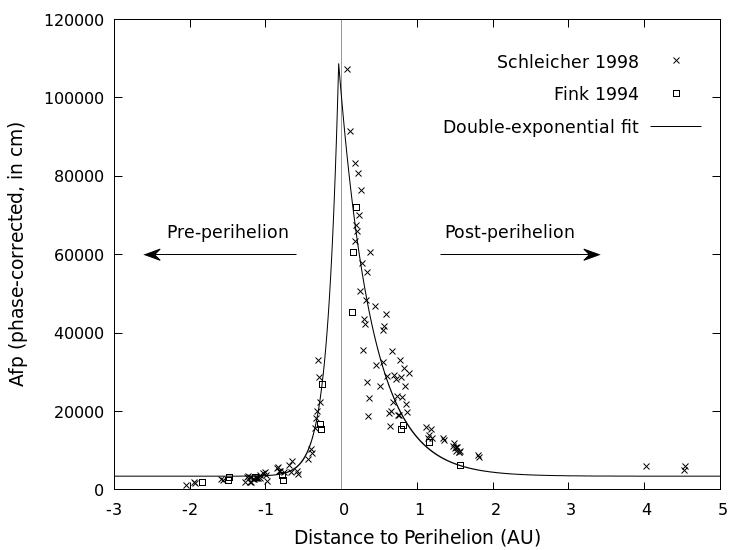}
     \caption{Raw (left panel) and phase-corrected (right) \Afrho measurements of comet 1P/Halley during the 1986 apparition presented in \cite{Schleicher1998} and \cite{Fink1994}. The abscissa (X) represents the comet's heliocentric distance to perihelion, arbitrarily negative for pre-perihelion measurements and positive for post-perihelion times ($X=(r_h-q)\frac{t-t(q)}{|t-t(q)|}$, with $t(q)$ the time at perihelion).}
     \label{fig:Afrho}
 \end{figure*}

 \subsection{Activity}
 
  \subsubsection{Location}
 
 Ice sublimation and the associated dust ejection from 1P/Halley's surface are generally confined to the sunlit hemisphere of the comet. Images taken by the Halley Multicolor Camera (HMC) on board the Giotto spacecraft revealed a concentration of the dust emission in a sector spanning an angle of 70 degrees of the solar direction \citep{Keller1987}.  Thermal infrared observations in March 1986 highlighted the existence of a sunward fan and the absence of a well-defined tail \citep{Campins1987}.
 
 The Giotto camera recorded several active jets on the comet's sunlit hemisphere, while the rest of the surface was largely inactive (Emerich 1987,  Sekanina 1987). The percentage of active area on the nucleus was estimated to be 10\% \citep{Keller1987} or 15\% \citep{Whipple1987}. Localized jet activity was found to be variable in time and location on the nucleus \citep{Whipple1987}. The onset of activity was first observed at 6 AU, well beyond the limiting heliocentric distance for water ice sublimation \citep{Wyckoff1985,Whipple1987,Hughes1987a}. 
 
  \subsubsection{Af$\rho$}
  
 Several dust production rates were estimated during the Giotto and Vega 1-2 flybys of the nucleus \citep[e.g.,][]{Edenhofer1987,Hanner1987,Mazets1987,Krasnopolsky1987}. However, these estimates are strongly dependent on the spacecraft's distance from the nucleus and do not necessarily reflect long-term variations in the cometary activity. 
 
 Cometary dust production is frequently characterized by the \Afrho parameter defined by \cite{AHearn1984}. \Afrho estimates depend on the comet's Bond albedo, the phase angle, the filling factor of the dust grains within the field of view on the plane of the sky and the aperture radius used for an observation.
 \Afrho measurements are independent of the aperture radius $\rho$ when the comet's radial brightness varies as $\rho^{-1}$ close to the nucleus. However, observations of 1P/Halley revealed a strong variation in \Afrho as a function of aperture size, implying that the inner coma brightness of the comet decreases more steeply than $\rho^{-1}$ before perihelion \citep{Schleicher1998}. 
 
 Comparisons of different \Afrho measurements of comet 1P/Halley therefore require a correction for the aperture radius. Examples of debiased \Afrho estimates can be found in \cite{Fink1994} and \cite{Schleicher1998}. In a second step, raw measurements need to be corrected for phase angle effects. 
 
 The variation of 1P/Halley's \Afrho as a function of the phase angle $\theta$ has been determined by \cite{Schleicher1998} for $\theta$ between 0$\degree$ and 60$\degree$. \cite{Schleicher1998}'s dust phase function of 1P/Halley became a reference relation to correct \Afrho measurements of other comets at small phase angles \citep[e.g.,][]{Blaauw2011}. Current cometary dust phase corrections are usually performed using a composite dust phase function, combining:

 \vspace*{-.5\baselineskip}
 \begin{enumerate} \label{enum:correction}
  \item \cite{Schleicher1998}'s relation for $\theta\leq42\degree$:   \\       
  \mbox{ }\hspace{0.2cm}$\text{log}_{10}Af\rho(\theta)=-0.01807\theta+0.000177\theta^2$\\[-.7\baselineskip]
  \item \cite{Marcus2007a,Marcus2007b} phase relation based on the Henyey-Greenstein function for $\theta>42\degree$
 \end{enumerate}
 \vspace*{-.5\baselineskip}
 
 The left panel of Figure \ref{fig:Afrho} presents the original \Afrho measurements of comet 1P/Halley, corrected for the aperture radius used, as presented in \cite{Fink1994} and \cite{Schleicher1998}. Phase-corrected estimates of \cite{Schleicher1998} are presented in the figure's right panel. Original estimates of \cite{Fink1994} were decoupled from phase angle effects via the composite dust phase function detailed above, and plotted along with \cite{Schleicher1998}'s measurements in Figure \ref{fig:Afrho}.

 Following the procedure of \cite{Egal2019}, we fit 1P/Halley's \Afrho profile of Figure \ref{fig:Afrho} with a double-exponential function of the form: 
 \begin{equation}\label{eq:afrho}
  \begin{aligned}
  & Af\rho(X)=K_1+Af\rho(X_{max})\times10^{-\gamma |X-X_{max}|}, \\
  & \gamma=\gamma_1 \text{ if } X\le X_{max} \text{ and } \gamma=\gamma_2 \text{ if } X\ge X_{max}, \\
    & X=(r_h-q)\frac{t-t(q)}{|t-t(q)|}\\
 \end{aligned}
\end{equation}
where $r_h$ is the comet's heliocentric distance at time $t$, $q$ the perihelion distance, and $\gamma_{1,2}$, $X_{max}$ and $K_1$ parameters which are to be determined by the fitting process. The best agreement with measurements of 1P's 1985 apparition gives $K_1\sim 3500$ cm, $Af\rho(X_{max})\sim105~285$ cm, $\gamma_1=2.97$ (pre-perihelion branch),  $\gamma_2=0.99$ (post-perihelion branch) and $X_{max}=-0.04$ AU as illustrated by the black line in Figure \ref{fig:Afrho}. The comet dust production is asymmetric, with a maximum around perihelion and a steeper pre-perihelion profile compared to post-perihelion. 

  \subsection{Dust} \label{sec:dust}
  
  In-situ measurements of the dust environment of comet 1P/Halley have provided insight into its meteoroids' composition, ejection speed, size and mass distributions. The Giotto spacecraft measured two distinct populations of extremely small dust grains, down to $10^{-16}$g in mass: one silicate-rich and the other referred as CHON (Carbon, Oxygen, Hydrogen and Nitrogen) particles \citep{Kissel1986a,Kissel1986b}. 
  
  The dust's bulk density was estimated to be 0.297 g~cm$^{-3}$ from Vega-2 measurements of the dust scattering \citep{Krasnopolsky1987}. From thermal infrared images, the estimated dust albedo decreased from 0.03 to 0.02 for phase angles between 20$\degree$ and 50$\degree$ \citep{Sekanina1987}.
  
  Measurements of the meteoroids' mass distribution were performed by impact detectors on-board both the Vega and Giotto spacecraft at different locations in the coma \citep{Mazets1986,Simpson1986,McDonnell1986,Zarnecki1986,Vaisberg1987}. The mass index $s$ was found to vary with the distance between the spacecraft and the nucleus in an irregular manner. In addition, a slope turnover in the mass distribution was identified for small masses ($\sim 10^{-8}-10^{-13}$ kg) but this value showed intersensor differences as well as variations with nuclear distance \citep{Hanner1987,McDonnell1987}.
  
  A summary of the mass distribution indices $s$ measured by the different spacecraft is presented in \cite{McDonnell1987} and \cite{Hughes1987b}. Estimates of $s$ (the differential mass distribution exponent in dN=Cm$^{-s}$dm, where dN is the number of grains between m and m+dm) ranged between 1.49 and 2.03 for masses between $10^{-6}$ and $10^{-13}$ kg, with an average value of 1.9 to 2.1 \citep{Hanner1987}. Measurement of the meteoroids' bulk mass density and the mass distribution index close to the nucleus are in general agreement with the values derived from meteor observations (see Section \ref{section:halleyids}). 
 
\subsection{Ephemeris}

A key parameter in modeling the Halleyid stream is the past orbital evolution of the parent comet, 1P/Halley. We expect uncertainties in the osculating orbit with time will have a significant influence on our ability to reproduce through modeling the present day streams. As a result, we critically examine all available information on 1P/Halley's previous orbital behaviour, with an eye toward identifying those elements of each stream which are most secure given the comet's ephemeris uncertainty.

\subsubsection{Since 1404 BCE} \label{Ephemeris_since_1404BCE}

Observations of Halley's comet could date back as far as 240 BCE in Chinese records \citep{Kiang1972} and 164 BCE in Babylonian records \citep{Stephenson1985}. Despite the existence of old observations which provide valuable constraints on the ephemeris of the comet, no single set of orbital elements successfully links all the recorded apparitions of 1P/Halley. Numerical integration of the comet's motion back in time from recent observations starts diverging from ancient observations around 218-374 CE \citep{Yeomans1981,Stephenson1985}. It has been suggested that variations of the comet's non-gravitational acceleration might be responsible for the differences between its modeled and observed past motion \citep{Sitarksi1987}.

Different sets of osculating elements of the comet have been proposed in the literature (see for example \citealt{Chang1979,Brady1982,Yeomans1981,Landgraf1984,Landgraf1986,Sitarksi1987}). To reproduce modern and ancient observations, \cite{Landgraf1984} introduced a linear variation of the comet's nongravitational parameters. \cite{Yeomans1981}, in contrast, considered a new determination of the comet's perihelion passage times in ancient Chinese observations to constrain the ephemeris. These authors assumed constant nongravitational coefficients with time, but had to make subjective changes in the comet's eccentricity during the Earth's close approach of 837 CE to match older observations. \cite{Yeomans1981} integrated the motion of the comet back to 1404 BCE, when an additional close encounter with Earth and the lack of older observational constraints on the comet's motion prevented extension of the analysis further back in time.

Despite the manual modification of the comet's orbital elements in 837 CE by \cite{Yeomans1981}, the resulting ephemeris provides the most accurate reproduction of ancient and modern observations of 1P/Halley \cite[e.g.,][]{Stephenson1985,Sitarksi1987}. \cite{Yeomans1981}'s solution for 1P/Halley's ephemeris between 1404 BCE and 1910 CE is usually taken as the definitive reference of the comet's orbital evolution for studies of its meteoroid streams \citep[e.g.,][]{McIntosh1988,Ryabova2003,Sekhar2014}. The evolution of 1P/Halley's nodal location since 1404 BCE, computed with \cite{Yeomans1981}'s ephemeris \citep[cf. Figure 1 in][]{Egal2020b}, is presented in Appendix \ref{appendix:ephemeris}. The Appendix also reproduces the variations of the Minimum Orbit Intersection Distance (MOID) between the Earth and 1P/Halley around each node \citep[cf. Figure 2 in][]{Egal2020b}, measuring the shortest Cartesian distance between the two osculating orbits when considering the pre-perihelion or the post-perihelion arcs of the comet's orbit.

\subsubsection{Long-term (pre-1404 BCE) evolution}  \label{sec:traceability}

A complete model of the Halleyids requires a thorough  understanding of the comet's orbital evolution prior to the 1404 BCE close encounter with Earth, as the streams are certainly much older. In the absence of earlier observations, the location of the comet becomes more unconstrained prior to 1404 BCE. However, the motion of 1P/Halley over longer periods of time has been investigated by several authors. As the orbital plane of the comet regresses backwards in time, the descending node of 1P/Halley approaches Jupiter's orbit. The date of this unobserved, but probable, close approach with the giant planet can be used to constrain the dynamical lifetime of the comet. 

Using this idea, \cite{Yeomans1986} estimated the dynamical age of 1P/Halley to be at least 16 000 years. \cite{Hajduk1985} estimated a maximum dynamical age within the inner solar system of 200 000 years. The theoretical model of the Halleyids by \cite{McIntosh1983} estimated the dynamical age of 1P/Halley at around 100 000 years. \cite{Sekhar2014}'s numerical integrations from \cite{Yeomans1981}'s 240 BCE orbital solution showed that close encounters with Jupiter significantly modified the comet orbit about 12 000 years in the past. 

\begin{figure}[!ht]
    \centering
    \includegraphics[width=0.49\textwidth]{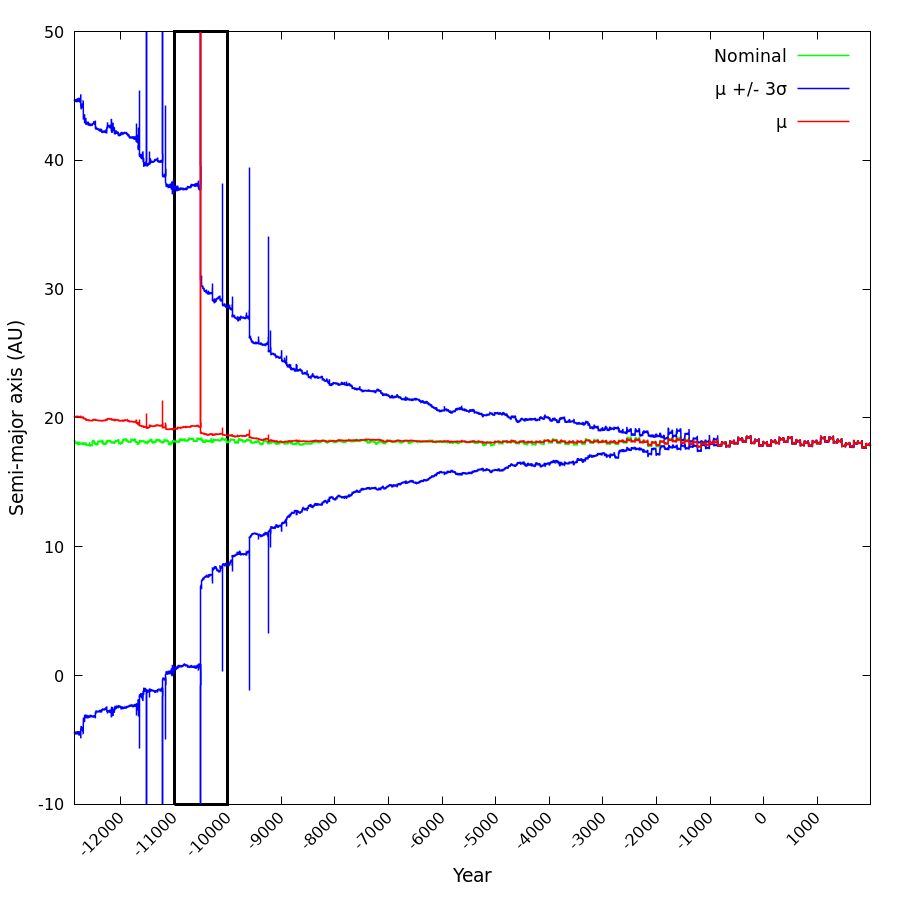} \\
    \includegraphics[width=0.49\textwidth]{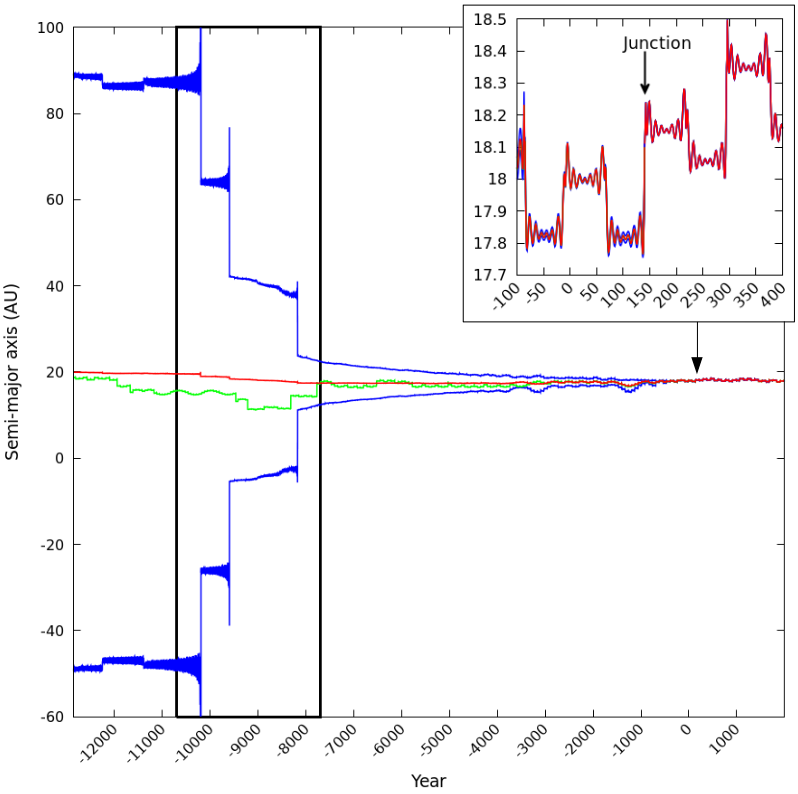}
    \caption{Time-evolution of the nominal (green) and average (red) semi-major axis of 1000 clones of comet 1P/Halley generated in 1994 CE (top) based on the JPL J863/77 solution and 141 CE (bottom) based on the solution of \cite{Yeomans1981}. The blue curves represent the three standard deviations range about the average semi-major axis. See text for details. }
    \label{fig:1P_SMA}
\end{figure}

We have simulated the past orbital evolution of the comet in a similar way. We use as starting conditions a recent orbital solution provided by the JPL, and assume constant nongravitational acceleration parameters with time. The comet was integrated from the JPL\footnote{https://ssd.jpl.nasa.gov/sbdb.cgi} solution until 15 000 BCE. A thousand clones of the comet's orbit were created according to the multivariate normal distribution defined by the solution's covariance matrix. The time evolution of the  nominal (green), average (red) and standard deviation (blue) of the clones semi-major axis is presented in Figure \ref{fig:1P_SMA} (top panel). After a slight dispersion of the clone swarm around 800 BCE, the standard deviation of the semi-major axis increases almost linearly into the past. A close encounter with Jupiter around 10 500 BCE (black box) accelerated the rate of dispersion of the comet's clones, significantly increasing the uncertainty in the comet's ephemeris at earlier times. The orbital evolution of the other osculating elements are presented in Appendix \ref{appendix:ephemeris}. 

Our integration is in good agreement with \cite{Sekhar2014}'s results, but does not take into account the manual corrections of 1P/Halley's ephemeris performed by \cite{Yeomans1981} around 837 CE. To check the validity of our analysis, we have also generated a second sample of clones from \cite{Yeomans1981}'s solution of the comet. The motion of 1P/Halley was integrated starting from 141 CE, at which time it matched the comet's observed perihelion passage with an accuracy of 0.08 days. Since no measurement uncertainties are available for the \cite{Yeomans1981} solution, the comet's clones were created using the standard deviation of the orbital elements determined from the first integration using the JPL solution in 141 CE ($\sigma_{e,~a~(AU),~q~(AU),~i~ (\degree),~\omega~ (\degree),~\Omega~ (\degree)}\sim[6\times 10^{-6},2\times 10^{-3},7\times 10^{-5},6\times 10^{-3},6\times 10^{-2},6\times 10^{-2}]$). The junction of the two samples of clones integrated over the periods 15000 BCE - 141 CE and 141 CE - 1994 CE, is presented in Figure \ref{fig:1P_SMA} (bottom panel).

The clones' evolution is similar for the two initial orbital solutions back to 8000 BCE, at which point the standard deviation of the clones generated in 141 CE from the \cite{Yeomans1981} solution starts to increase much faster than that of the clones created in 1994 CE based on the JPL solution. The abrupt dispersion of the clones around 8000 BCE, 9600 BCE and 10 200 BCE in the \cite{Yeomans1981} 141 CE starting conditions integration are all attributable to close encounters of $\sim$40\% of the test-particles with Jupiter. Our simulations are consistent with the hypothesis of a close encounter of 1P/Halley with Jupiter about 8000 to 12 000 years ago that significantly modified the orbital evolution of the comet. The dynamical lifetime of 1P/Halley is probably much lower than its physical lifetime, which is estimated to be fewer than 1000 revolutions \citep{Whipple1987}. 

Since the date of this close approach depends sensitively on the initial conditions used for the integration, we are not able to set a clear upper time limit for our meteoroid stream simulations. We suggest that if the exact location of 1P/Halley is uncertain prior to 1404 BCE, the comet orbit also becomes uncertain after 10 000 years of backward integration. This affects our choice of timeframe chosen in our meteoroid simulations; we will return to this topic in Section \ref{section:discussion}.

 \subsubsection{Resonances} \label{section:resonances}
 
 Mean motion resonances (MMR) with Jupiter play a major role in modulating the level of activity of the $\eta$-Aquariid and Orionid showers. In particular, the 1:6 MMR with Jupiter is likely responsible for ancient and recent Orionid outbursts \citep{Sato2007,Rendtel2007,Rendtel2008,Sekhar2014,Kinsman2020}. \cite{Sekhar2014}'s analysis of the resonant behavior of the Orionid meteoroid stream highlighted the influence of the 1:6 and 2:13 MMR with Jupiter. The authors examined the evolution of the comet's 1:6 and 2:13 resonant arguments over 6 000 years, and concluded that 1P/Halley was trapped in the 1:6 and
2:13 MMR with Jupiter from 1404 BCE to 690 BCE and 240 BCE to 1700 CE respectively.
 
\begin{figure}[!ht]
    \centering
    \includegraphics[trim=0 0.7cm 0 0,clip,width=.49\textwidth]{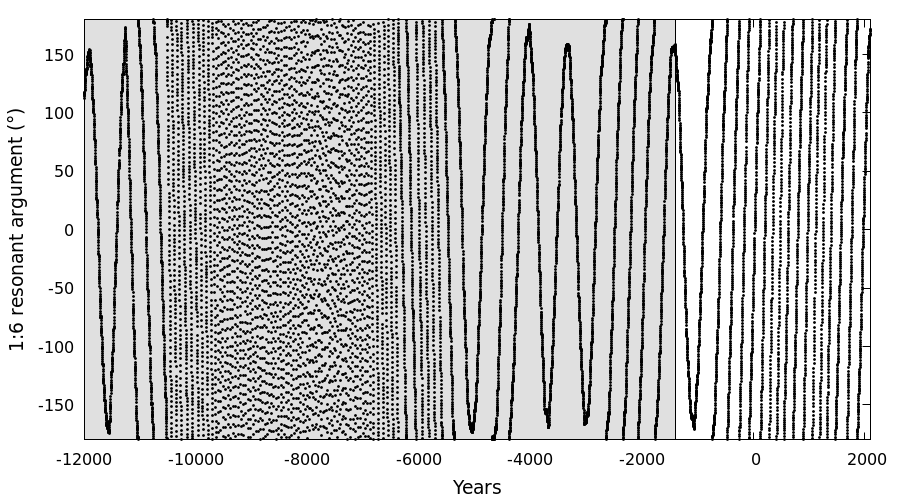}\\
    \includegraphics[trim=0 0.2cm 0 0.5cm,clip,width=.49\textwidth]{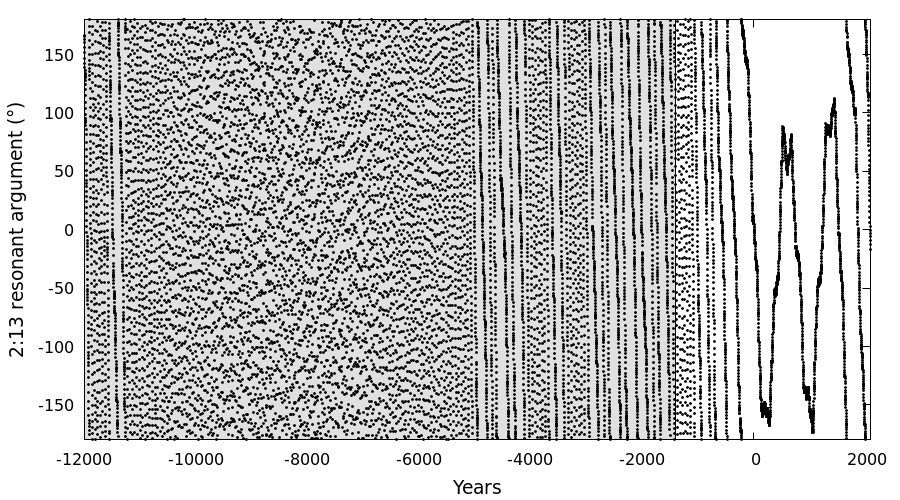}
    \caption{The 1:6 (top) \& 2:13 (bottom) resonant arguments for 1P/Halley based on integrations of the nominal orbit using the orbital solutions of \cite{Yeomans1981}. The comet is in the resonance during times when the resonant argument librates as opposed to when it circulates. The time period when the orbit is less constrained, prior to 1404 BCE, is shown in grey.}
    \label{fig:resonant_angles}
\end{figure}

 Figure \ref{fig:resonant_angles} presents an extension of \cite{Sekhar2014}'s computation of the resonant arguments of 1P/Halley between 12 000 BCE and 1986 CE. Here the comet's motion was integrated for 14 000 years using the orbital solutions of \cite{Yeomans1981}. For this ephemeris solution, though it is uncertain prior to 1404 BCE (see Section~\ref{Ephemeris_since_1404BCE}), we can still see that the comet spent of order a few thousand years within the 1:6 MMR with Jupiter since 12 000 BCE. From our analysis in Figure \ref{fig:resonant_angles}, we find that the times when 1P/Halley is in the 1:6 and 2:13 resonances since 1404 BCE are the same as found by \cite{Sekhar2014}. 
 
 In addition to being influenced by mean motion resonances, 1P/Halley is known to have undergone Kozai resonances during its long term evolution \citep{Kozai1962,Kozai1979,EmelYanenko2001}. The comet's perihelion argument librates between 47$\degree$ and 133$\degree$, while the longitude of the ascending node varies between -18$\degree$ and 82$\degree$. Kozai librations of the comet's orbit might be helpful in constraining the formation of the Halleyids meteoroid streams \citep{McIntosh1983}.

\section{Halleyids} \label{section:halleyids}

Comet 1P/Halley produces two annual meteor showers on Earth; the $\eta$-Aquariids in early May and the Orionids in late October. The $\eta$-Aquariids occur at the descending node of the comet, while the Orionids occur at the ascending node. Modern observations of the $\eta$-Aquariids and Orionids date back to the late 19th century \citep{Egal2020b}. However, analysis of ancient observations of meteor showers suggests that both showers have been active for at least a millennium, and likely longer \citep{Ahn2003,Ahn2004,Imoto1958,Zhuang1977}.  

The quantitative characteristics of the Halleyids meteor showers since 1985 are described in detail in \cite{Egal2020b}. That work documented the showers' average activity, duration, profile shape, annual variations and individual apparitions based on visual, video and radar data collected since 1985, 2011, and 2002 respectively. The purpose of \cite{Egal2020b} was to provide a consistent and multi-instrumental monitoring of the two meteor showers over several decades, in order to constrain models of 1P/Halley's meteoroid streams. That study is the main source we use to validate our numerical model of the Halleyids in the present work. 

The next subsections briefly review the general characteristics of the Halleyid showers as determined in \cite{Egal2020b}. The results presented here, along with additional figures presented in that paper (annual variations and intensity profiles), are used throughout the rest of this study to calibrate our model with meteor observations. As usual, the activity of the showers are characterized by the zenithal hourly rate (ZHR), which represents the number of meteors a single observer would observe per hour under standard reference conditions.  

 \subsection{$\eta$-Aquariids}

The $\eta$-Aquariids are usually active between 35$\degree$ and 60$\degree$ solar longitude (SL), with the highest meteor rates recorded between 44$\degree$ and 50$\degree$. The average maximum zenithal hourly rate lies around 65 to 70 meteors per hour, but in the interval since 1985 the shower displayed two notable outbursts (ZHR $>$ 100) in 2004 and 2013. The average profile of the $\eta$-Aquariids and Orionids was found to be variable from year to year. Over the period 2002 to 2019, the $\eta$-Aquariids average activity profile as a function of time is asymmetric, with a rise of intensity steeper than the subsequent decline.  

The mass index $s$ of the shower is also variable each year as a function of time during the shower's period of activity. Lower mass indices are generally measured close to the main peak of activity, while higher values are found at the beginning and end of the shower. Estimates of $s$ derived from visual observations range between 1.72 and 1.94 \citep{Rendtel1997,Dubietis2003}, and from 1.75 to 1.95 in radar records \citep{Blaauw2011,CB2015}. The average mass index of the shower lies around 1.87 to 1.9 \citep{Dubietis2003,Egal2020b}. 

 \subsection{Orionids}
 
The Orionids are generally observed between 195$\degree$ and 220$\degree$ SL, and present a broad maximum between 206$\degree$ and 211$\degree$. The average profile of the Orionids is more symmetric than the $\eta$-Aquariids around the maximum of activity. 

Maximum meteor rates range between 20 to 40 meteors per hour, with occasional outbursts reaching 2 to 4 times these values (e.g., in 2006 and 2007). Recent CMOR measurements reinforce the hypothesis of a $\sim$12 years cycle in the Orionids activity. However, additional and longer term observations are required to definitively confirm this periodicity. 

As with the $\eta$-Aquariids, the Orionids mass index varies with the solar longitude and by year. Estimates of $s$ were found to vary between 1.46 or less (in 2006) and 1.96 in most years, with an average value of 1.87 \citep{Dubietis2003,Rendtel2008}. Estimates of the mass index from radar observations range between 1.65 and 1.93 from CMOR measurements \citep{Blaauw2011}, and around 1.95 in MAARSY data \citep{Schult2018}. 

The comparison of visual, video and radar observations since 2002 confirm the existence of multiple peaks of activity during a given year's apparition in both showers, variable in location and intensity. The consistency between measurements of different systems supports the idea that there is no significant size sorting of the particles within the stream. However, the changing activity profiles indicate the existence of a complex fine structure within the meteoroid stream, as found in previous work \citep{Egal2020b}.

\section{Previous models} \label{sec:models}

 Many characteristics of the $\eta$-Aquariid and the Orionid meteor showers have challenged researchers for decades. In particular, the similar activity level and duration of the two showers (with such different nodal distances of the comet at the present day) can not be explained with a toroidal meteoroid stream model. If the meteoroid streams were simply centered on the present orbit of the comet, the shower intensity and duration should be much higher for the $\eta$-Aquariids than for the Orionids. The toroidal model also can not explain the gradual displacement of maximum activity observed for some Orionid apparitions \citep[see review in][]{Egal2020b}. 
 
 \subsection{The shell or ribbon-like model}
 
 Perhaps the most successful model of the Halleyids is the theoretical 'shell' model developed by \cite{McIntosh1983}. This model assumes that the evolution of meteoroids results in belts or "ribbon-like" structures intersecting Earth's orbit at the time of the $\eta$-Aquariid and Orionid meteor showers.
 
 The formation and the thickening of each meteoroid belt is explained by gravitational perturbations induced by Jupiter. In \cite{Yeomans1981}'s ephemeris, the variations of 1P/Halley's longitude of the ascending node $\Delta \Omega$ induced by Jovian perturbations range between 0.2$\degree$ and 2.2$\degree$. If we assume that meteoroids are spread all along the comet's orbit, then about 1/6 ($\sim P_\text{Halley}/P_\text{Jupiter}$) of the meteoroid stream is subject to the full range of $\Delta \Omega$ perturbations induced by Jupiter at each revolution of the planet. After a few revolutions, the meteoroid stream will be distributed along a belt characterized by different $\Omega$ values. In other words, dust particles tend to diffuse to where the comet was a long time ago, along the present comet orbit, as well as on orbits that will be likely reached by the comet in the future.  The thickness of each individual belt is expected to be relatively uniform along the intersection of the meteoroid's nodes with Earth's orbit \citep{McIntosh1983}. 
 
 The shell model offers a natural explanation for the similar duration and intensity of the contemporary Halleyids showers. The libration of the argument of perihelion of the orbit suggested by \cite{Kozai1979} creates a superposition of the meteoroid belts over time, also explaining the presence of multiple peaks of activity in the $\eta$-Aquariid and Orionid annual profiles. To reproduce the five Orionids activity peaks reported by \cite{Hajduk1970}, this model would require five libration cycles of the comet representing a period of about 100 000 years \citep{McIntosh1983}. 
 
 The theoretical shell model was investigated via numerical simulations by \cite{McIntosh1988}. Several sets of about 500 particles were ejected from the comet in 1404 BCE using \cite{Yeomans1981}'s orbital solution, and integrated until 1986. The authors observed a quick elongation of the particles' nodal locations, as well as the formation of fine structure within the stream. This work resulted in global confirmation of the ribbon-like structure suggested by \cite{McIntosh1983}, but the similar duration and intensity of the meteor showers were not explained from the numerical simulations. In addition, this work suggested that particles ejected after 1404 BCE and with masses larger than 1 g would not intersect Earth's orbit today. Therefore, the model suggests that such large meteoroids, recently ejected from 1P/Halley, would not contribute to the present day Orionid meteor shower \citep{McIntosh1988}.
 
The results of \cite{McIntosh1988} show that the fine-structure of the stream can be created much faster than the 100 Kyr suggested by \cite{McIntosh1983}, possibly in as little as a few thousands of years. The authors suggested that the meteoroid stream observed today at Earth developed after a close approach with Jupiter about 20 000 years ago, but that no conclusive evidence of this close encounter exists because of the uncertain comet ephemeris prior to 1404 BCE. In a further analysis of the Orionids using radar, TV and visual observations \cite{Jones1989} estimated a total mass of the Halleyids meteoroid stream of $1.3\times10^{13}$ kg, resulting in age of 2 500 to 62 000 years. The author's preferred age estimate lies around 23 000 years. 

 \subsection{Previous Dust trail simulations}
 
 \subsubsection{General characteristics}
 
Using a similar approach to \cite{McIntosh1988}, \cite{Ryabova2003} integrated the motion of discrete simulated meteoroid trails ejected in 1404 BCE, 141 CE, 837 CE and 1910 CE. Each trail was composed of 5000 particles, ejected with a tenth of the velocity predicted by \cite{Whipple1951}'s method and were within a 70$\degree$ sunward cone. Particles of masses 1 g and 0.001 g were simulated, with a density of 0.25 g/cm$^{-3}$. The orbital solution of \cite{Yeomans1981} was used for the comet ephemeris. 

The simulations predicted a meteoroid spatial density structure closer to layers than filaments.  From the trails' nodal locations, \cite{Ryabova2003} suggested that no simulated meteoroid of mass $>$0.001g would contribute to the present day Orionid meteor shower. Particles producing the contemporary $\eta$-Aquariids were found to have been ejected prior to 837 CE at least, and only trails older than 1404 BCE could explain the observed duration of the shower. In these simulations, the nodal footprint of particles of 1g in mass are shifted compared to particles of 0.001g. The mass distribution of the two showers was therefore expected to be dissimilar. 

 \subsubsection{Resonant trails}

More recently, classical dust trail simulations of 1P/Halley's meteoroid streams were undertaken with the goal of identifying the years when past and future outbursts of the $\eta$-Aquariid and the Orionid meteor showers may occur. After the observation of the unexpectedly strong Orionid activity in 2006, the role of mean motion resonances with both Jupiter and Saturn in producing Halleyids outbursts was investigated by several authors. 

Following the methodology described in \cite{Kondrateva1985}, \cite{McNaught1999} and \cite{Sato2003}, \cite{Sato2007} simulated dust trails released from the comet at each perihelion passage since 1404 BCE. They identified the 1:6 MMR to be responsible for the 2006 Orionid outburst. From similar calculations, \cite{Sato2014} predicted an enhanced $\eta$-Aquariids activity on May 6, 2013 due to the encounter with meteoroids ejected in 910 BCE and 1197 BCE. The outburst was recorded by radio, visual and video techniques as predicted \citep[e.g.,][]{Molau2013b,Cooper2013,Steyaert2014,CB2015}, confirming the validity of the numerical model. 

Adopting the same approach, \cite{Kinsman2017} found a correlation between probable past outbursts of the $\eta$-Aquariids in CE 400-600 and Maya records of meteor activity. The model applied to the Orionids revealed the existence of probable significant outbursts in 417 CE and 585 CE, due to 1:6 and 1:7 mean motion resonances with Jupiter \citep{Kinsman2020}.

The role of mean motion resonances with Jupiter on the occurrence of Orionid outbursts was investigated more precisely by \cite{Sekhar2014}. The authors found the 2:13 and especially the 1:6 resonances to be responsible for enhanced meteor activity on Earth over several consecutive years. From the analysis of these resonant regions and specific dust trail calculations similar to \cite{Sato2007}, \cite{Sekhar2014} explained the existence and date of outbursts observed in 1436-1440, 1933-1938 and 2006-2010 as being due to the encounter by Earth of meteoroids trapped in the 1:6 MMR. The simulations showed that enhanced activity in 1916 and 1993 was caused by 2:13 MMR meteoroids, which they also predicted to cause a future Orionid outburst in 2070. 

\cite{Sekhar2014}'s simulations highlighted the importance of resonant structures in the meteoroid stream to explain the majority of the Orionids outbursts. Unfortunately the analysis was not extended to the $\eta$-Aquariids, due to the paucity of observations of the shower available at that time. With the publication of multiple decades of $\eta$-Aquariids observation \citep[e.g., in][]{CB2015,Egal2020b}, the time is now ripe for development, calibration and validation of a more comprehensive model of 1P/Halley's meteoroid stream.

 \subsection{The need for a new model}
 
 Several broad characteristics of the $\eta$-Aquariid and Orionid meteor showers have been successfully described by different studies detailed previously. \cite{McIntosh1983}'s shell model, for example, offers a theoretical explanation of the similar intensity and duration of the two showers, as well as a general explanation for the existence of fine structure within the meteoroid streams. Past numerical simulations of 1P/Halley's meteoroid stream generally agree with the stream evolution described by the shell model. In addition, the more recent recognition of the importance of resonant structures within the stream provide a natural explanation for outbursts and serve as an efficient tool to predict the occurrence of future Orionid and $\eta$-Aquariid outbursts. 

However, several aspects of the streams remain unexplained. In particular, models developed so far do not provide quantitative estimates of the showers' intensity variations, nor are they able to reproduce their comparable duration and strength or details of the radiant structure of the streams. While the rare major outbursts of the Halleyids' are understood as being associated with MMRs, the mechanism behind the observed 11-12 year modulation of annual shower activity previously reported for both streams is unknown. 

Moreover, the range of ages of meteoroids comprising the stream for a given year's apparition is only crudely constrained as is variations in either the age or mass of particles as a function of time during any one  return. Finally, the mass distribution as a function of particle size has not been predicted or explained by published models to date. 

The broad aim of this project is to develop a new comprehensive numerical model of 1P/Halley's meteoroid streams. Custom-made simulations of meteoroids ejected from the comet are calibrated using the 35 years of Halleyids observations compiled in \cite{Egal2020b}, with the same model parameters applied to both the $\eta$-Aquariid and Orionid meteor showers. In particular, this model aims to reproduce the intensity profiles and annual ZHR variations of both showers in an automated way, in order to forecast Halleyids activity until 2050. We also wish to estimate each shower's age, radiant structure and the presence (if any) of mass sorting within the streams. Finally, we seek to produce a model mass distribution for both streams over a large range of sizes appropriate to existing observations.
 
\section{Simulations} \label{sec:model}

  Meteoroid stream simulations performed in this work follow the methodology described in \cite{Egal2019}. Using an ephemeris of the comet's past orbital behaviour as its basis, millions of test particles are ejected from the comet nucleus and numerically integrated forward in time. The characteristics of  Earth-impacting particles, as well as the dynamical evolution of the meteoroid stream, are recorded and analyzed. In order to calibrate the model with meteor observations, our primary simulation outputs are: 
  
  \begin{enumerate}
      \item the date and duration of potential meteor activity on Earth
      \item the characteristics (age, size, etc.) of the trails involved
      \item the stream's structure close to Earth 
      \item the radiants of Earth-intercepting meteors 
      \item the approximate shower intensity
  \end{enumerate}
  
  The purpose of this section is to summarize the main model parameters chosen for our Halleyids meteoroid simulations. For additional information about the model, the reader is referred to \cite{Egal2019}.
  
  \subsection{Nucleus}
  
  In our model we take 1P/Halley to have a spherical nucleus of 11 km diameter \citep{Lamy2004}, corresponding to an effective cross-section of about 95 km$^2$. We assume a nucleus density of 0.3 g~cm$^{-3}$, which matches that of  \cite{Rickman1986}. The geometric albedo of the nucleus is taken as 0.04 (cf. Section \ref{sec:1P_nucleus}).
  
  The motion of the comet was integrated between 1404 BCE and 2100 CE with an external time step of one day as described in Section~\ref{section:integration}. As in \cite{Egal2019}, the ephemeris of each apparition of the comet is computed from the closest available orbital solution. Apparitions prior to 1986 CE are built from the set of osculating elements provided by \cite{Yeomans1981}, converted into ecliptic J2000 coordinates. The comet coordinates since 1986 CE are determined from the JPL J863/77 solution\footnote{https://ssd.jpl.nasa.gov/sbdb.cgi}. The model parameters are summarized in Table \ref{table:parameters}. 
  
        \begin{table*}[!ht]
         \centering
        \begin{tabular}{lcc}
         Comet parameter & Choice & Reference \\
            \hline
            \hline
          Shape & Spherical & .\\
          Radius & 5.5 km & \cite{Lamy2004}\\
          Density & 0.3 g~cm$^{-3}$ & \cite{Rickman1986}\\
          Albedo & 0.04 & \cite{Hughes1987a} \\
          Active surface & 10\% & \cite{Keller1987} \\
          Active below & 6 AU & \cite{Whipple1987}\\
          Variation index $\gamma_1$, $\gamma_2$ & 2.97, 0.99 &   \\
              \hline
                  & & \\        
         Particle parameter & Choice & Reference \\
              \hline
              \hline
          Shape & Spherical &  \\
          Size & $10^{-4}$ to $10^{-1}$ m & \\
          Density & 0.3 g~cm$^{-3}$ & \cite{Krasnopolsky1987} \\
          Ejection velocity model & ``CNC'' model & \cite{Crifo1997} \\ 
          Ejection rate & Each day & \\
          N$_{particles}$ / apparition & 95 000 & \\
            \hline
        \end{tabular}
        \caption{\label{table:parameters} Comet and meteoroid characteristics considered by the model. See the text for more details.}
      \end{table*}  
  
  \subsection{Stream simulation} \label{section:trails_simulated}
  
  As described in \cite{Egal2019}, particles are ejected from the comet at each time step where the heliocentric distance is below 6 AU \citep{Whipple1987}. Simulated meteoroids are released each day from the sunlit hemisphere of the nucleus, following the ejection velocities model of \cite{Crifo1997}. \cite{Crifo1997}'s model is restricted to the sublimation of frozen water, responsible for $\sim$90\% of the volatiles flowing out of a comet below 3 AU \citep{Combi2004}.  Our simulations extend meteoroid ejection beyond this distance, in order to mimic the activity of 1P/Halley due to the sublimation of other volatiles further from the Sun \citep{Epifani2007,Womack2017}. Particles are taken to have a constant density of 0.3 g~cm$^{-3}$ \citep{Krasnopolsky1987}. 
  
  Around 95 000 dust particles were ejected for each apparition of the comet between 1404 BCE and 2100 CE, for a total of about 4.48 million particles simulated. The particles are equally divided amongst the following three size/mass/magnitude bins:
  \begin{enumerate}
  \item $[10^{-4},10^{-3}]$ m, $[10^{-9},10^{-6}]$ kg, [+10 to +3] mag
  \item$[10^{-3},10^{-2}]$ m, $[10^{-6},10^{-3}]$ kg, [+3 to -4] mag
  \item $[10^{-2},10^{-1}]$ m, $[10^{-3},1]$ kg, [-4 to -10] mag
  \end{enumerate}
  These bins approximately correspond to meteors detectable by radar (bin 1),  visual/video means (bin 2) and as fireballs (bin 3). 
  
  Despite the unreliability of the comet's location prior to 1404 BCE \citep{Yeomans1981}, additional simulations were conducted to assess the contributions of older trails, as some parts of the stream are almost certainly populated by ejecta older than 1404 BCE. 1P/Halley's motion was integrated back to 7000 BCE using the oldest orbital solution of \cite{Yeomans1981}. For these older ejections, we released a few thousand particles at each apparition of the comet at heliocentric distances inside 6 AU, covering the period [6982 BCE, 1404 BCE]. Between 18 000 and 6000 meteoroids were ejected at each return of the comet, for a total of about 1 million simulated particles. The number of particles ejected decreased arbitrarily as a function of the ejection epoch (cf. Table \ref{table:Np_ejected}) to reduce the integration time of the oldest streams.
  
  \begin{table*}[!ht]
     \centering
     \begin{tabular}{cccc}
         \hline
         Epoch & Number of & Np/app & Np total \\
          & apparitions & & \\
         \hline
         \hline
         ]1404 BCE - 2100 CE] & 46 & 95 000 & 4 370 000\\ 
         1404 BCE & 1 & 113 000 & 113 000\\
         {[}2946 BCE - 1404 BCE[ & 21 & 18 000 & 378 000 \\
         {[}4292 BCE - 2946 BCE[ & 28 & 15 000 & 420 000\\
         {[}6982 BCE - 4292 BCE[ & 28 & 6000 & 168 000 \\
         \hline
         {[}6982 BCE - 2100 CE]  & 124 & - & 5 449 000\\
         \hline
     \end{tabular}
     \caption{Number of simulated meteoroids as a function of the ejection epoch. The last row contains the sum of all the particles simulated between  6982 BCE and 2100 CE.}
     \label{table:Np_ejected}
 \end{table*}
  
  These older ejected particles were integrated forward until 2050 CE to investigate the influence of older trails on the Orionids and $\eta$-Aquariids. Because of the more uncertain ephemeris of the comet in this time period compared to the era post-1404 BCE, results of these simulations should be viewed with caution. The number of particles released at each comet apparition is summarized in Table \ref{table:Np_ejected}. 
  
  Despite the large number of particles generated in all our simulations, the number of model meteoroids is negligible compared to the real number of meteoroids released by comet 1P/Halley at each perihelion return. To bridge the gap between our simulations and the real meteor shower activity, each simulated particle is assigned a weight representing the actual number of meteoroids released by the comet under similar ejection circumstances. This weight depends upon, among other things, the particle's radius and the comet's dust production at these sizes at the heliocentric distance of ejection. In this work, we use the weighting scheme described in \cite{Egal2019}. Details of the calibration of the weighting factors incorporating observations will be presented in Section \ref{section:weights}. 
  
  \subsection{Integration and data analysis} \label{section:integration}
  
  The comet and meteoroids are integrated with a 15$^\text{th}$ order RADAU algorithm \citep{Everhart1985}, with a precision control parameter LL of 12 and an external time step of one day. The integration is performed considering the gravitational influence of the Sun, the Moon, and the eight planets of the solar system. General relativistic corrections and non-gravitational forces are also taken into account. 
  The meteoroids are integrated as test particles in the stream, under the influence of solar radiation pressure and Poynting-Robertson drag. The Yarkovsky-Radzievskii effect is neglected since the size of our simulated particles does not exceed 10 cm \citep{Vokrouhlicky2000}.

 The integrations are performed in a Sun-centered, ecliptic J2000 coordinate system. However, the physical motion of the comet (and the particles) is well described with heliocentric elements only when the body is close to perihelion (typically within Jupiter's orbit). Indeed, the periodic motion of the Sun around the solar system's barycenter induce oscillations of the comet's heliocentric elements when computed far from perihelion. These oscillations are an artifact of the non-inertial heliocentric frame and only apparent in the osculating elements. They can be removed by estimating the body's orbital elements in the (inertial) barycentric frame. For this reason, the results presented in this work (particles orbits, nodes location) are determined and presented in the solar system's barycentric frame for convenience.
 
 In order to avoid confusion, we employ some specific terminology in our simulations analysis. The terms "impactors" and "impacting particles" are applied to simulated particles approaching Earth within the distance $\Delta X$ and time $\Delta T$ defined in Section \ref{section:impactors}. Physically, these represent test particles whose actual position in space best represent meteoroids which may actually collide with the Earth. The term "nodal crossing location" will refer to the particle's position when physically crossing the ecliptic plane. "Nodes", "node location" or "nodal footprint" refer to the ecliptic position of the particle's ascending and descending nodes as determined by their osculating orbit at the epoch of interest, even when the particle itself is far from the ecliptic plane. 
   
 \section{Stream evolution} \label{sec:stream_evolution}
 
 Before moving to an in-depth investigation of the simulation results, we present in this section a reconnaissance analysis of the meteoroid stream's evolution. For this broad overview of general stream characteristics, we ejected one thousand particles at each apparition of the comet between 1404 BCE and 2050 CE as described in Section \ref{sec:model}. Each particle's position, orbit and node location is recorded at a regular time step of 10 years during the whole integration period, and every year between 1980 CE and 2030 CE. Three independent sets of simulations were conducted for particles in our three bins, representing sizes of 0.1-1 mm, 1-10 mm and 10-100 mm. 

  \subsection{Current structure}
  
\begin{figure}[!ht]
     \centering
     \includegraphics[trim={0cm 1cm 1.5cm 0cm}, clip,width=0.49\textwidth]{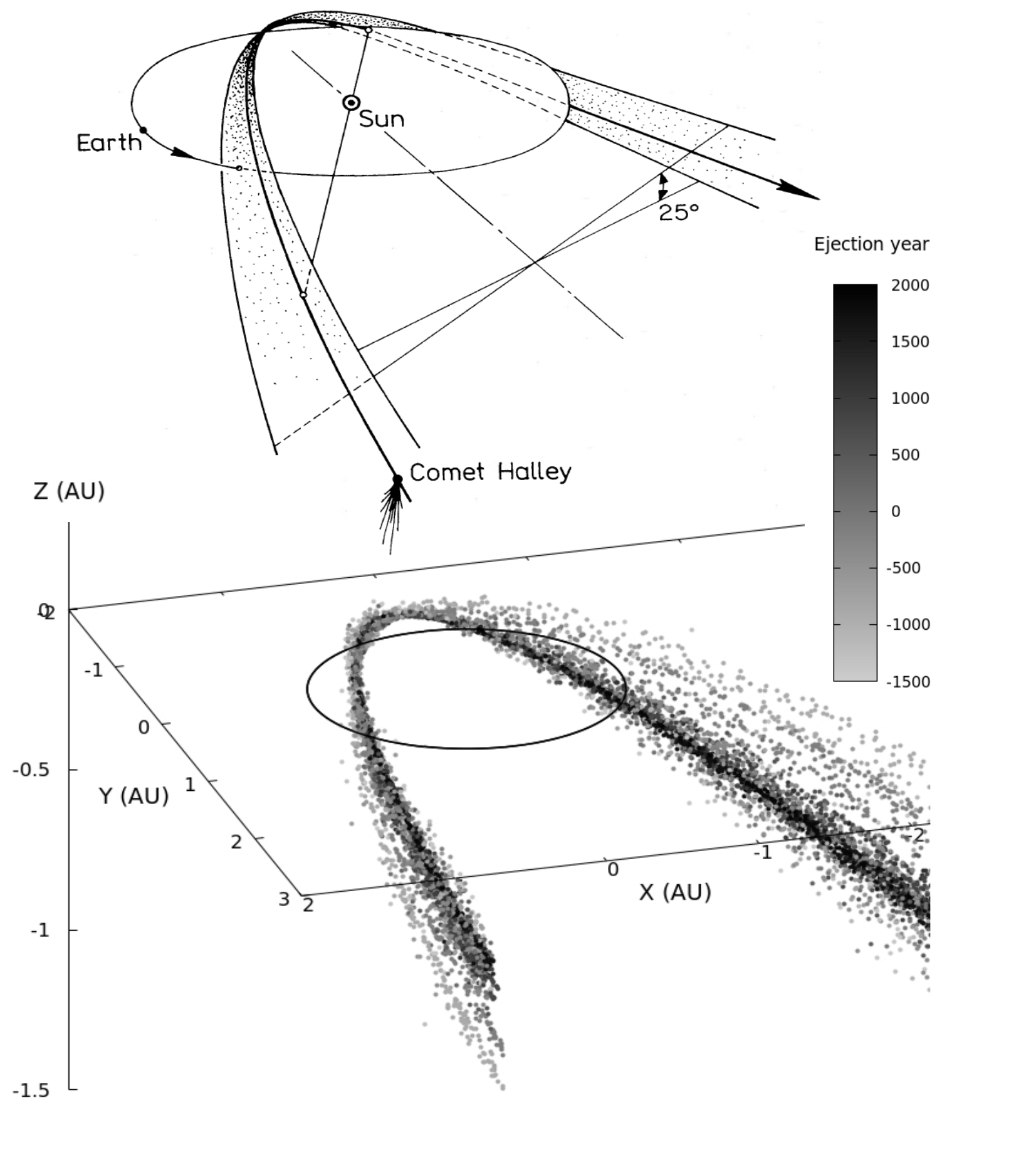}
     \caption{Top panel: schematic representation of the Halleyids shell model as presented in \cite{McIntosh1983} and \cite{McIntosh1988}. Bottom panel: 3D location of the simulated particles between 1990 CE and 2020 CE, color coded as a function of the particles' ejection year.}
     \label{fig:shell}
   \end{figure}

  Figure \ref{fig:shell} presents the current location of the simulated meteoroid stream, including all meteoroids having nodal crossings between 1990 CE and 2020 CE to enhance its visibility using this reconnaissance survey. The theoretical shell model presented in \cite{McIntosh1983} and \cite{McIntosh1988} is reproduced at the top of Figure \ref{fig:shell} for comparison.
  
  The simulated meteoroid stream intersects the Earth's orbit at two locations, and reproduces the ribbon-like shape predicted by \cite{McIntosh1983}. When looking at the edges of the meteoroid belt, the fine structure of the stream is clearly noticeable. As a first approximation, the gross behaviour of the meteoroid stream is in good agreement with the \cite{McIntosh1983} and \cite{McIntosh1988} shell model. 
  
 \begin{figure}[!ht]
    \centering
     \includegraphics[width=0.49\textwidth]{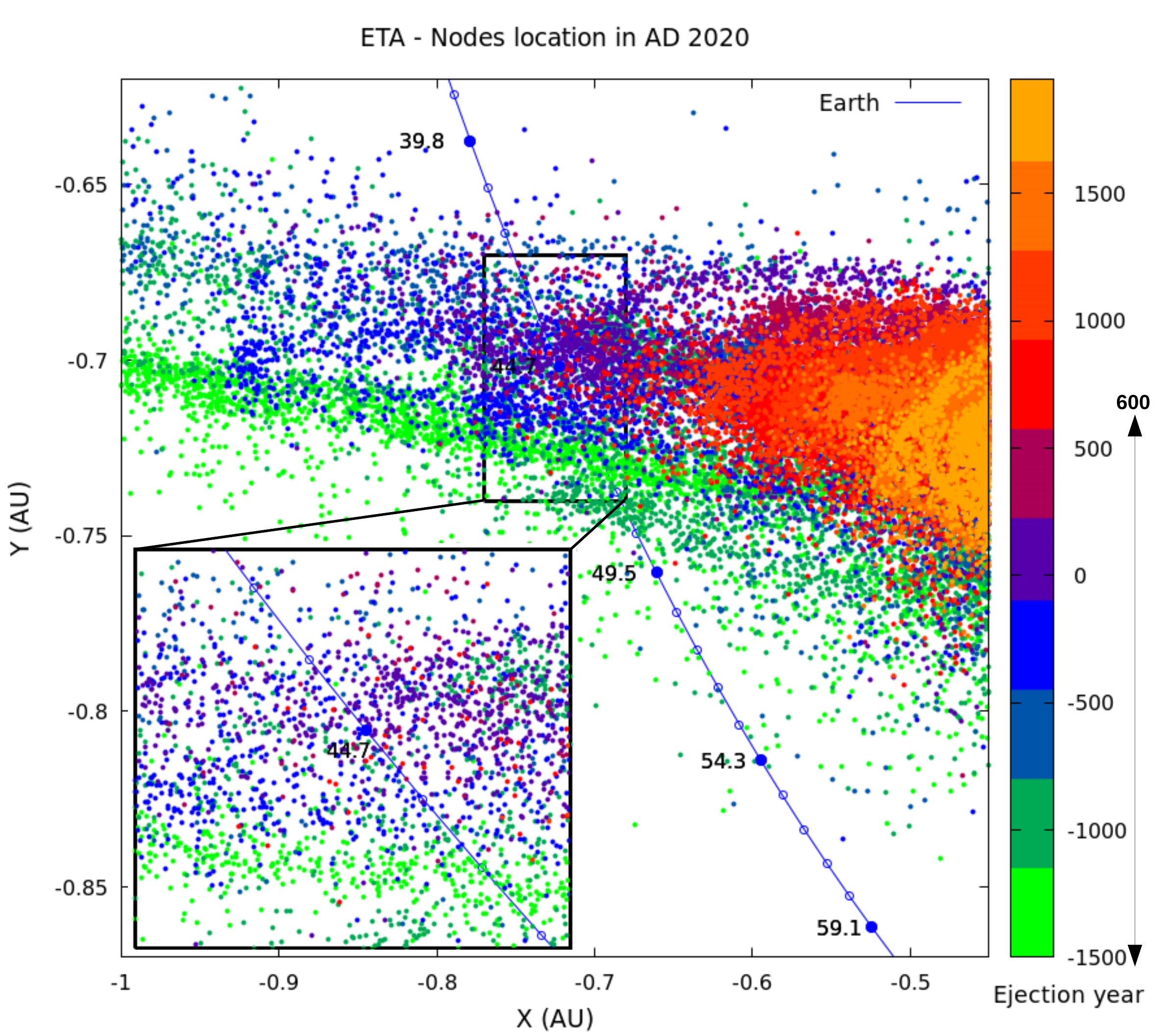}\\
     \includegraphics[width=0.49\textwidth]{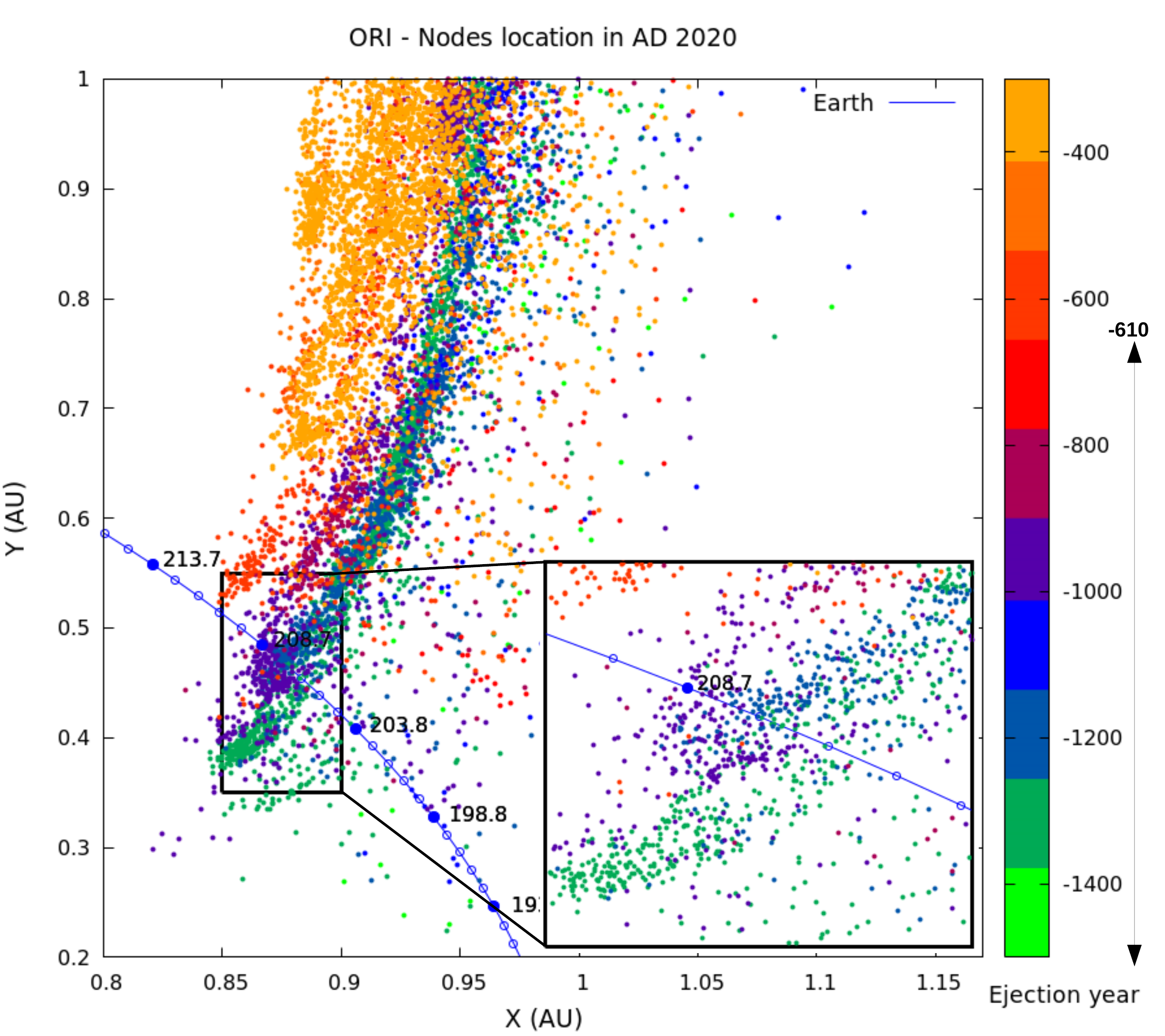}\\
     \caption{Node locations in 2020 CE for all particles as a function of their ejection date. The daily positions of the Earth on its orbit are indicated by blue circles. Numbers along the orbit correspond to the solar longitude with a time step of 5 days. This distribution represents a crude estimate of the overall duration of the shower and peak activity location expected at the return in 2020 CE.}
     \label{fig:nodes_2020_all_sizes}
  \end{figure}
  
  Figure \ref{fig:nodes_2020_all_sizes} shows the nodal footprint of all the simulated particles in 2020 CE, at the location of the $\eta$-Aquariid (top panel) and the Orionid showers (bottom panel). The stream nodes intersect Earth's orbit between 40$\degree$ SL and 54.5$\degree$ SL at the time of the $\eta$-Aquariids, and between 194$\degree$ SL and 211$\degree$ SL at the time of the Orionids.
  Trails ejected between 1404 BCE and 600 CE reproduce much of the current  $\eta$-Aquariid activity. The present Orionids, observed between 195$\degree$ SL and 220$\degree$ SL, are in contrast not as well reproduced. The Orionids seem to be composed of a larger percentage of older particles, ejected before 600 BCE (see Figure \ref{fig:nodes_2020_all_sizes}). Thus we conclude that simulations of meteoroid trails ejected since 1404 BCE can reproduce the showers' peak date, but probably underestimate their observed duration. 
  
\begin{figure}[!ht]
    \centering
     \includegraphics[trim=0 0.2cm 0 0, clip,width=0.49\textwidth]{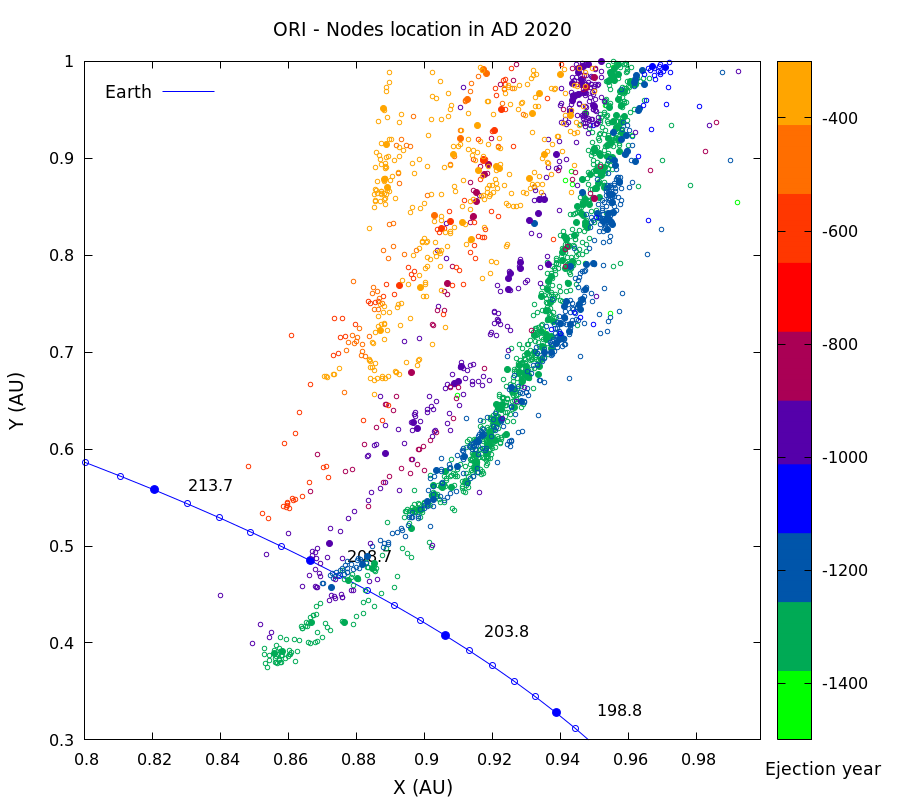}\\
     \caption{Node locations in 2020 CE for particles of masses $10^{-6}-10^{-3}$ kg (empty circles) and $10^{-3}-1$ kg (filled dots) as a function of the year of ejection. }
     \label{fig:nodes_2020_1_100mm}
  \end{figure}

In Figure \ref{fig:nodes_2020_all_sizes}, we see that particles ejected between 1404 BCE and 600 BCE contribute to the present Orionid meteor shower. Simulations of \cite{McIntosh1988} and \cite{Ryabova2003} implied that meteoroids ejected after 1404 BCE and of mass higher than 1 g and 0.001 g respectively do not approach Earth's orbit in recent times. To compare our model with these results, we present in Figure \ref{fig:nodes_2020_1_100mm} the node locations of particles with masses higher than 1 g (filled circles) and between 0.001 and 1 g (empty circles). As expected, the nodal footprint is reduced when removing the smallest particles and the width of our simulated stream does not reproduce the shower's observed duration. However, and in contrast with \cite{Ryabova2003}, we find that particles belonging to these mass ranges may still approach Earth's orbit at the current epoch. Differences between our model and these earlier efforts  may be due to different assumed ejection conditions (location, direction and velocity).
  
 From Figures \ref{fig:nodes_2020_all_sizes} and \ref{fig:nodes_2020_1_100mm}, we see that in the case of the $\eta$-Aquariids, the simulations predict the onset of the shower at low solar longitudes with a mixture of trails ejected at different epochs. However, by the end of the shower activity period it consists primarily of the oldest trails. The simulated Orionids display a weaker gradient in ejection age as the Earth crosses the meteoroid stream, but the existence of such a gradient is hard to assess because of the overlapping locations of the trail's nodes.
  
 \subsection{Time evolution}
 
\begin{figure}[!ht]
    \centering
     \includegraphics[trim=0 0.7cm 0 0, clip,width=0.47\textwidth]{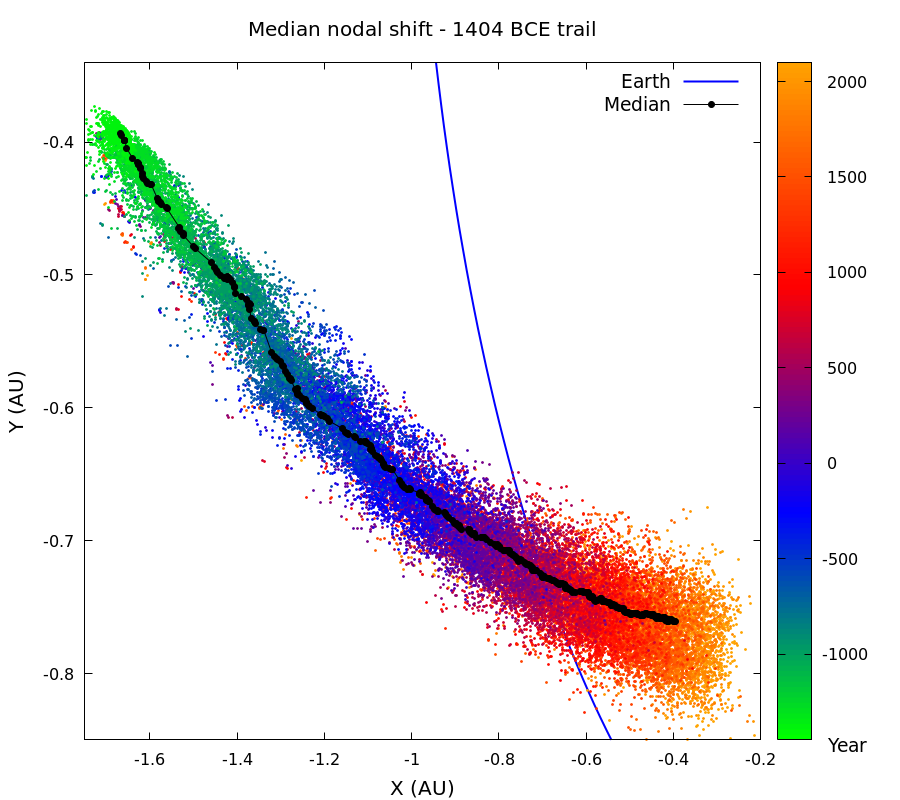}\\
     \includegraphics[trim=0 0.1cm 0 1.6cm, clip,width=0.47\textwidth]{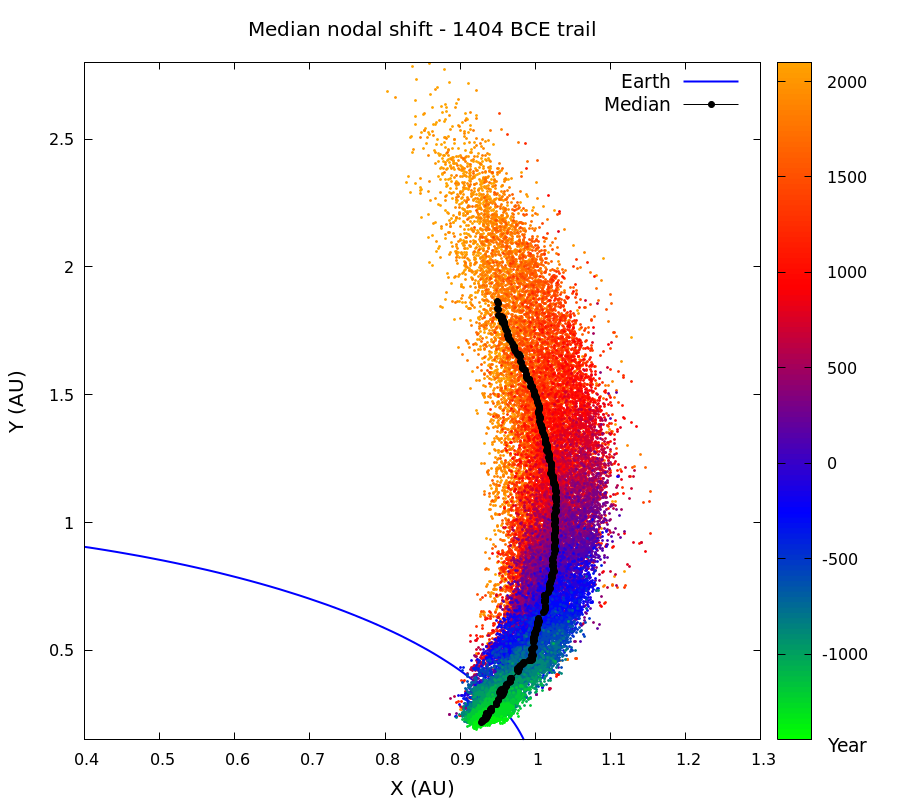}\\
     \caption{The evolution as a function of time (color coded by ejection year) of the descending nodal footprint representing the Eta Aquariids (top) and ascending nodal footprint representing the Orionids (bottom) of a simulated trail ejected in 1404 BCE. The particle sizes are between 0.1 mm and 1 mm. The Earth's orbit in 2020 CE, computed in the barycentric frame, is represented with a blue line. }
     \label{fig:1404BC_trail}
  \end{figure}

 To illustrate the evolution through time of the simulated streams, we independently analyze the location of the nodes of each ejected trail as a function of time. As an example, we show the evolution of the meteoroid stream ejected around 1404 BCE until 2050 CE in Figure \ref{fig:1404BC_trail}. In this figure, we compile the locations of the particles' descending (top panel) and ascending (bottom panel) nodes at different epochs from their ejection time (in green) to 2050 CE (in orange). 
 
 As noted in \cite{McIntosh1988}, the nodes of the ejected trail (in green) quickly spread, exceeding 2 AU in extent within 3000 years.  The median location of the nodes from the 1404 BCE ejection at each epoch is indicated with black dots in Figure \ref{fig:1404BC_trail}. The median node location of the 1404 BCE trail has clearly changed considerably since the particles were ejected. The median descending node (top plot) is located well outside the Earth's orbit in 1404 BCE, and migrates inside the planet's orbit after a few thousand years, intersecting the Earth's orbit and contributing to Eta Aquariid activity roughly from 0 - 1000 CE.
  
  In contrast, the median ascending node of the 1404 BCE trail in Figure \ref{fig:1404BC_trail} (bottom) starts much closer to the Earth's orbit at the time of ejection, but drifts outward with time to a present day distance of about 1 AU from the Earth. Here we see that the trail would significantly contribute to Orionid activity circa 1404 - 1000 BCE but not as much today.
  The evolution of the median location of every meteoroid trail ejected since 1404 BCE is presented in Appendix \ref{appendix:precession}. Each of the other simulated trails behaves very much like the 1404 BCE trail, with little difference as a function of the particle sizes.
 
\begin{figure}[!ht]
    \centering
     \includegraphics[width=0.5\textwidth]{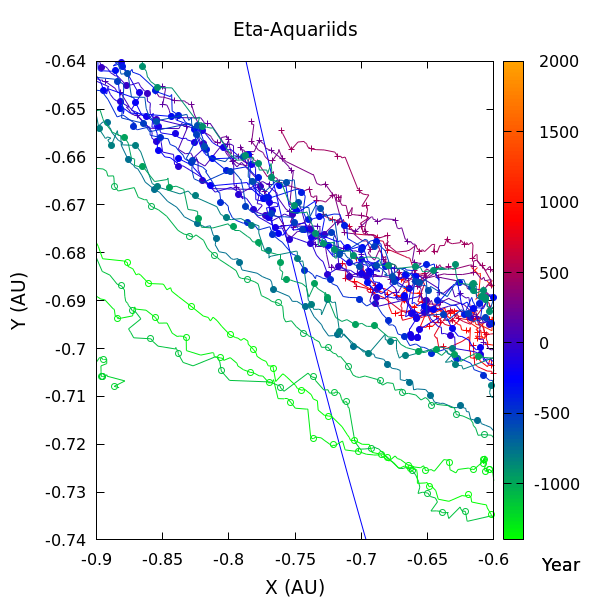}\\
     \includegraphics[width=0.5\textwidth]{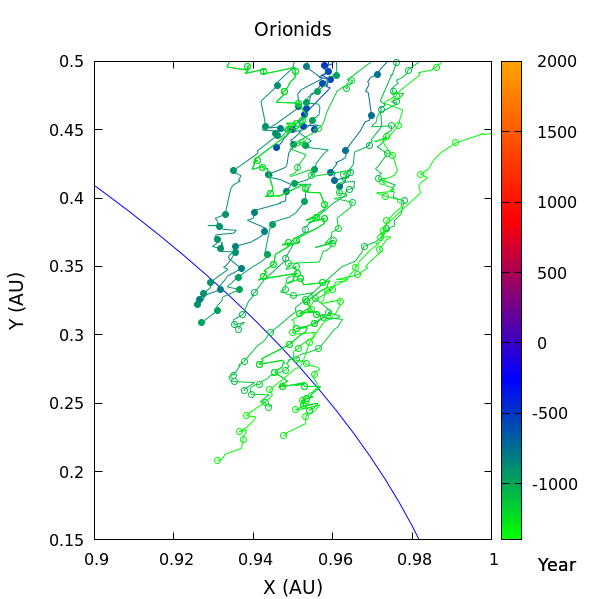}\\
     \caption{The evolution of the median location of the descending (top panel) and ascending (bottom panel) nodes for each trail ejected between 1404 BCE and 1986 CE, color coded as a function of the epoch. Empty circles refer to trails ejected prior to 1000 BCE, filled circles to trails ejected between 1000 BCE and 1 CE, crosses and plus symbol to trails ejected after 1 CE. Symbols mark the median location with a time step of 70 years ending in 2020 CE. Most trails have median locations outside the plot by 2020 CE. Particle sizes are 0.1 - 1 mm.}
     \label{fig:01_1mm_age}
  \end{figure}

  Figure \ref{fig:01_1mm_age} presents a zoom around Earth's orbit of the median node locations for trails composed of 0.1 mm to 1 mm particles. The median node trajectories are in agreement with the 2020 CE nodal footprints observed in Figure \ref{fig:nodes_2020_all_sizes}, confirming the picture that some particles remain on orbits which the comet was on a long time ago. When observing the intersection of the median ascending nodes with Earth's orbit, the gradient in age of the Orionids is not clear. Instead, we observe a separation between trails ejected after or prior to 1000 BCE. The existence of such an age gradient needs to be re-examined including the simulation of older meteoroid trail ejections from 1P/Halley. 
  
  \subsection{Time to disperse around the orbit}
  
  Up to this point, we have investigated the evolution of the meteoroid streams by looking at the particles' nodes at different epochs, which because of their long orbital periods are often somewhat different from the locations at which they  actually cross the ecliptic plane (their "nodal-crossing positions" as defined in Section \ref{section:integration}). 
  In the previous section, we found that trails ejected prior to 600 CE and 600 BCE could contribute to the present $\eta$-Aquariids and Orionids activity, respectively. However, this lower age-limit of the stream is valid only if these trails have had time to stretch along the whole orbit before the present epoch, which we verify below.
  
  For several ejection epochs and particle sizes, we estimate the time $T_S$ required for a newly ejected trail to entirely spread along the orbit. After ejection, we identify the average orbital period ($\mu$) and standard deviation ($\sigma$) of the simulated particles and consider two particles of periods $P_1=\mu + 5\sigma$ and $P_2=\mu - 5\sigma$. After $n$ revolutions, particle 1 lags particle 2 by $\Delta T=n(P_1-P_2)$. We therefore approximate $T_S$ to the time required to satisfy $\Delta T = P_1$, i.e. $T_S=P_1 P_2/(P_1-P_2)$. In this formulation the influence of radiation and gravitational perturbations, that could accelerate the particles' spread, is not taken into account. 

 For several trails ejected between 1404 BCE and 2050 CE, we estimate an average $T_S$ time of 310 years for 0.1 to 1 mm particles, 640 years for 1 to 10 mm particles and 1400 years for particles of 10 to 100 mm in size. The dependence of $T_S$ with meteoroid size is not surprising. Indeed, the combined influence of the radial radiation pressure and the meteoroids velocity relative to the nucleus (higher for small particles) tend to disperse the meteoroids immediately after ejection; the large meteoroids remain close to the nucleus for a longer period of time, while small particles extend ahead and behind the comet more quickly, a general result of the modification of the orbits of meteoroids ejected from parent comets \citep{Pecina1997}. 

  In our simulations, every trail ejected prior to 600 CE (for each size bin) has had time to spread along the whole orbit. The implication of these stretching time calculations is that it is reasonable to investigate the general evolution of the stream through the position of the particle nodes instead of the nodal-crossing locations for trails ejected before  600 CE. However this approach might not be valid to analyze the fine structure of the stream, which is conditioned by planetary perturbations.
  
 \section{Extracting particles crossing their nodes and/or impacting Earth} \label{section:impactors}
 
 The goal of our simulations is to convert our individual ejected test particles into equivalent shower activity at Earth. To accomplish this conversion we must decide which particles represent meteoroids which can intersect the Earth. Since many of our test particles are from very old ejections, we don't expect the exact location of such particles to be highly accurate except in a statistical sense. In this case, the osculating elements of the particle may be a less accurate but more robust measure of the intersection of trails of a certain age with the Earth's orbit. Using the nodal location based on the osculating orbit as described earlier can therefore provide a good global view of the age and extent of the overall trail, in particular since we have good number statistics. 
 
 In contrast, to achieve the highest fidelity in terms of actual particle numbers encountering the Earth in a given year, we must choose a volume box around the Earth in which to declare a particle an actual impactor. This population will have far fewer numbers, but better reproduce year to year variations in activity as this approach captures features such as meteoroids residing in mean motion resonances. We next describe the details of how we declare a simulated particle an impactor. 
 
   \subsection{Impacting particles}
 
 During the integration, particles that approach Earth within a distance $\Delta X = V_r \Delta T$ are considered impactors. Here  $V_r$ is the relative velocity between the planet and the particle and $\Delta T$ a time parameter depending on the accuracy of the shower prediction. For this study, $\Delta T$ was fixed at 12h, leading to a distance criterion of $1.88\times 10^{-2}$ AU for impactor selection. 
 
   \begin{figure*}[!ht]
     \centering
     \includegraphics[width=0.99\textwidth]{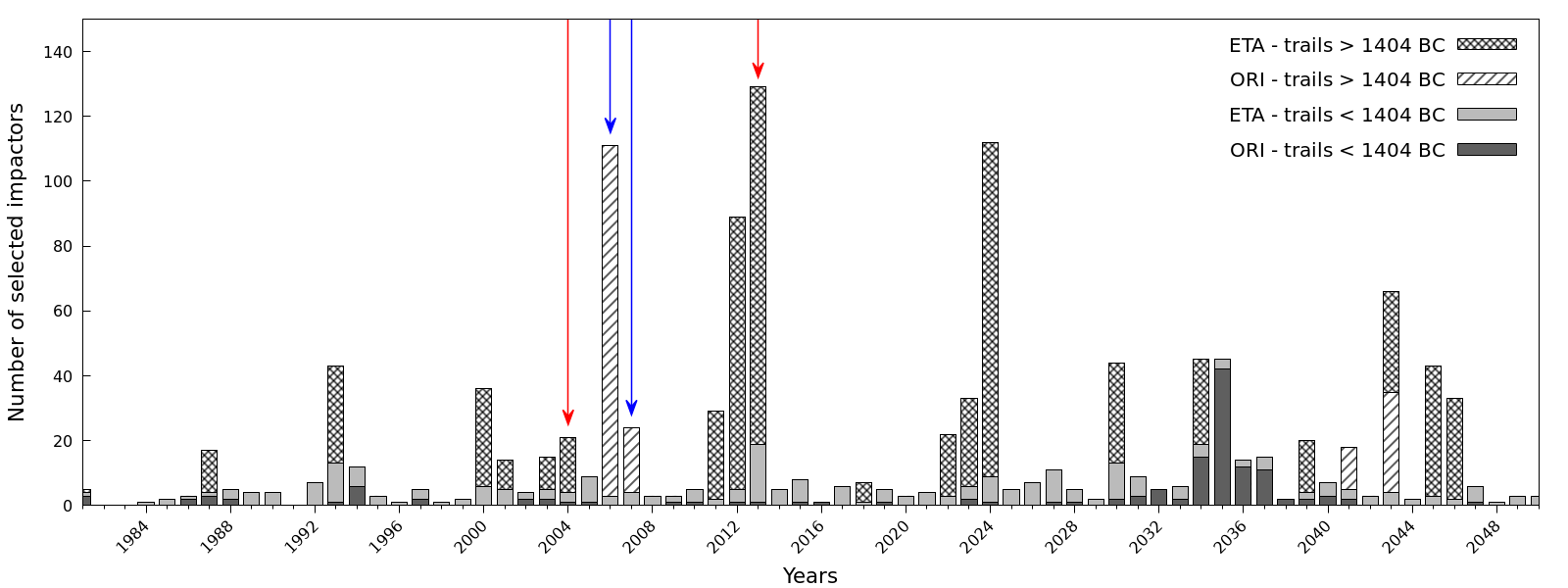}
     \caption{Number of particles identified as impactors between 1980 and 2050. Striped and dashed boxes refer to $\eta$-Aquariids and Orionids ejected after 1404 BCE while the solid-colored boxes represent particles ejected before 1404 BCE. Vertical arrows show the years of recent $\eta$-Aquariid (red) outbursts in 2004 and 2013 and Orionid (blue) outbursts in 2006/2007.}
     \label{fig:impactors}
 \end{figure*}
 
  \subsubsection{Statistics}

 Overall, about 4\% of the particles were flagged as impacting Earth and removed from the integration. Most of the impactors struck Earth when the MOID between the comet and the planet was at its lowest range ($<$0.02 AU), i.e. around 800 BCE for the Orionids and 500 CE for the $\eta$-Aquariids. Only 222 of the $\sim$4.5 million particles simulated were impactors over the period 1980 to 2019. 
 Among them, 62 particles approached Earth at the time of the Orionid meteor shower, and 160 at the time of the $\eta$-Aquariids. Most years, fewer than 5 particles met the criteria of a impactor. With such low numbers, no comparison of the simulations with observed meteor rates can be attempted directly; other methods to increase our statistics must be considered.
 
 When arbitrarily increasing the selection criteria $\Delta X$ to 0.05 AU (a factor of two increase), as commonly adopted for meteor shower forecasting \citep[cf. review in][]{Egal2020}, 1268 particles are retained as impactors over the period 1980 to 2050. The unweighted number of particles selected as a function of time for both showers is presented in Figure \ref{fig:impactors}. Despite the increase in $\Delta X$, most of the $\eta$-Aquariids and Orionids annual apparitions in this period contain less than 10 impacting particles. Due to the small number of particles retained and the lack of realistic weights applied to the simulations (cf. Section \ref{section:weights}), again no reliable intensity prediction can be generated from Figure \ref{fig:impactors}. So we again conclude that our statistics must be increased by other means. But first we will validate our simulations by comparing these initial impactors with observations.

 \subsubsection{Validation}
 
\begin{figure*}[!ht]
     \centering
     \includegraphics[trim={0cm 0 1cm 0.4cm},clip,width=\textwidth]{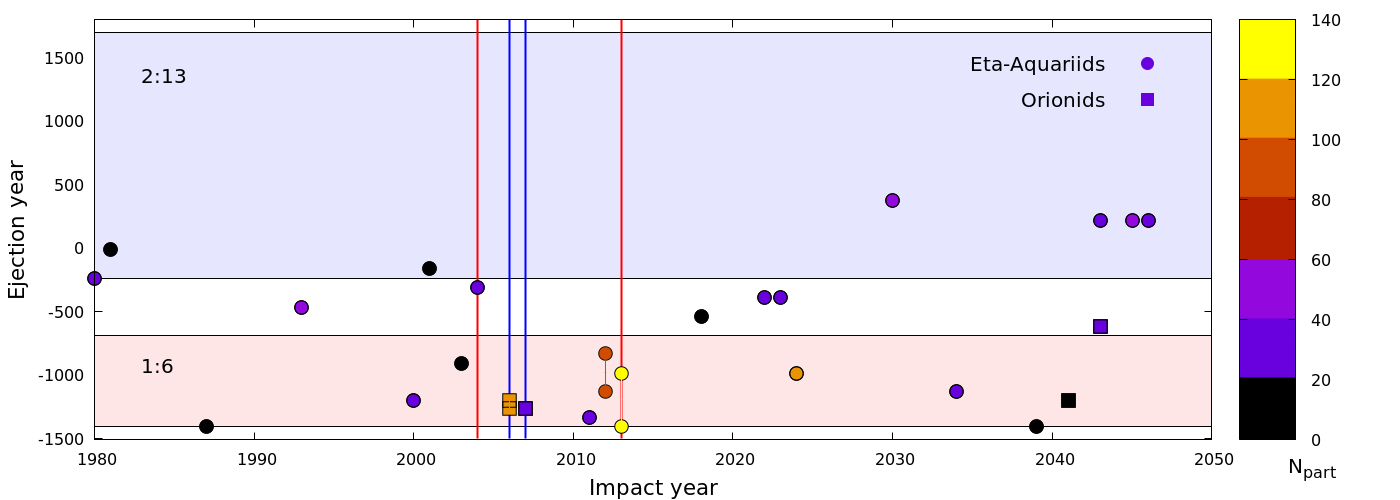}
     \caption{The particle ejection year from 1P/Halley as a function of impacting year at the Earth for the $\eta$-Aquariids (circles) and the Orionids (squares). Symbols are color coded as a  function of the number of impactors (see sidebar). When several trails from different ejection epochs are involved in producing impactors  (e.g., in 2012 and 2013), connected symbols frame the ejection period. The rectangles illustrate the periods when the ejection epoch occurred with 1P/Halley trapped in the 1:6 (red) and 2:13 (blue) MMR with Jupiter. Meteoroids ejected during these intervals may remain in the MMRs and form comparatively denser "clumps" in the stream. Vertical lines show the years of recent $\eta$-Aquariid (red) outbursts in 2004 and 2013 and Orionid (blue) outbursts in 2006/2007.}
     \label{fig:impactors_Yejec}
 \end{figure*}

 Figure \ref{fig:impactors} shows impactors between 1980 and 2020. To zeroth order, the larger the number of model impactors in a given year, the higher we expect $\eta$-Aquariid and Orionid activity. Almost all the particles impacting Earth in a given year were found to be ejected during a single apparition of the comet. The ejection year of the impactors since 1980 is shown in Figure \ref{fig:impactors_Yejec}. For each year (abscissa), the ejection dates of particles impacting Earth are marked with circles for the $\eta$-Aquariids and squares for the Orionids. The color of the symbol represents the number of simulated impactors for that year (low activity in black, strong activity in yellow - see sidebar). The red and blue rectangles illustrate time periods when comet 1P/Halley was trapped in the 1:6 and 2:13 MMR with Jupiter. 
 
 Our model reproduces the basic features of the streams. In particular, no model particles ejected after 600 CE and 800 BCE become $\eta$-Aquariids or Orionids (respectively) on Earth at the current time, a finding we expect because of the evolution of 1P/Halley's node away from the Earth after these times. In addition, the years with higher numbers of model impactors  (i.e. 2013-2014 for the $\eta$ Aquariids and 2006-2007 for the Orionids) correspond to ejecta potentially trapped in the 1:6 resonance, as predicted by \cite{Sato2007}, \cite{Rendtel2008}, \cite{Sato2014} and \cite{Sekhar2014}. 
 
These first comparisons are encouraging and suggest our basic model captures the important details of the stream. However no quantitative comparison with meteor observations can be performed without weighting the simulated meteoroids and increasing the number of particles selected. Section \ref{sec:postdiction} addresses this issue. 

  \subsection{Annual nodal crossings distribution}

One approach to increase particle statistics in our analysis is to increase the temporal criterion $\Delta T$. The temporal $\Delta T$ selection amounts to replacing each simulated particle $p_0$ by a swarm of particles on the same orbit, centred around $p_0$, and crossing the ecliptic plane at $p_0$'s nodal location during a period of $2\Delta T$.  Small $\Delta T$ values (of a few hours or a few days) are required to reproduce short-lived activity variations of a simulated meteor shower. On the other hand, high $\Delta T$ parameters (of several months or even years) are useful for the investigation of the general evolution of a meteor shower's activity, incorporating good statistics but at the expense of temporal resolution.
 
 In this section we consider a large $\Delta T$ criterion of half of a year, meaning that we retain all the test particles crossing the ecliptic plane less than 0.05 AU from Earth's orbit over an entire year (i.e., $\pm$ 6 months either side of the time when the Earth intersects the shower node). With this relaxed selection threshold, our number statistics improve dramatically. 
 
 We find a few hundred to a few thousand particles belonging to the $\eta$-Aquariids are retained every year (at least 200). For the Orionids, at least a few tens (about 40) and sometimes a few hundred particles per year encounter the Earth over the period 1980 to 2050. In some specific years (e.g., 2002, 2003, 2004, 2014 and 2015) no particles ejected after 1404 BCE approach Earth at the time of the Orionids. 
 
  \begin{figure*}[!ht]
      \centering
      \includegraphics[trim={0cm 0.2cm 0cm 0.4cm},clip,width=\textwidth]{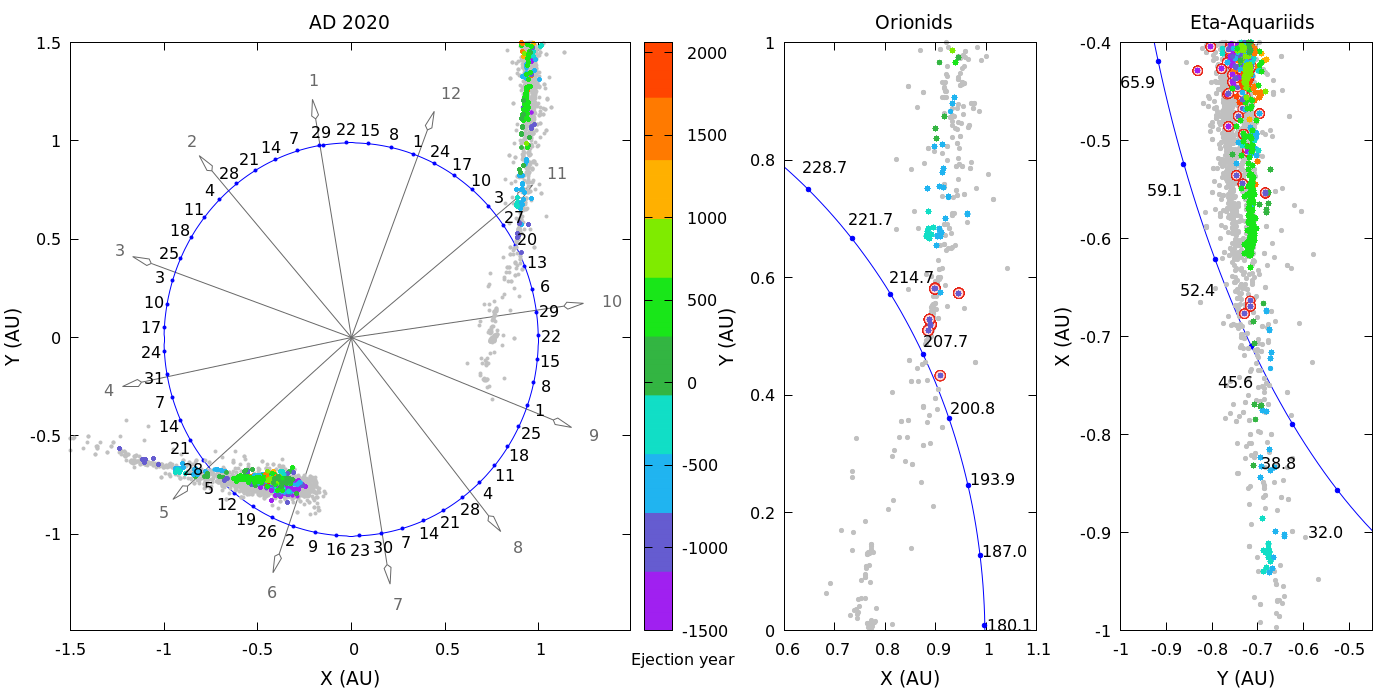}
      \caption{[Animated Figure] An example of the distribution of nodal locations of particles when crossing the ecliptic plane in 2020 CE as a function of their time of ejection from 1P/Halley. The regions where meteoroids have their ascending node (Orionids) and their descending node ($\eta$-Aquariids) are shown in magnified form in the middle and right panel of the figure. The position of Earth along its orbit is represented by blue dots and dates (gray arrows indicating month and dots the day of each month) or solar longitudes (middle and right panel). Particles are color coded as a function of the ejection epoch. Meteoroids ejected when the comet was trapped in the 1:6 resonance are highlighted with red circles. Trails ejected between 3000 BCE and 1404 BCE are presented in gray as these are associated with the time frame when 1P/Halley has a less certain ephemeris. The animated figure is available at \url{http://astro.uwo.ca/~wiegert/Halleyids/Halleyids_animated_nodes.mp4}.}
      \label{fig:2020_nodal_crossing}
  \end{figure*}
  
 Figure \ref{fig:2020_nodal_crossing} shows the stacked nodal crossing locations of particles traversing the ecliptic plane in 2020. An animation of the nodal crossing locations of our simulated particles ejected between 1980 and 2050 is available at \url{http://astro.uwo.ca/~wiegert/Halleyids/Halleyids_animated_nodes.mp4}. 
 The  particles are color coded as a function of the ejection epoch. Meteoroids ejected when the comet was trapped in the 1:6 resonance are highlighted with red circles. 

 These plots show an overview of the years of potentially enhanced activity together with the age of the trails involved. From the raw number of nodes in the simulation located around Earth's orbit, enhanced Orionid rates are expected in 2006 and 2007, along with relatively lower rates in 1997-1998 and 2010. The annual number of particles reaching the descending node suggests increased activity for the $\eta$ Aquariids in  1994, 2003-2004, 2012-2013, 2023-2024, and 2045-2046. 
 
 These probable outburst years based on the nodal-crossing locations are in good agreement with years showing higher numbers of impactors (cf. Section \ref{section:impactors}). 
 From the animation, we also see that the density of the nodal footprint around Earth is sufficient to produce meaningful activity profiles of the $\eta$-Aquariid meteor shower every year. With the selected $\Delta X$ and $\Delta T$ values of 0.05 AU and $\pm$ 6 months, several Orionid activity profiles around the outburst years have sufficient particle numbers for good profiles, but further expansion of number statistics using different methods are required to produce reliable predictions of Orionid profiles and peak intensity in other years.
 
\section{Postdictions} \label{sec:postdiction}

The final step in our modeling is to calibrate our model and determine particle weighting for the Halleyids meteoroid streams through direct comparison with meteor observations. We first compute synthetic intensity profiles (model shower activity as a function of time around the maximum of each shower, each year) from our simulations. This provides a quick analysis of the meteor showers' shape, duration, average activity and year to year variations to compare to the measurements provided in \citet{Egal2020b}. In addition, the simulated radiants and particles' mass and size distributions at Earth can be compared to meteor observations, all of which both constrain and validate the simulations.   

 \subsection{Simulated activity profiles}
 
 \subsubsection{Computation} \label{section:weights}
 
 With the $\Delta X$ and $\Delta T$ values selected in the previous sections, tens to thousands of particles crossing the ecliptic plane every year are included for activity profile computations. Each particle is assigned a solar longitude, corresponding to the date when the Earth is the closest to the particle's node. Within a given solar longitude bin, the total number of particles is converted into an equivalent flux $\mathcal{F}$, which is then transformed into a ZHR by \cite{Koschack1990}'s relation: 
 \begin{equation}
 \textrm{ZHR}=\frac{\mathcal{A_\text{s} F}}{(13.1r-16.5)(r-1.3)^{0.748}}
 \end{equation}
  \noindent where $\mathcal{A_\text{s}}$ is the typical surface area for meteor detection by a visual observer in the atmosphere at the ablation altitude ($\mathcal{A_\text{s}}\sim 37200\; \mathrm{km}^2$) and $r$ is the measured differential population index. We assume a common population index of 2.5 for the two showers. This lies in between the values of 2.46 and 2.59 generally measured for the $\eta$-Aquariids and the Orionids \citep{Egal2020}.
  
  For the ZHR model estimates to be meaningful we first require a realistic, physical simulated flux. This is accomplished by weighting each particle as discussed in Section~\ref{section:trails_simulated} through calibration against measurements. 
  
  The initial weighting scheme applied to our simulation follows the methodology detailed in \cite{Egal2019}. In this approach, each particle weight depends on: 
  \begin{enumerate}
      \item The initial number of particles of a given radius ejected at a given epoch,
      \item 1P/Halley's dust production at the heliocentric distance of ejection, measured via Af$\rho$,
      \item The differential size frequency distribution of the particles radii after ejection, characterized by the differential size index $u$.
  \end{enumerate}
  In practice, the final weight is normalized by the number of particles sharing the same ejection characteristics (epoch of ejection and size) and is increased for low heliocentric distances as the comet dust production peaks around the perihelion.  The weighting varies significantly as a function of the differential size index $u$ at ejection which is used. For additional details about the weighting scheme, the reader is referred to \cite{Egal2019}.
  
  Given our spatial and temporal accuracy, the aforementioned weighting scheme proves to be effective when the raw number of simulated particles in a shower profile exceeds a few hundred (i.e. in the case of the $\eta$-Aquariids analysis). For cases with fewer simulated particles, additional empirical weights were found to improve the quality of our postdictions. These additional weights are somewhat ad hoc and were found semi-empirically but are physically motivated. They are necessary to obtain sufficient statistics. The three complementary weights applied to our simulations are: 
  
  \begin{enumerate}
      \item A coefficient $C_1$ proportional to the inverse distance of the particle's node to the Earth's orbit. The somewhat large $\Delta X$ criterion of 0.05 AU, which increases the number of simulated particles selected, is corrected by this coefficient that gives more importance to nodes close to Earth's orbit. This weight is an approximation of the impact parameter described in \cite{Moser2008}.
      \item A coefficient $C_2$ proportional to $1/n$, with $n$ the number of revolutions completed by the trail since its ejection. This parameter, similar to the $fM$ coefficient of \cite{Asher1999},  gives more weight to young trails that are known to be more dense than older streams.  The inclusion of this corrective factor has been considered in past meteoroid stream simulations, for example  by \cite{Watanabe2008} to model the Draconid meteor shower.
      \item A coefficient which is a function of the time $\delta T_p$ between the time a particle crosses the ecliptic plane and the date of the prediction (by definition, $\delta T_p <\Delta T$). As noted in the previous section, a $\Delta T$ of 0.5 years leads to no simulated particles available for certain Orionids apparitions. so this parameter is increased to 1.5 years in some case (amounting to concatenation of three consecutive years of simulations at the epoch being investigated). The importance of these additional particles is however decreased by a weight of $\gamma^{\delta T_p}$, with $\gamma$ a positive integer to be determined in our calibration fits. 
  \end{enumerate}
  
   The inclusion of these additional weights has only a small effect on years with high predicted activity, but considerably improved our results for years with only a few simulated particles approaching Earth's orbit. Of these three complementary weights, the last correction is arguably the least justified. However, we found that the inclusion of these additional particles and their relative weighting when determined empirically does improve model agreement with measurement of several apparitions of the Orionids. We also note that stacking several years of nodal crossing is a common practice in meteor shower simulations; this method has been used by several authors to study the general structure of streams when limited by small-number statistics \citep[e.g.,][]{Jenniskens2008,Segon2017}. 
  
In summary, the weighting scheme applied to our raw model simulations has three tunable parameters, namely the particle size distribution index at ejection, $u$, the contribution of particles crossing the ecliptic plane at different dates $\gamma$, and a final normalization coefficient $K_2$ used to scale all the ZHR predictions to the observed activity level of the meteor showers as in \citep{Egal2019}. Since our simulations aim to model the Halleyids meteoroid streams self-consistently, we apply the exactly same weighting scheme to the simulated particles comprising both the $\eta$-Aquariid and the Orionid showers. In this way we can test the robustness of the model fits by limiting the weighting coefficients to be the same for the two showers as expected to be the case on purely physical grounds. 
  
 \subsubsection{Calibration}
 
 The best weighting solution is determined by calibrating the activity profiles on past observations of the meteor showers. At present, the longest and most consistent continuous monitoring of the Halleyids is provided by the Canadian Meteor Orbit Radar (CMOR), operational since 2002 \citep{Egal2020b}. Our simulated activity profiles are calibrated to CMOR observations at 29 MHz, first for the $\eta$-Aquariid meteor shower (which has more particles) and then refined with the Orionids simulations. 
 
 The determination of the weights was performed in a semi-automated way. We first estimated the value of the particles' size distribution index $u$ using a Particle Swarm Optimization (PSO) algorithm \citep{PSO}. The PSO was run with two different cost functions, related respectively to the profiles main peak ZHR and shape. The first cost function compared the computed and observed maximum ZHR reached for all the apparitions between 2002 and 2019. The second cost function reflected the difference between the observed and the simulated activity profiles when the main peaks are scaled to the observed maximum at each apparition of the shower. 
 
 With the PSO, we found the $u$ value that best reproduces the shower activity profiles' shape to be 3.4 (corresponding to a mass index of 1.80), in good agreement with Giotto in-situ measurements close to the nucleus (see Section \ref{sec:dust}). On the other hand, we found the interannual variations of the main peak intensity are better modeled with a size index of 3.9 (i.e. a mass index of 1.97). The best $u$ was also found to vary both as a function of the number of apparitions considered in the calibration and the specific years investigated. Since one of the main goals of this work is to predict future $\eta$-Aquariids and Orionids outbursts, we adopted a size index $u$ of 3.9 (which provided a better match to the peak intensities) for the rest of this study.
 
 The application of the three additional weights ($C_1$, $C_2$, and $\gamma$) was performed manually in a second step. 1) The weight modulation as a function of the particle's nodal distance or ejection epoch was selected so as to improve specific Orionid apparitions. 2) Each particle is divided by the approximate number of revolutions it has performed since its ejection $n$. This operation was found to improve the average level of the $\eta$-Aquariids maximum activity between 1985 and 2020. 3) The $\gamma$ coefficient mainly affects the year to year variations of the main peak ZHR, which are reduced as $\gamma$ decreases. We find the best value to lie around $\gamma=3$, which does not affect the $\eta$-Aquariid activity variations but improves the statistics of lower-activity Orionid apparitions. 
 
 \subsection{Results}
 
  In the following subsections, we investigate the agreement of our simulated activity profiles with video, visual and radar observations of the $\eta$-Aquariid and Orionid meteor showers presented in \cite{Egal2020b}. In this analysis, we consider every simulated trail ejected since 3000 BCE as described in Section \ref{section:trails_simulated}. The inclusion of trails ejected before the 1404 BCE apparition, that helps improve our results in some cases, is discussed in Section \ref{section:discussion}.

 \subsubsection{Average activity profile} \label{sec:average}

\begin{figure}[!ht]
  \includegraphics[width=0.48\textwidth]{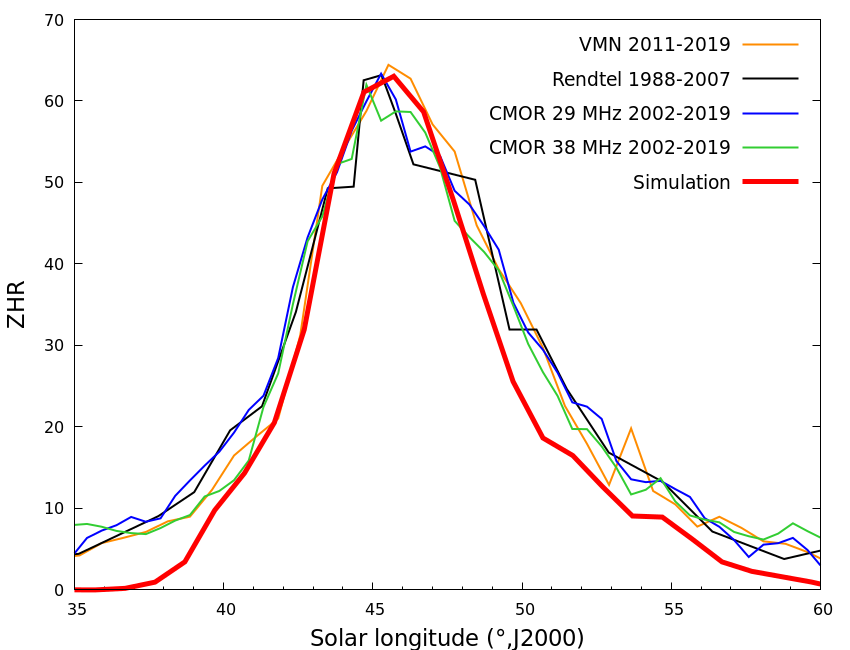}
  \includegraphics[width=0.48\textwidth]{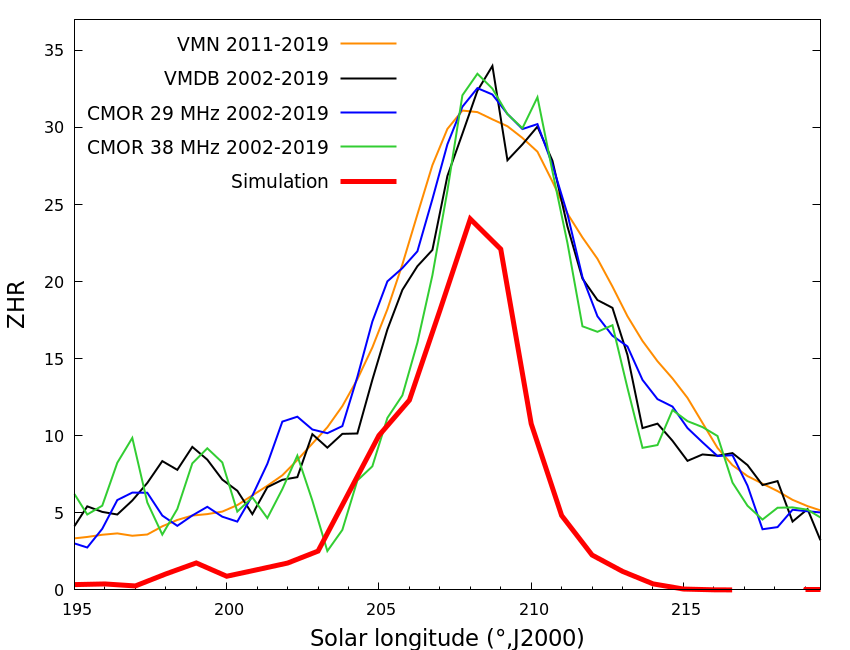}
   \caption{Average activity profile of the $\eta$-Aquariid (top) and Orionid (bottom) meteor showers, as measured with the CMOR 29 MHz system (blue) and the 38 MHz system (green) between 2002 and 2019, the VMN (orange) between 2011 and 2019, and visual observations from the IMO between 1988 and 2007 \citep{Rendtel2008b} or 2002 and 2019 (VMDB in black). Video and radar observations were scaled to visual measurements as described in \cite{Egal2020b}. The best-weighted simulated average profile using one degree solar longitude binning is drawn in red. Note the factor of two difference in the ordinate scale between the top and bottom plots.}
   \label{fig:average_profile}
 \end{figure}

   The average activity profile of the simulated $\eta$-Aquariid and Orionid apparitions between 2002 and 2019 is presented in Figure \ref{fig:average_profile}. The average profiles as measured with the CMOR 29 MHz and 38 MHz systems, with the IMO Video Meteor Network (VMN) and visual observations are plotted for comparison. The simulated activity profiles were scaled to the observations of the $\eta$-Aquariids shower, and the same scaling factor was applied to the modeled Orionids profile. 
  
   The average $\eta$-Aquariid profiles generated from the simulations are in good agreement with the observations. The modeled activity peaks around 45.5$\degree$ SL, with the rise of activity more sudden than the post-maximum decrease in intensity in accordance with measurements. The simulation's post-maximum activity matches that of the observed profiles well, while the pre-maximum shows some  divergence at solar longitudes below 42$\degree$. The simulated profile underestimates the total shower duration (of around 25 days) by about 5 days, a potential indicator that the shower contains material even older than the oldest included in our simulations (3000 BCE). 
  
  With the same scaling factors, the average Orionid activity deduced from our simulations peaks around 25 meteors per hour, i.e. about a factor 2.5 times less than the $\eta$-Aquariids. Though our simulations do not exactly reproduce the observed activity ratio between the two showers (of about 2), we obtain for the first time comparable meteor rates with a self-consistent stream model. The simulated profile peaks around the 208$\degree$ to 209$\degree$ SL at the observed location of the first peak of activity, but underestimates the plateau and falling portion of the activity curve. The total duration of the shower is also underestimated by about 12 days. These mismatches between the model and shower are an even stronger indication that a more significant fraction of the meteoroids comprising the Orionids are from ejections earlier than 3000 BCE compared to the $\eta$-Aquariids.
  
  In summary, the good match between our simulated profiles and the observations suggest that most of the $\eta$-Aquariids activity can be reproduced with trails ejected since 3000 BCE. On the other hand, as expected from the discussion in previous sections, the Orionids average profile can not be modeled completely with the same trails, and require the addition of older material. Trails ejected after 3000 BCE contribute to the core of the Orionid activity profile, but the width of the profile and the second maximum of activity reported around 210$\degree$ SL are probably caused by material ejected before 3000 BCE. 
  
    \subsubsection{Average mass index}
    
\begin{figure}[!ht]
  \includegraphics[trim=0 0.4cm 0 0,clip,width=0.46\textwidth]{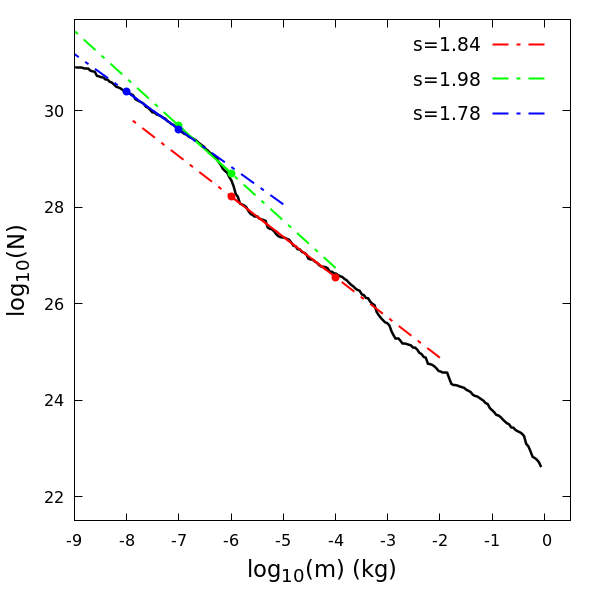}
  \includegraphics[trim=0 0.4cm 0 0.4cm, clip, width=0.46\textwidth]{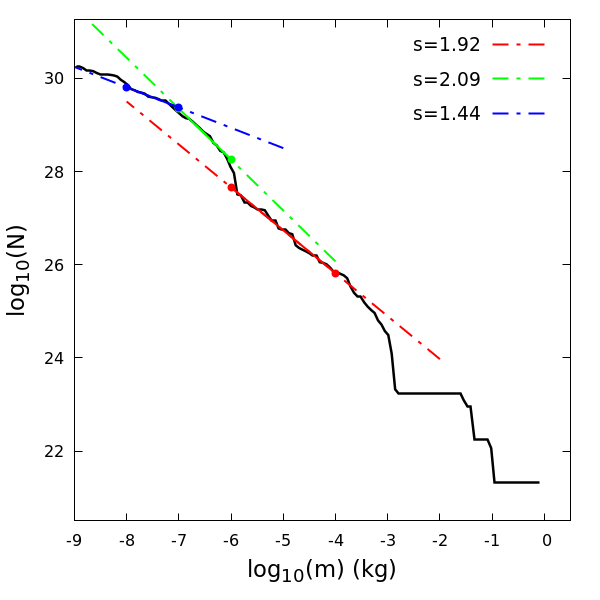}
   \caption{The modeled mass distribution of the $\eta$-Aquariids (top) and Orionids (bottom) based on simulated meteoroids encountering the Earth between 2002-2019 near the time of the main peak of activity. The logarithm of the cumulative number of simulated meteors (weighted) is drawn as a function of the logarithm of the particles mass (in kg). Estimates of the differential mass index $s$ in the visual (red), video (green) and radar (blue) mass ranges are provided for comparison. See text for details.}
   \label{fig:average_s}
 \end{figure}

  As we generate absolute fluxes from our model scheme, we are also able, for the first time for any shower model to our knowledge, to generate predictions of the synthetic mass distribution of a shower at the Earth. The cumulative mass distribution of the simulated showers, computed based on weightings of all the particles encountering Earth between 2002 and 2019 is presented in Figure \ref{fig:average_s}. Only particles contributing to the main peak of activity of the $\eta$-Aquariids (44$\degree$-46$\degree$ SL) and Orionids (208$\degree$-2010$\degree$ SL in the simulations) were included. The simulated mass distributions display breaks and changes in the slope, though an overall rough power-law trend is evident. This is particularly true for the Orionids, because of the smaller number of simulated particles in that sample. 
  
 The differential mass index $s$ of each shower is estimated
 over the mass ranges $[10^{-4},10^{-6}]$ kg, $[10^{-6},10^{-7}]$ kg and $[10^{-7},10^{-8}]$ kg, corresponding to visual (M$\sim[-2,+3]$), video (M$\sim[+3,+6]$) and radar (M$\sim[+6,+8]$) magnitudes. Average $s$ values estimated from the model span from 1.78 to 1.98 for the $\eta$-Aquariids and from 1.44 to 2.09 for the Orionids, in relatively good agreement with observations of these showers (cf. Table \ref{table:average_s}). 

 Estimates in the visual range, of respectively 1.84 and 1.92, are particularly consistent with the average value of 1.87 reported by \cite{Dubietis2003} and \cite{Rendtel1997}. 
 A key prediction of the modeled mass distribution is that while the overall distribution is roughly a power law, the specific power law value which is obtained depends very much on the mass interval used for measurements. In particular, in the radar and video mass ranges, our modeling suggests significant kinks in the distribution. This naturally can lead to quite different mass index values depending on the exact lower and upper bound to the mass range being measured and may account for some of the variance in mass index for the showers reported in the literature. 
 
  \begin{table}[!ht]
     \centering
     \begin{tabular}{lllll}
     \hline
     \hline
     Shower & Optical & Ref. & Radar & Ref. \\
     \hline
     $\eta$-Aquariids & 1.84 - 1.98 & [1] & 1.78 & [1]\\
     & 1.72 - 1.94 & [2, 3] & 1.75 - 1.95 & [4, 5] \\
     \hline
    Orionids & 1.92 - 2.09 & [1] & 1.44 & [1]\\
     & 1.46 - 1.96 & [3, 6]& 1.65 - 1.95 & [4, 7]  \\
     \hline     
     \hline    
     \end{tabular}
     \caption{Compilation of simulated and observed mass indices of the $\eta$-Aquariid and Orionid meteor showers, in the optical (M$\sim[-2,+6]$) and radar (M$\sim[+6,+8]$) magnitude ranges. This table summarizes the measured mass indices presented in Section \ref{section:halleyids}. For more details, the reader is referred to the review in \cite{Egal2020b}.
     References: [1] simulated (this work), [2] \cite{Rendtel1997}, [3] \cite{Dubietis2003}, [4] \cite{Blaauw2011}, [5] \cite{CB2015}, [6] \cite{Rendtel2008} and [7] \cite{Schult2018}.}
     \label{table:average_s}
 \end{table}    

  \subsubsection{Annual variation}

The ZHR of the main peak of each simulated $\eta$-Aquariid and Orionid apparition between 1985 and 2020 is presented in Figure \ref{fig:past_annual_variation}. The showers' individual activity profiles between 2002 and 2019 are provided in Appendix \ref{appendix:individual_profiles}. As noted in Section \ref{sec:average}, our model slightly underestimates the observed average activity level of the Orionids, potentially a consequence of it being an older shower compared to the $\eta$-Aquariids. In Figure \ref{fig:past_annual_variation} and Appendix \ref{appendix:individual_profiles}, the overall modeled and observed profiles have been rescaled so they reach the same maximum level. Because of this correction, the relative intensity of the two simulated showers which from section \ref{sec:average} is about 2.5 is not respected anymore. However, we find this scaling necessary to compare the predicted and observed activity profiles shapes and maximum rates with clarity and is likely due to the model not including older ejections which make up proportionately more of the Orionids.

 \begin{figure*}[!ht]
  \includegraphics[width=0.99\textwidth]{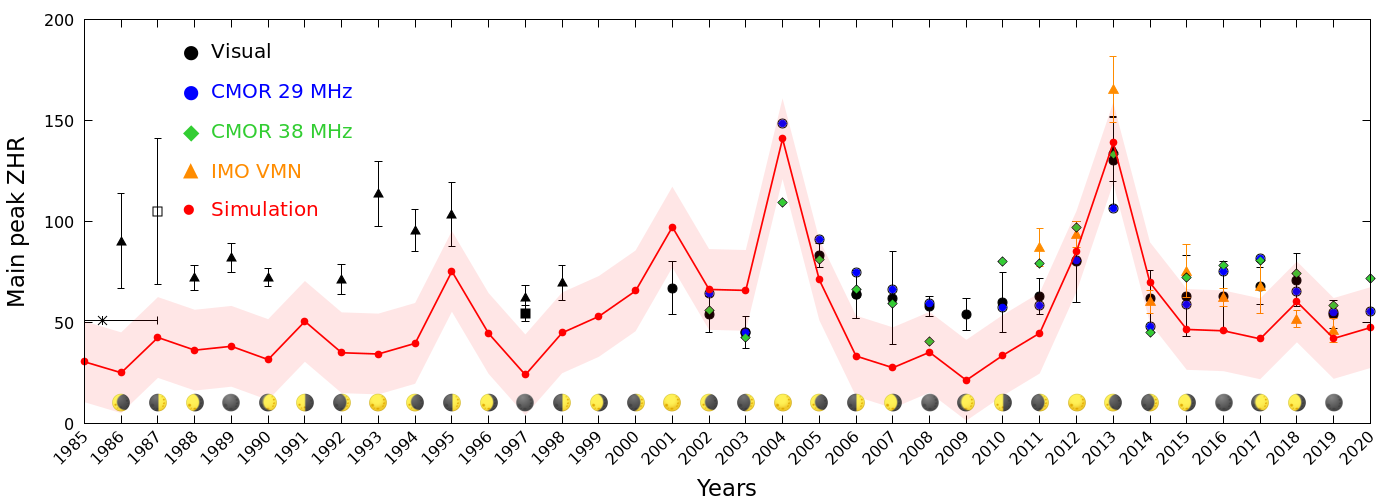}\\
  \includegraphics[width=0.99\textwidth]{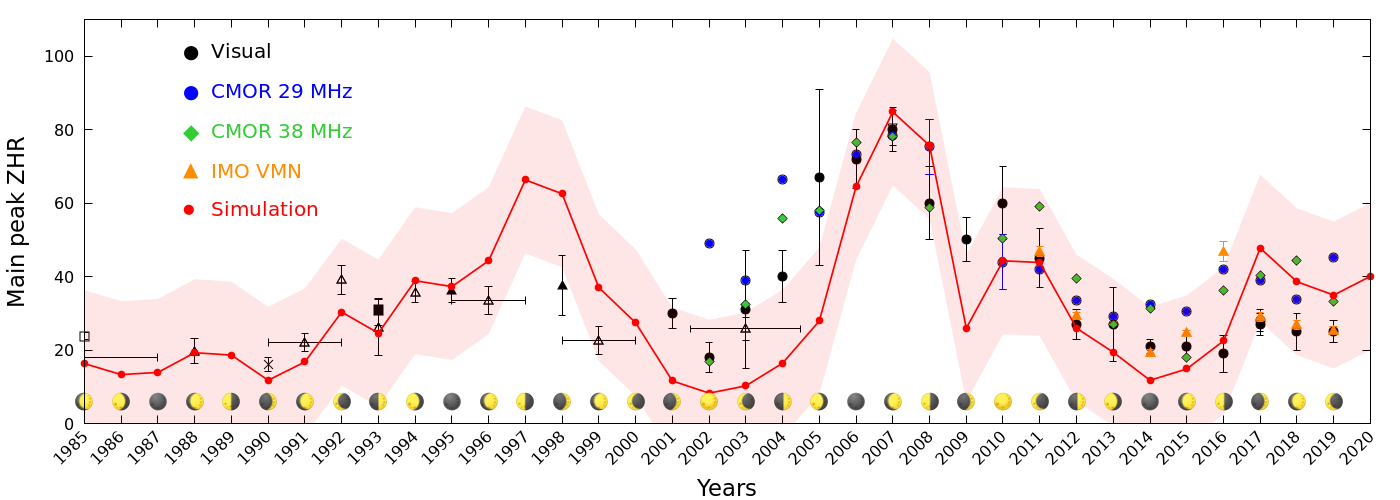}
   \caption{Annual variations of the $\eta$-Aquariids (top) and Orionids (bottom) peak ZHR for the years 1985 - 2019. The maximum rates recorded by CMOR 29 MHz (blue dots) and 38 MHz (green diamonds) are plotted along with the results of the VMN system (orange triangles) and visual observations (black dots) as in \cite{Egal2020b}. The envelope of estimated simulated peak variations, with an arbitrary uncertainty of $\pm$20 meteors per hour allowing to frame most Orionids yearly maximum ZHR rates, are presented in red. The phase of the Moon for each year at the time of the shower peak is also shown as this significantly affects the quality and quantity of visual and video data.}
   \label{fig:past_annual_variation}
 \end{figure*}
 
The yearly variations of the simulated $\eta$-Aquariid maximum ZHR in Figure \ref{fig:past_annual_variation} are in general good agreement with the observations. The selected weighting scheme reproduces the observed 2004 and 2013 outbursts, as well as the overall ZHR variations observed since 2000. On the other hand, this model underestimates the shower activity prior to 1998, between 2006 and 2010, and in 2017. In our simulations, these apparitions are more dependent to the presence of trails ejected before 1404 BCE, which we recall are in era where the ephemeris for 1P/Halley is less certain. 

We can improve our  agreement with the maximum $\eta$-Aquariid ZHRs observed prior to 1995 by increasing the weights of older trails in our simulations. However, a consequence of this improvement is to raise our models' 2003 and 2014 ZHR by a factor of about 1.5. Since the purpose of this work is to predict future Halleyids outbursts and their peak activity, we prefer to underestimate the usual intensity level of the shower than to lose reliability around the years of expected enhanced activity. This situation demonstrates that our weighting scheme is subjective and not unique but is conditioned by which observations are used and most emphasized in the calibration. 

For the Orionids, despite the incompleteness of the activity profiles above 210$\degree$ SL, our simulations reproduce the variations of the Orionids main peak intensity prior to 1996, around the 2006 and 2007 outbursts, and since 2010. 
ZHR values at other times may be underestimated, with the exception of 1998 apparitions that is significantly overestimated. In the IMO Visual Meteor DataBase (VMDB), we find the observed maximum Orionid rate in 1997 reached 27 meteors per hour\footnote{https://www.imo.net/members/imo\_live\_shower?shower=ORI\&year=1997}, suggesting that our model also overestimates the shower activity for that particular year.
As was the case for the $\eta$-Aquariids, modifications of the weighting scheme (and especially the corrective factor depending on the particles nodal distance) would allow us to remove the excess intensity in 1997 and 1998. However, such modifications also worsen the fit for other apparitions of the two showers, and were not retained here.

For most apparitions since 1985, our chosen model reproduces the reported maximum ZHR to within 30\% or less. This translates to an absolute error of about 20-30 meteors per hour for each shower, depending on the meteor observation considered (this is indicated by the red region of Figure \ref{fig:past_annual_variation}). We therefore consider this model successful in reproducing the global yearly activity variations of the $\eta$-Aquariids and Orionids.

  \subsubsection{Individual apparitions} \label{sec:individual_apparitions}

  The strengths and limitations of our model  are well described by the comparison of the simulated average profiles and year to year ZHR variations with meteor observations discussed previously. However, individual apparitions of the $\eta$-Aquariids and Orionids are also of interest, and some of them will be briefly detailed in this section. 
  
  As expected from the average activity profile, our simulated apparitions of the $\eta$-Aquariids (see Appendix \ref{appendix:individual_profiles}) are generally most consistent with observations. Except for 2019, the simulated maximum of activity matches the reported main peak, though with some dependence on the observations considered. In several years the model even reproduces the shape of the observed profile in some detail (e.g., in 2012, 2015, 2016, 2017, etc.). 
  
  The model is less successful in reproducing the Orionids profiles, especially for solar longitudes above 210$\degree$. We interpret this as a likely consequence of it being more composed from ejections predating our oldest modeled ejections. For this reason, the $\sim$12 year periodicity of the Orionids main peak location \citep{Egal2020b} can not be reliably investigated with our simulations. The pre-maximum branch of activity is in general in good agreement with the observations, with some notable exceptions (e.g., in 2002-2003, 2008 and 2014), suggesting it is relatively composed of younger material. However, we note that the observed profiles from the various detection networks often differ as well, so varying observational sensitivities may also account for some of the differences.
  
  The characteristics of the modeled outbursts ($\eta$-Aquariids in 2004 and 2013, Orionids in 2006 and 2007) are detailed in Appendix \ref{appendix:outbursts}. The activity profiles of the showers are presented along with the trails involved and the cumulative mass distributions. In these four cases, the simulated profile reproduces well the intensity, duration and shape of the observed activity. The mass distributions are variable and $s$ indices also vary with the range of masses considered (cf. figures in Appendix \ref{appendix:outbursts}).
  
  In agreement with \cite{Rendtel2007}, \cite{Sato2007} and \cite{Sekhar2014}, we find several trails ejected while 1P/Halley was in the 1:6 resonance to be responsible for the 2006 and 2007 Orionid outbursts.  In our model, trails ejected in 1196 BCE and 1264 BCE are particularly strong contributors to these years. We also find multiple trails ejected after 3000 BCE to have moderate contributions to the 2004 and 2013 $\eta$-Aquariids outbursts. The main contributor to the 2004 outburst is the 314 BCE trail, and to a lesser extent, the 910 BCE and 1964 BCE streams. In our model, the 1128 BCE trail is the principal source of the 2013 outburst, together with particles ejected in 2731 BCE, 1196 BCE, 835 BCE and 762 BCE (among others). This result slightly diverges from \cite{Sato2014}, who identified the 1196 BCE and 910 BCE trails as the main cause of this outburst. 
  
 \subsection{Radiant}

\begin{figure}[!ht]
     \centering
     \includegraphics[trim=0 1cm 0 1.5cm, clip, width=.46\textwidth]{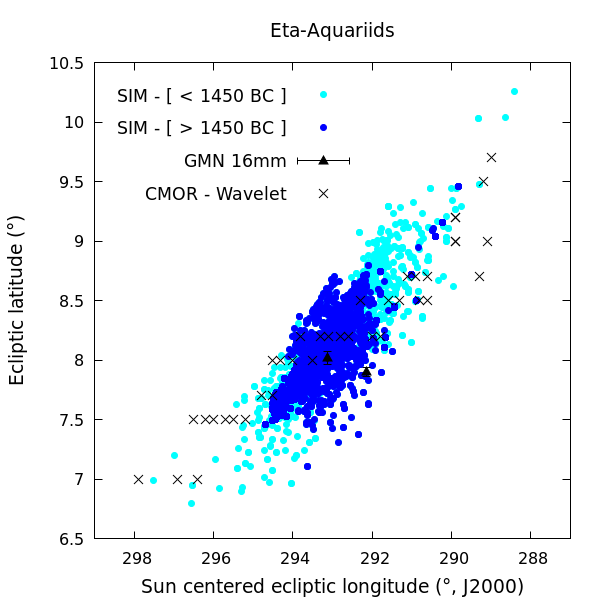}
     \includegraphics[trim=0 0.1cm 0 1.5cm, clip, width=.46\textwidth]{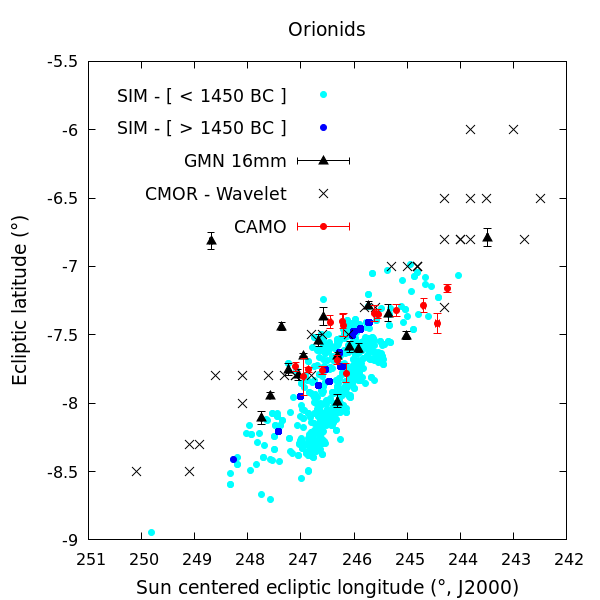}
    \caption{The radiants of the $\eta$-Aquariids (top) and Orionids (bottom) measured with CAMO (red circles) and the GMN (black triangles) video networks in 2019.  The average shower radiant location in one degree solar longitude bins determined from the wavelet analysis of CMOR data is identified with black crosses. For GMN and CAMO the formal radiant uncertainty was found using the method of \citet{vida2020a}. Simulated radiants computed for particles ejected before and after 1450 BCE are plotted with cyan and dark blue circles, respectively.}
    \label{fig:radiants}
 \end{figure}

 Another way to validate our model with real observations is through comparing the meteor radiants. Measurements of the 2019 $\eta$-Aquariids (top panel) and Orionids (bottom panel) recorded by video cameras of the Canadian Automated Meteor Observatory \citep[CAMO, see][]{VidaPHD} and the Global Meteor Network \citep[GMN, cf.][]{Vida2018,Vida2019} are presented in Figure \ref{fig:radiants}. The figure includes the general radiant location computed from CMOR using stacked data from 2002-2016, as well as the modeled 2019 radiants of particles ejected from 1P/Halley before (cyan) or after (dark blue) 1450 BCE.  
 
 For both showers, the simulated radiants of trails ejected after 1450 BCE are located close to the observed radiants. The area delineated by the radiant dispersion of all the modeled $\eta$-Aquariids covers the area of average CMOR radiants for longitudes below 295$\degree$, but poorly reproduces radiants located above this longitude. The simulated Orionid dispersion reproduces well the radiant structure around CAMO measurements, but the simulated meteors are shifted or absent for higher and lower latitudes. A complete modeling of the radiant structure probably requires the inclusion of meteoroids trails ejected before 3000 BCE, particularly for the parts of the shower away from the core. For example, the simulation of older trails might be necessary to explain the formation of the observed Orionid radiant tail reported by \cite{Jenniskens2016}. 

\section{Forecast} \label{sec:forecast}
 
The comparison of our new Halleyids model with existing  meteor observations in the previous section provides us with confidence that the model can reproduce the overall profile of the showers and their interannual peak activity. With the inclusion of trails ejected since 3000 BCE, our model satisfactorily reproduces the average activity profile, mass distribution, radiant dispersion and yearly intensity variations of the $\eta$-Aquariid meteor shower. The same model applied to the Orionids leads to a 30\% error of the shower's average activity level and underestimates its duration, but reproduces the first observed peak of activity and the year to year maximum ZHR variation. Our simulations indicate that the Orionids are probably composed of much older material, as highlighted in \cite{McIntosh1988} or \cite{Ryabova2003}, particularly during the descending branch of the shower. 

Our simulations do not reproduce the totality of the observed activity profiles or radiant structure, perhaps not surprising given that the comet's ephemeris and dust production rates are uncertain and/or completely unknown for much of its history. Despite these limitations, the general trends of the main peak ZHR annual variations are reasonably well reflected in our model. The models' success in the postdiction of the $\eta$-Aquariids and Orionids characteristics since 1985 gives us enough confidence to forecast the future activity of these showers. In the following sections, we present the predictions of our model for Halleyid activity in the period 2020 to 2050. 

\subsection{Annual variation}
 
\begin{figure}[!ht]
  \includegraphics[width=0.49\textwidth]{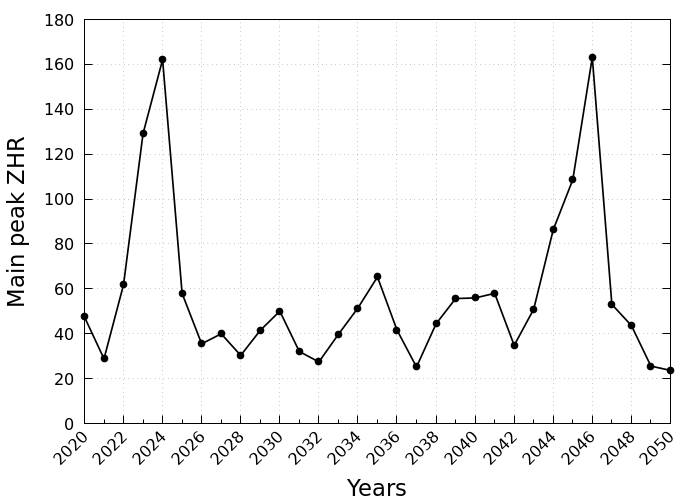}
  \includegraphics[width=0.49\textwidth]{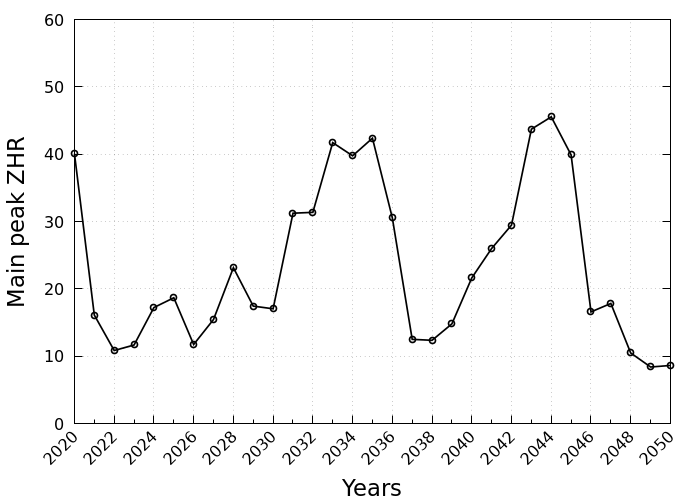}
   \caption{Predicted variations of the $\eta$-Aquariids (up) and Orionids (down) main peak ZHR between 2020 and 2050. }
   \label{fig:future_annual_variation}
 \end{figure}

 The main peak ZHR of each $\eta$-Aquariid and Orionid apparition between 2020 and 2050 based on our calibrated model is shown in Figure \ref{fig:future_annual_variation}. Consistent with \cite{Sekhar2014}, we expect no strong Orionid activity over the analyzed period. The predicted meteor rates vary between 20 and 40 meteors per hour, corresponding to the usual range of ZHR reported for the shower \citep{Egal2020b}. 
In contrast, Figure \ref{fig:future_annual_variation} shows  enhanced $\eta$-Aquariid activity in 2023-2024 and 2045-2046, reaching peak rates similar to those observed in 2004 and 2013. These will be examined below.

\begin{figure*}[!ht]
  \includegraphics[width=0.33\textwidth]{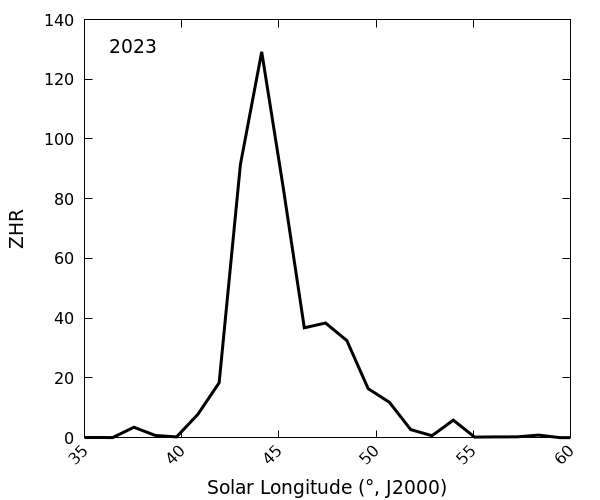} 
  \includegraphics[width=0.33\textwidth]{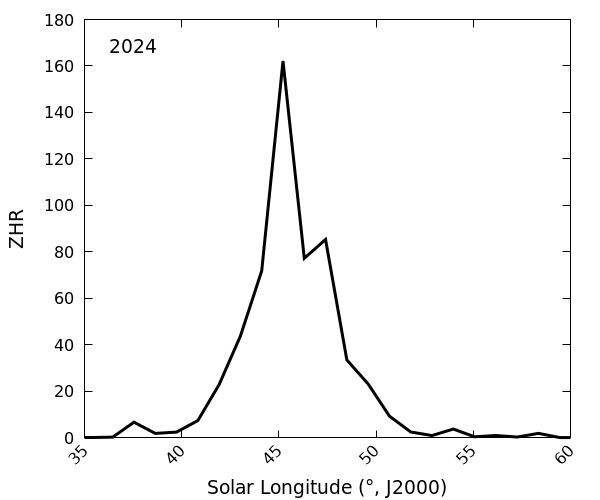}
  \includegraphics[width=0.33\textwidth]{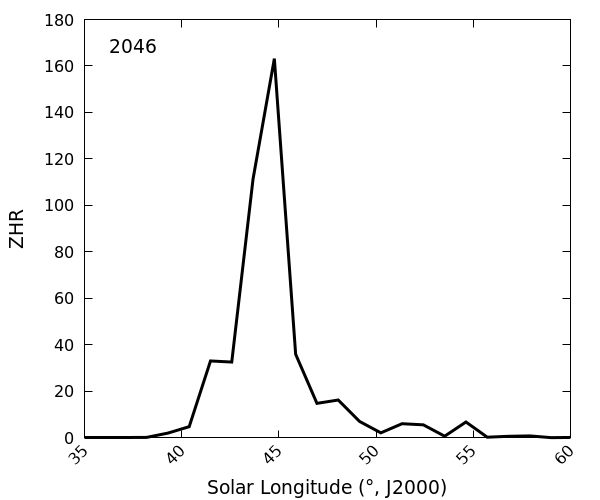}\\
  \includegraphics[width=0.33\textwidth]{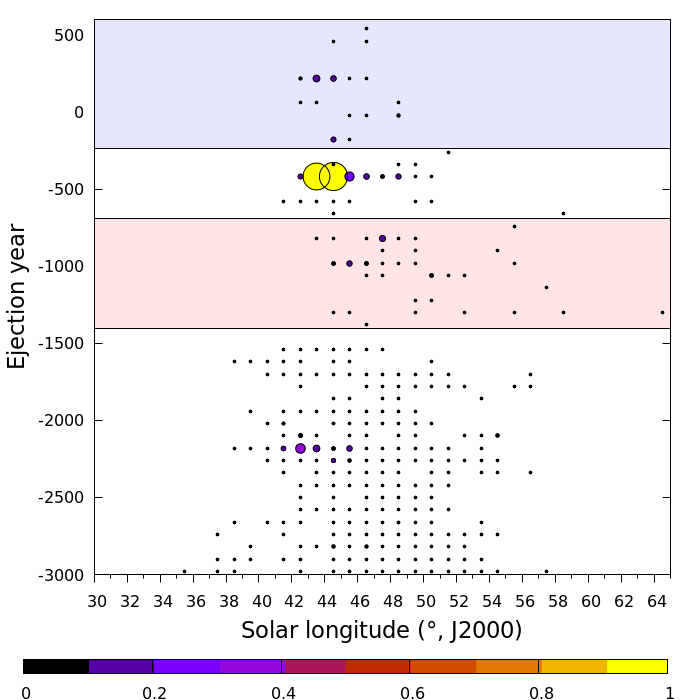} 
  \includegraphics[width=0.33\textwidth]{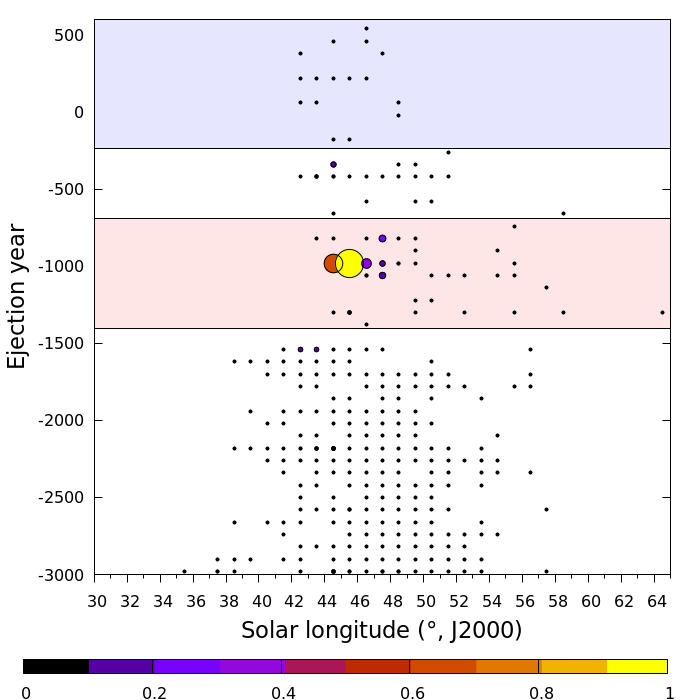}
  \includegraphics[width=0.33\textwidth]{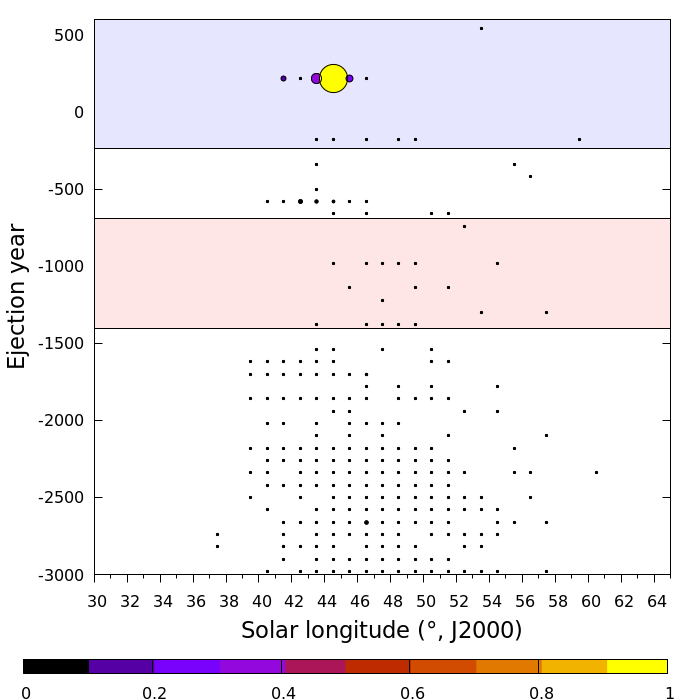}\\
  \includegraphics[width=0.33\textwidth]{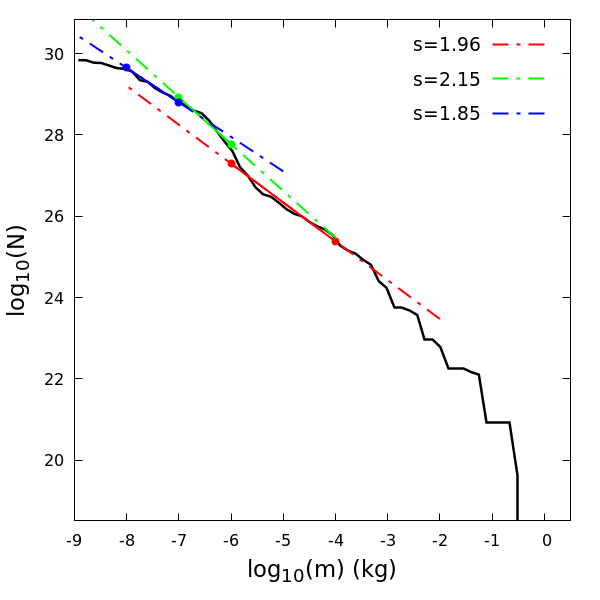} 
  \includegraphics[width=0.33\textwidth]{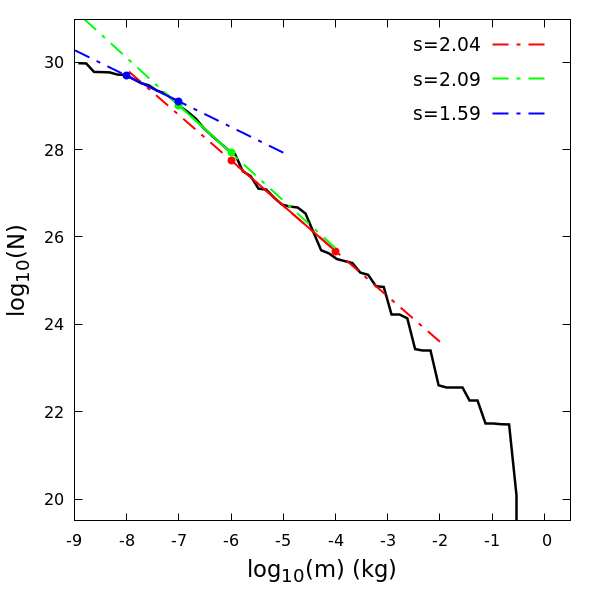}
  \includegraphics[width=0.33\textwidth]{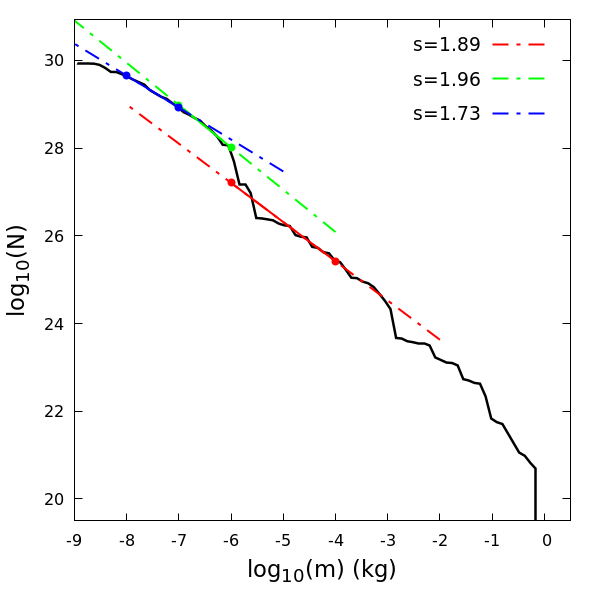}\\
  \caption{Details of model predictions for future $\eta$-Aquariid outbursts. From top to bottom: predicted model ZHR profile, relative contribution of simulated trails as a function of solar longitude for a given year and the model cumulative mass distribution. The middle trail plots show each ejections relative contribution by circle size (small to large) and color following the scale bar with black showing trails with lowest contribution to yellow trails which provide the highest contribution. The colored horizontal periods show the intervals when 1P/Halley was in the 1:6 MMR in red and in the 2:13 MMR in blue. The mass distribution plots follow the conventions from Figure \ref{fig:average_s} with estimates of the differential mass index $s$ in the visual (red), video (green) and radar (blue) mass ranges provided for comparison.}
  \label{fig:future_distributions}
\end{figure*}

\subsection{Outbursts characteristics}

Details of the predicted future $\eta$-Aquariids outbursts are presented in Figure \ref{fig:future_distributions}. The 2045 and 2046 predictions are similar (in profile shape, trails involved and mass distribution) and therefore only the 2046 outburst is detailed in the figure. The four showers are predicted to display noticeable activity between 40$\degree$ and 55$\degree$ in solar longitude, with a maximum around 44.1$\degree\pm0.5\degree$ (2023), 45.2$\degree\pm0.5\degree$ (2024) and 44.8$\degree\pm0.5\degree$ (2045 \& 2046). 

Most of the expected activity in 2023 is produced by particles ejected from 1P/Halley around 390 BCE, with the presence of potentially resonant particles ejected in 829 BCE. Additional trails (e.g., in 2215 BCE or 218 CE) have only a moderate influence on the profile's shape. The main peak of activity in 2024 arises from the 985 BCE trail, with smaller contributions of particles ejected during the 835 BCE, 1058 BCE, and 314 BCE apparitions. The combined contributions of several trails ejected while the 1P/Halley was trapped in the 1:6 MMR with Jupiter increases the expectation of enhanced $\eta$-Aquariid rates in 2024. The 2045 and 2046 enhancements are caused by more recent material, ejected around 218 CE. In contrast with the $\eta$-Aquariid outburst in 2013, these two years will mainly show enhancement due to material ejected outside the 1:6 MMR making them more similar to the situation in 2004. 

As discussed earlier, the modeled mass distribution of the predicted outbursts do not strictly follow a power law, showing slope variations for different mass ranges. Observations of the mass index $s$ therefore need to be computed with caution, carefully taking into account the mass range involved. Predicted $s$ values span from 1.85 to 1.96 in 2023, 1.59 to 2.09 in 2024, 1.72 to 2.1 in 2045 and 1.73 to 1.96 in 2046 over typical radar, video and visual mass intervals. 

\section{Discussion} \label{section:discussion}

\subsection{Limitations}

Most of the limitations of the proposed model stem from the modifications implemented to increase the number of particles retained. The larger $\Delta X$ and $\Delta T$ thresholds selected can lead to increasing prediction errors, artificially increasing ZHR in years around an outburst year or confusing the identification of the dominate contributing trail for a particular apparition.
Future meteor observations will help to update the weighting constraints developed in this work, which in turn should lead to refinement in modeled future activity of the Halleyids. 

The inclusion of trails ejected between 3000 BCE and 1404 BCE in our analysis also introduces some uncertainty. Section \ref{sec:traceability} suggests that the comet's orbit is known with some accuracy over the past several millennia, even if its exact location is not precisely determined before 1404 BCE \citep{Yeomans1981}. To investigate the influence of 1P/Halley's ephemeris prior to 1404 BCE in our simulations, we have ejected 5000 particles of 1-10 mm in size from different clones of the comet around  3000 BCE. Five clones of the comet were selected from the analysis of Section \ref{sec:traceability} to cover the full range of possible semi-major axis (from 16.6 to 18.8 AU). The 25 000 simulated meteoroids were ejected as described in Section \ref{sec:model} and integrated until 2020 CE, when the particles nodal location and average activity profiles were compared. 

This analysis revealed that the choice of the comet ephemeris around 3000 BCE has a moderate impact on the showers average activity profile (mainly on the modeled total duration), with a potential variation of 20 to 40\% of the meteor flux in periods away from the main peak of activity.  The selection of different comet clones prior to 1404 BCE  leads to only a small variation of the modeled peak ZHR for a given year, and does not modify the characteristics of the predicted future $\eta$-Aquariid outbursts. This reflects the fact that the majority of the core of both streams is composed of material ejected more recently than 1404 BCE, though this is more true of the $\eta$-Aquariids than the Orionids.  In addition, we found that the differences in the modeled radiants using different clones prior to 1404 BCE still produce radiants which lie within the observed radiant dispersion.

We therefore consider the results presented in Sections \ref{sec:postdiction} and \ref{sec:forecast} as valid despite the inclusion of trails ejected from the nominal comet orbit between 3000 BCE and 1404 BCE where the comet ephemeris is more uncertain. Despite the model's limitations, our simulations reproduce the main characteristics of the $\eta$-Aquariids apparitions since 1985, as well as the general trends in ZHR variations measured for the Orionids, including the years of outburst activity, the mass distribution of the showers and the location of the shower peaks. Our model shower durations are smaller than is observed, consistent with the notation that much of the material at the extreme start and end of the activity periods are older than 3000 BCE.  

These successes increase our confidence that our model is able to accurately  predict activity of the two showers through 2050, and allows for examination of their potential behavior over a longer time period. In particular, we can apply the model to investigate the long-standing question as to the cyclic variation of the $\eta$-Aquariid and Orionid activity which we discuss below.  

\subsection{Long term periodicity}

The existence of a periodicity in either Orionid peak activity or timing of maximum (or both) has been proposed by several authors in the past \citep[cf.][]{Egal2020}. In particular, a cyclic variation of the shower's maximum rates of about $\sim$12 years was suggested by \cite{McIntosh1983} and also by \cite{Dubietis2003}. The analysis of CMOR measurements between 2002 and 2019 by \cite{Egal2020} found a possible periodicity in the solar longitude of the Orionids peak timing with a period of 11.88 years. However, no strong conclusion about the shower's cyclic variations in either peak time or activity could be established without additional and longer term meteor observations. \citet{Egal2020b} found no clear periodicity in the annual activity level of the $\eta$-Aquariids from analysis of CMOR data. With our new model reproducing calibrated fluxes as a function of time for the Halleyid streams, we are now in a position to examine this question in more detail.  

\begin{figure}[!ht]
 \includegraphics[width=.49\textwidth]{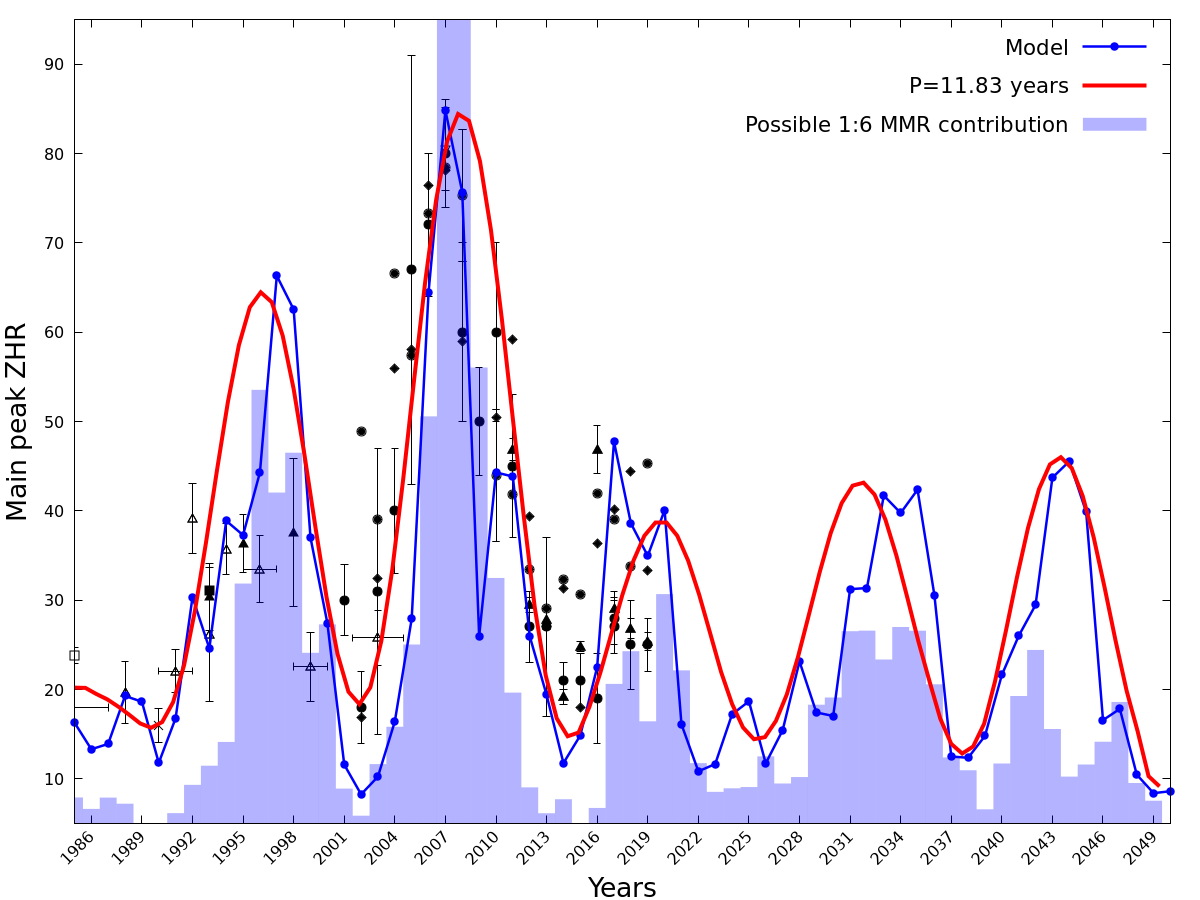}\\
 \includegraphics[width=.49\textwidth]{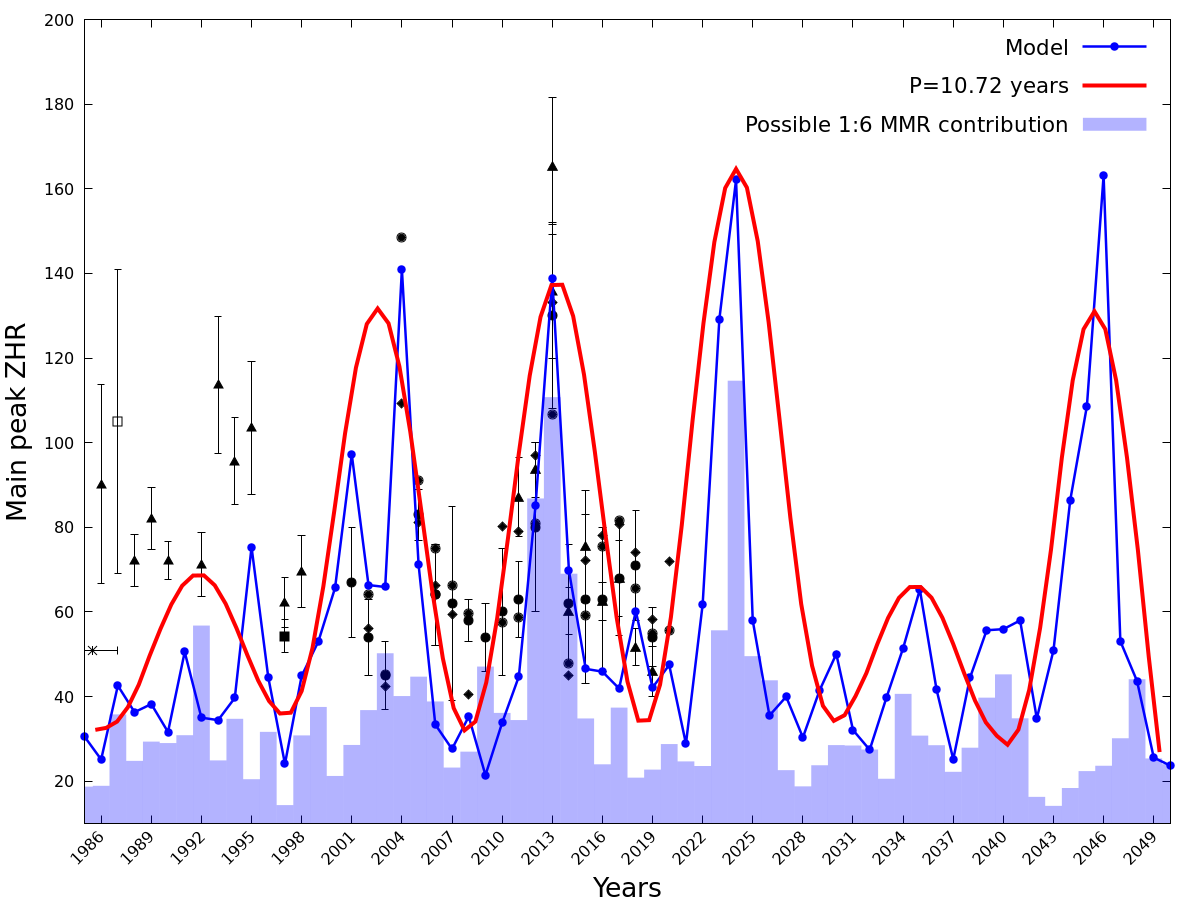}
\caption{\label{fig:model_sine} The annual variation of the observed (in black) and modeled (dark blue line) Orionids (top) and $\eta$-Aquariids (bottom) peak ZHR between 1985 and 2050. This figure represents a combination of Figures \ref{fig:past_annual_variation} and \ref{fig:future_annual_variation}.  A sinusoidal curve of variable amplitude was fit to the modeled and observed meteor rates. This fit result is presented in red in the figure. The best match with the model yields a period of 11.83 years in the case of the Orionids, and 10.72 years for the $\eta$-Aquariids. The relative contribution of particles intercepted by the Earth in a given year which were ejected while 1P/Halley was trapped in the 1:6 MMR with Jupiter are shown with blue boxes.}
\end{figure}

The annual variation in the observed (black line) and modeled (dark blue line) $\eta$-Aquariids and Orionids peak ZHR between 1985 and 2050 is given in Figure \ref{fig:model_sine}. A sinusoidal signal was fit to both the predicted and observed ZHR estimates using the PSO method. The amplitude of the sine wave was allowed to vary to improve the fit with the modeled ZHR.
This best fit solution to the model for each shower is presented in red in Figure \ref{fig:model_sine}.

The best PSO solution for the peak Orionid activity has a period of 11.83 years, close to the value 11.88 found in \cite{Egal2020} and Jupiter's orbital period of 11.86 years. Though the agreement between the PSO fit and observations showing outburst activity is not perfect, the sine function fit generally reproduces the observed and modeled intensity variations of the shower. 

To investigate the underlying physical cause of this periodic behavior, we explored the contribution to each annual return from particles ejected when 1P/Halley was trapped in the 1:6 MMR with Jupiter. The result is illustrated with the blue boxes in Figure \ref{fig:model_sine}. There is a clear correlation between the activity level of the shower and the number of particles reaching the Earth in a given year which were ejected when 1P/Halley was in the 1:6 MMR. This suggests that the 1:6 mean motion resonance induces a $\sim$12-year cyclic variation of Orionid activity. 

The annual $\eta$-Aquariids activity is better described with a sinusoid having a period of 10.72 years. The sinusoidal signal fit is in good agreement with the modeled peak ZHR since 2005, but does not reproduce well earlier measurements of the shower. When examining the possible contribution of 1:6 resonant particles, we note that these particles are involved in some observed and predicted outbursts (like in 2013 and 2024) but not all of them (e.g., in 2004 or 2045-2046). The existence of a periodic variation of the $\eta$-Aquariids annual activity is suggested by our analysis but not as convincing as the Orionid fits. This periodicity needs to be confirmed by future radar and optical observations of the shower. 

\section{Conclusions}\label{sec:conclusion}

We have developed a new model of the meteoroid streams released by comet 1P/Halley and used it to characterise the associated $\eta$-Aquariid and Orionid meteor showers. Through simulation of more than 5 million particles and the application of a custom-made weighting solution, we have modeled the showers' activity profiles, radiants, mass distributions and maximum ZHR variations between 1980 and 2050. The simulation results were calibrated using the peak activity and profiles of 35 years of Halleyid observations, self-consistently applying a unique weighting scheme to the two showers.

From the model we find that the $\eta$-Aquariids characteristics (activity profile, radiants and annual variations) can be largely reproduced by material ejected between 3000 BCE and 600 CE. In particular, the core of the stream is almost entirely made up of ejecta from this interval. The wings of the activity are less well fit and suggest that even older meteoroids are present in the stream, particularly near the start and end of the activity period. 

The model fit to the Orionids activity profile, in contrast, leads to an underestimate of the activity. This indicates that the Orionids are older than the $\eta$-Aquariids on average. The model is generally able to reproduce the first (ascending) part of the broad Orionid maximum but fails to reproduce the reported second activity peak around 210$\degree$ SL and the descending branch of activity. Based on our model, the core and early portion of the Orionid activity are primarily populated by material ejected between 3000 BCE and 610 BCE.  Despite this limitation, the simulations reproduce the general trends of the maximum interannual ZHR variations, providing us with some confidence in our activity forecasts. 

From our calibrated model we have produced estimates of the mass distribution of meteoroids in the core of both streams. While the overall distribution roughly follows a power-law, our modeling predicts significant "kinks" in the distribution, notably in the mass range 10$^{-5}$ to 10$^{-7}$ kg. This may explain the variability in measurements of the mass index reported by different instruments for the Halleyid streams as the values will depend sensitively on the upper and lower mass ranges used for a given fit.

Our simulations suggest that the age of the present day $\eta$-Aquariids slightly exceeds 5000 years, while the Orionids are older. Future simulation efforts could better constrain the upper range to the Orionids age by simulating meteoroids ejected from several clones of 1P/Halley's nominal orbit before 3000 BCE. This also implies that the minimal physical age for 1P/Halley is 5000 years. 

Our model provides predictions for the peak ZHRs expected from the Halleyids in the period 2020 to 2050. These predictions suggest no significant Orionid outbursts in this interval and four potential $\eta$-Aquariid outbursts in 2023, 2024, 2045 and 2046. The expected maximum ZHRs for the $\eta$ - Aquariid outbursts vary from 120 to 160 meteors per hour, with a 30\% confidence on the predicted rates. Our model suggests that most of the $\eta$ - Aquariid activity expected in 2024 originates from particles ejected while comet 1P/Halley was trapped into the 1:6 MMR resonance, similar to several previously observed $\eta$-Aquariid and Orionid outbursts.

Among the mean motion resonances with Jupiter and Saturn that can shape the long term evolution of the Halleyids meteoroids stream, the 1:6 MMR with Jupiter has the strongest effect on the stream’s structure. The 1:6 MMR  is a significant, but not the only, source of Halleyid outbursts. This is consistent with the results of other authors \citep[cf.][]{Sato2007,Rendtel2008,Sato2014,Sekhar2014} and also suggests that the resonance induces a cyclic variation of the Orionids average ZHR levels with a period of about 11.8 years. A potential periodicity of 10.7 years for the $\eta$-Aquariids needs to be confirmed by future observations of the shower. Visual, video and radar observations of each $\eta$-Aquariid and Orionid apparition are strongly encouraged to provide additional constraints on numerical models. 

\begin{acknowledgements}
  This work was supported in part by NASA Meteoroid Environment Office under cooperative agreement 80NSSC18M0046 and by the Natural Sciences and Engineering Research Council of Canada (Grants no. RGPIN-2016-04433 \& RGPIN-2018-05659), and by the Canada Research Chairs Program. We are thankful to Aswin Sekhar for his careful review that helped improve the manuscript.
\end{acknowledgements}

 
  \bibliographystyle{model2-names.bst} 
  \bibliography{References}


\begin{appendix} 

\onecolumn

 \section{1P/Halley orbital motion}\label{appendix:ephemeris}

\begin{figure}[!ht]
    \centering
    \includegraphics[trim={0.6cm 2.1cm 0.5cm 0.5cm},clip,width=0.45\textwidth]{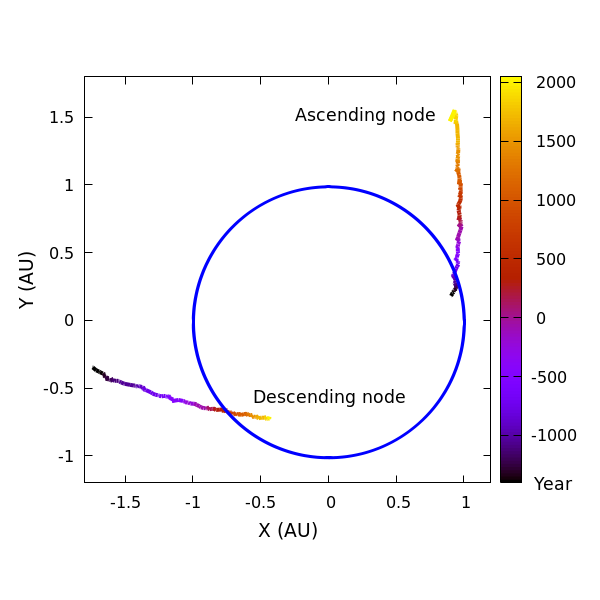}
    \includegraphics[width=0.5\textwidth]{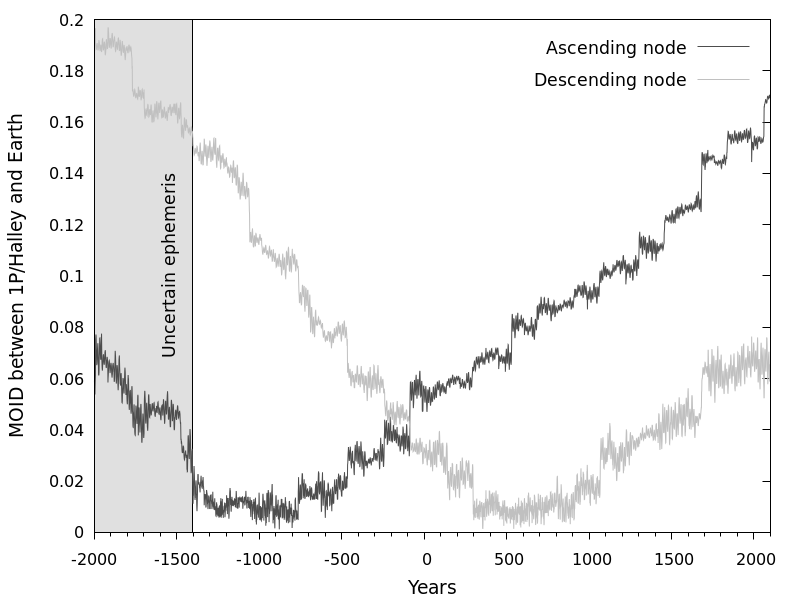}
    \caption{Left: ecliptic location of 1P/Halley's ascending and descending node as a function of time, in the barycentric frame (ecliptic J2000). The Earth's orbit is represented in blue. Right: Time variations of the Minimum Orbit Intersection Distance (MOID) of comet 1P/Halley and Earth close to the comet's ascending and descending node.}
    \label{fig:Nodes_and_MOID}
\end{figure}
 
 \begin{figure}[!ht]
    \centering
    \includegraphics[width=\textwidth]{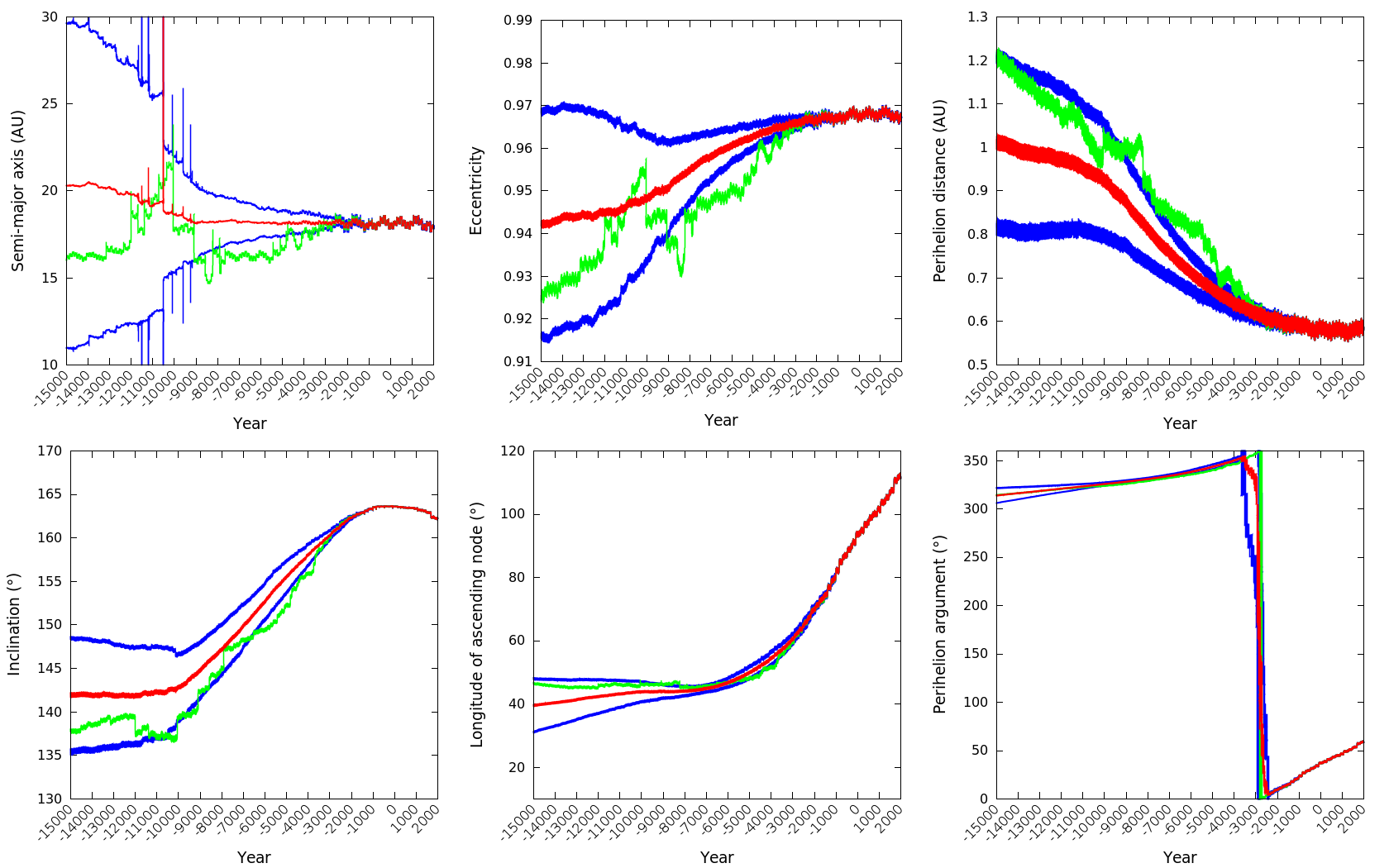}\\
    \caption{Time-evolution of the nominal (green) and average (red) orbital elements of 1000 clones of comet 1P/Halley generated in 1994 CE and based on the JPL J863/77 solution. The blue curves represent the one standard deviation range about the average semi-major axis.}
    \label{1P_traceability}
\end{figure}
 
 \newpage
 \section{Average precession} \label{appendix:precession}
 
      \begin{figure*}[!ht]
    \centering
     \includegraphics[width=0.95\textwidth]{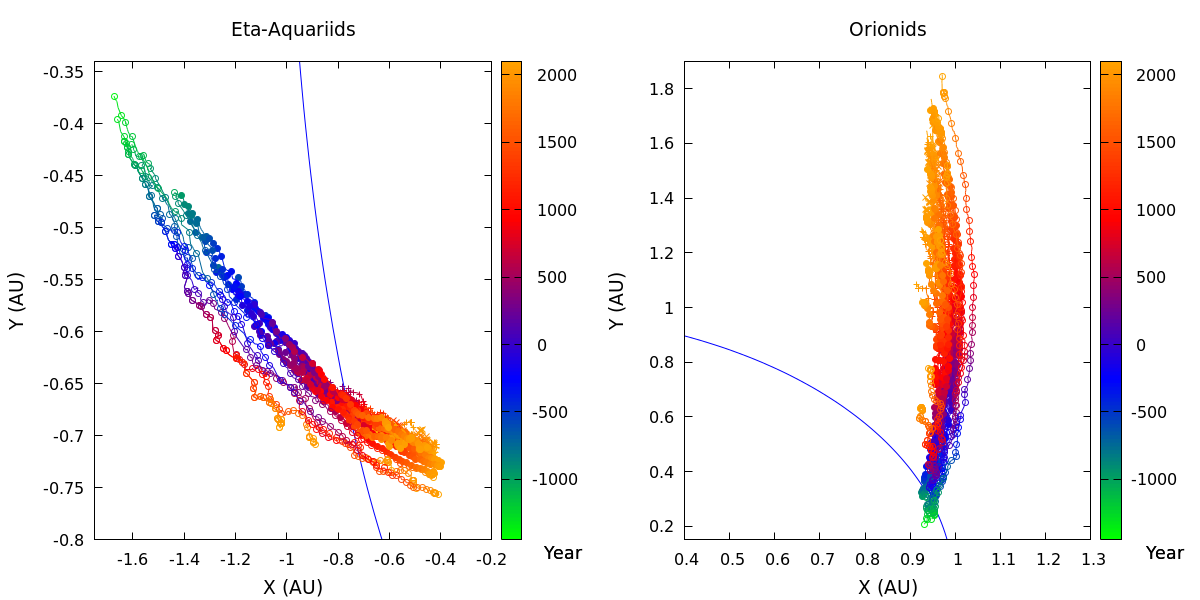}\\
     \includegraphics[width=0.95\textwidth]{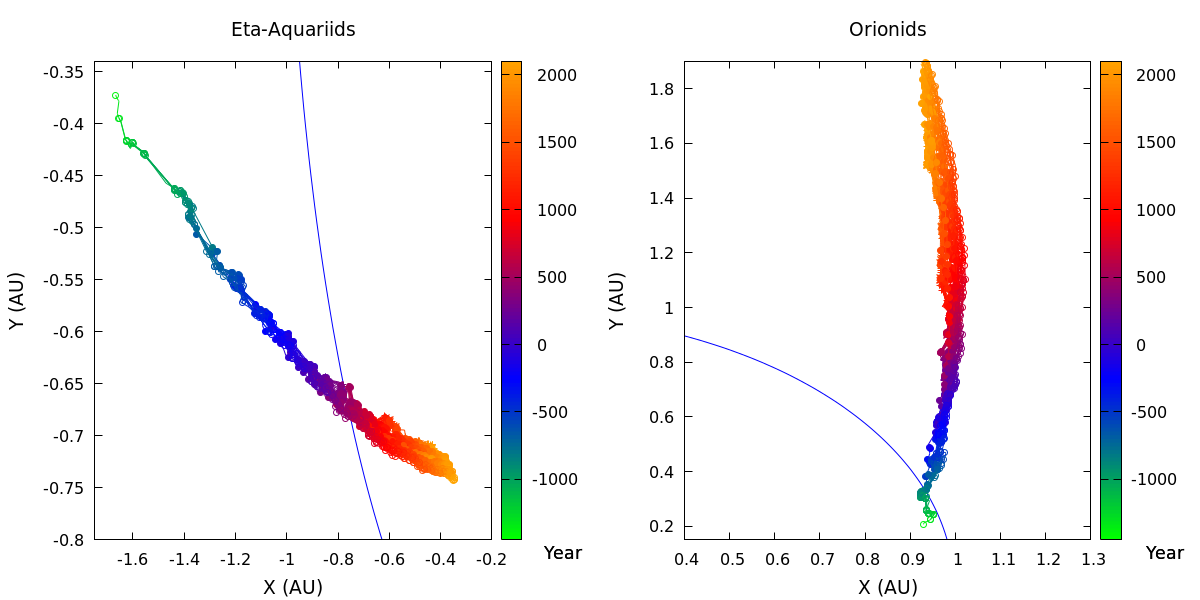}\\
     \caption{Median location of the trails descending node (left panels)  and ascending node (right panels) in function of time for particles of sizes 0.1-1mm (top panels) and 10-100 mm (bottom panels). Each line represents a meteoroid trail ejected during a single apparition of the comet, between 1404 BCE and 1910 CE.}
     \label{fig:precession_fsize}
  \end{figure*}

   \newpage
 \section{Simulated activity profiles} \label{appendix:individual_profiles}
 
     \begin{figure*}[!ht]
    \centering
      \includegraphics[width=.24\textwidth]{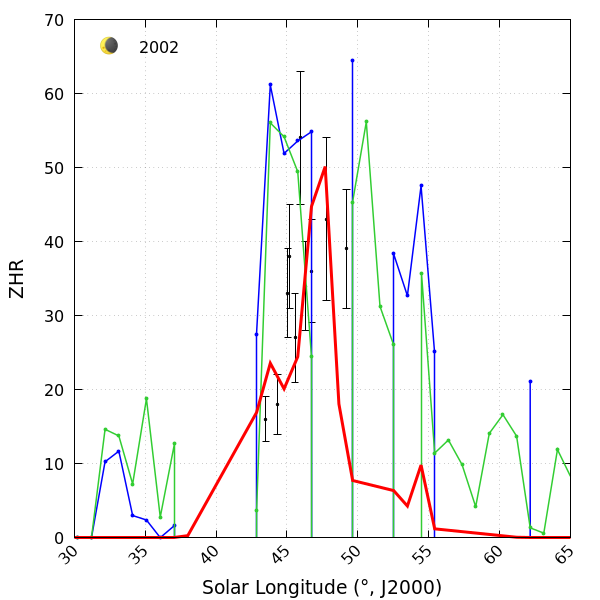}
      \includegraphics[width=.24\textwidth]{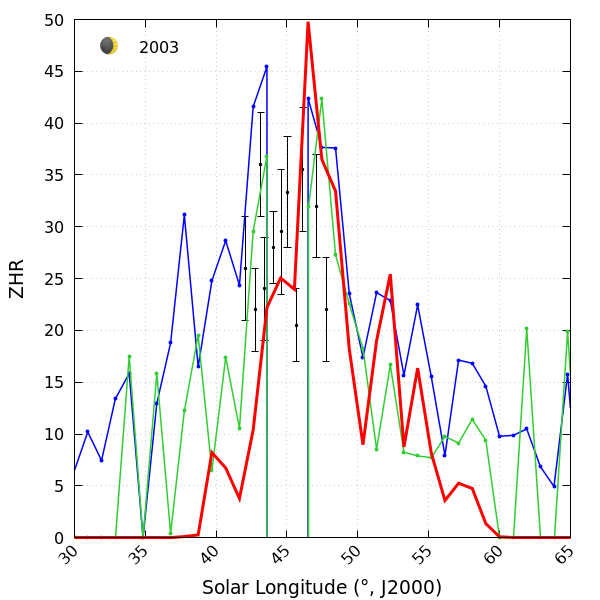}
      \includegraphics[width=.24\textwidth]{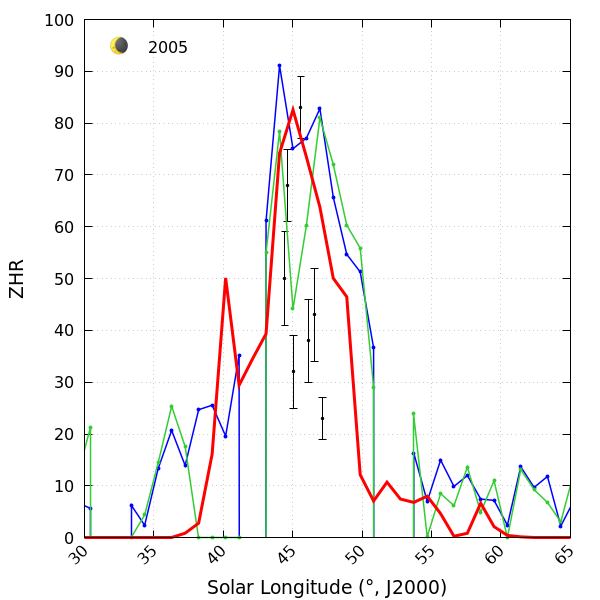}
      \includegraphics[width=.24\textwidth]{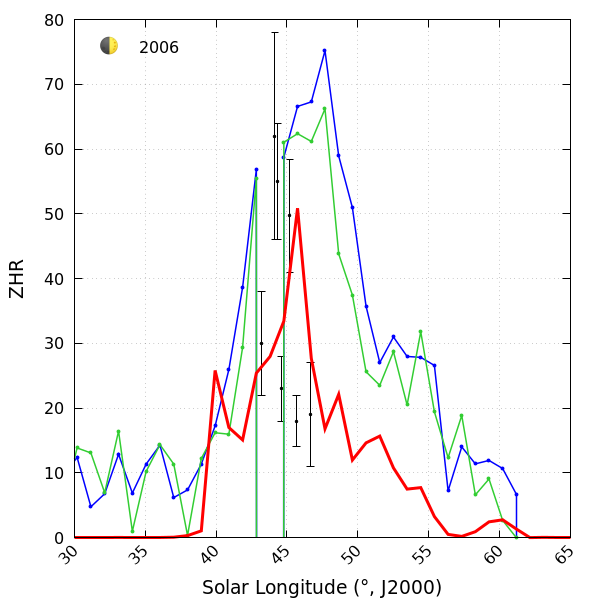}\\
      \includegraphics[width=.24\textwidth]{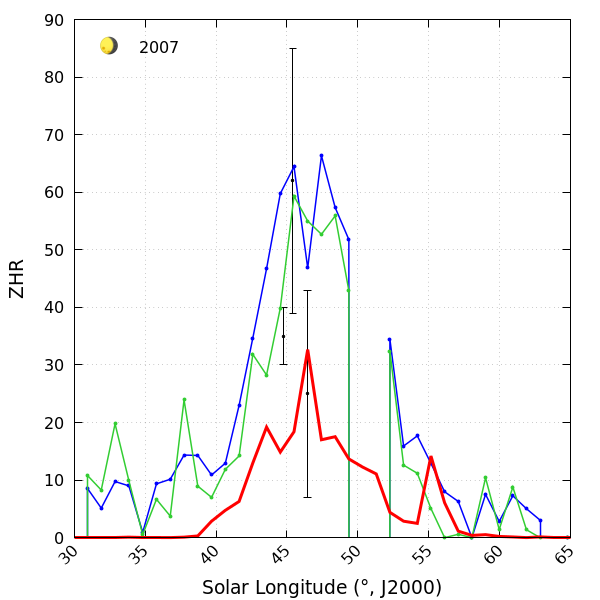}
      \includegraphics[width=.24\textwidth]{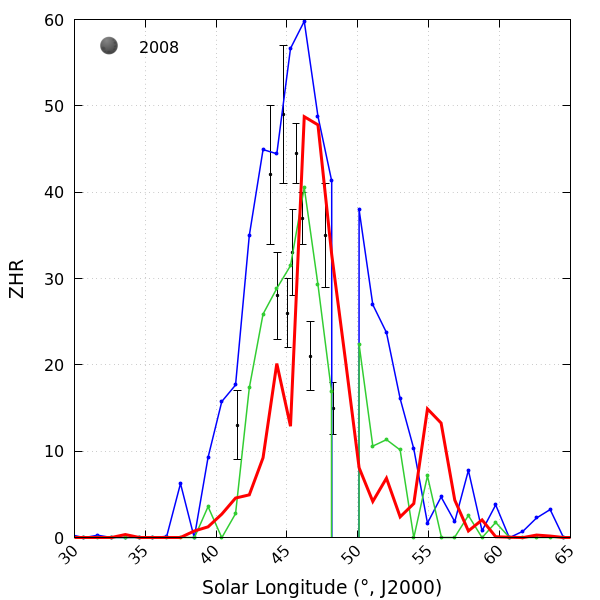}
      \includegraphics[width=.24\textwidth]{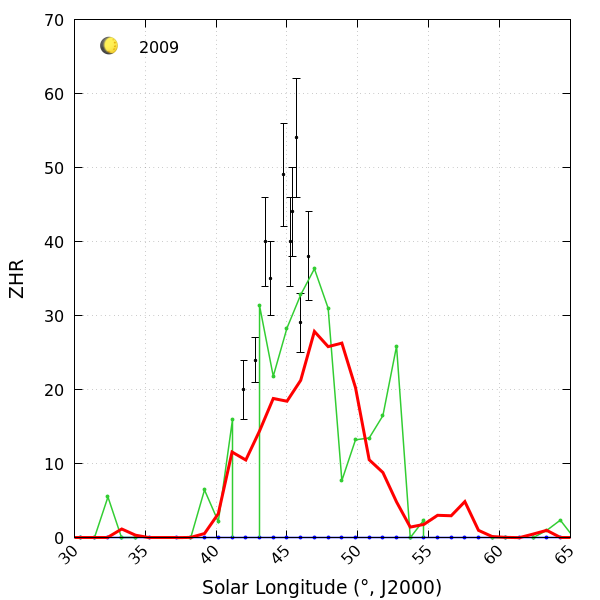}
      \includegraphics[width=.24\textwidth]{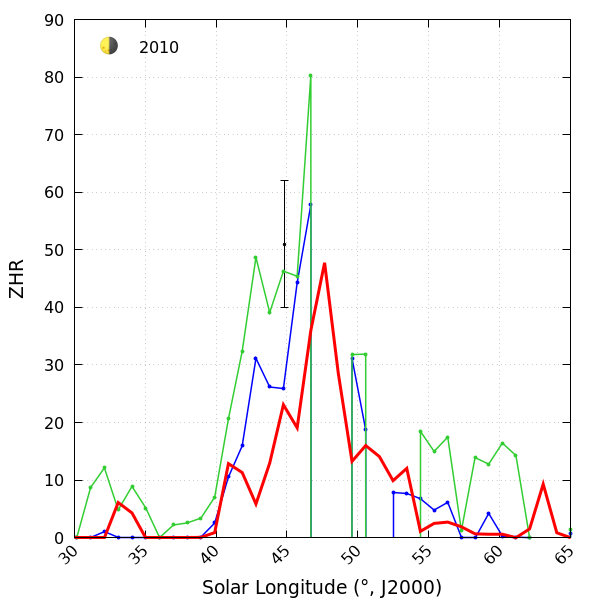}\\
      \includegraphics[width=.24\textwidth]{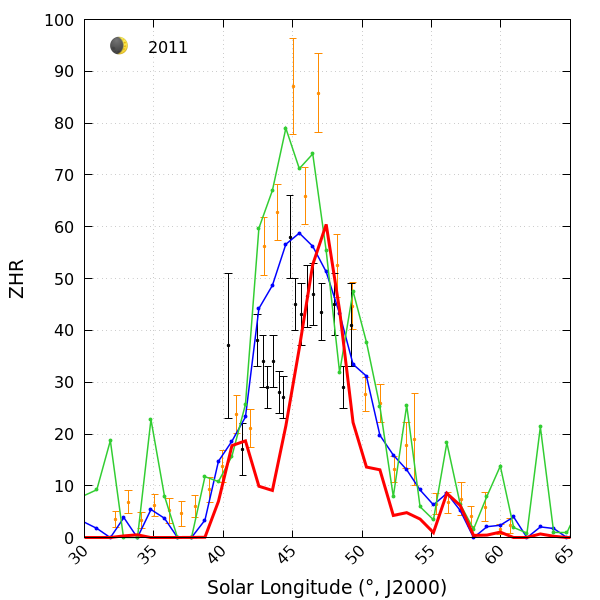}
      \includegraphics[width=.24\textwidth]{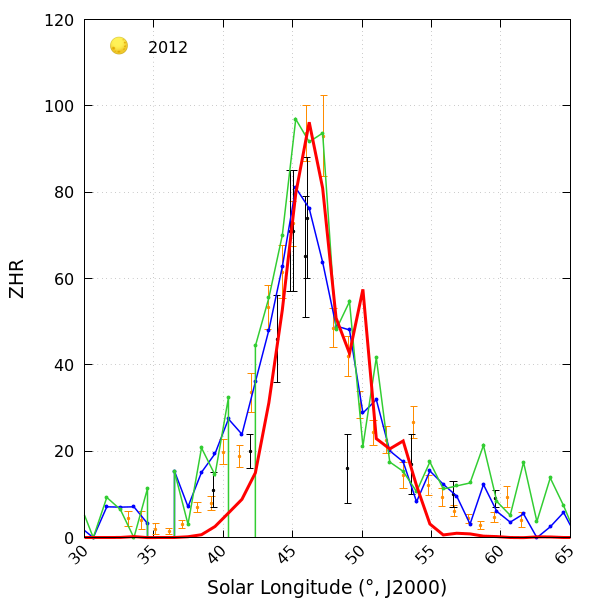}
      \includegraphics[width=.24\textwidth]{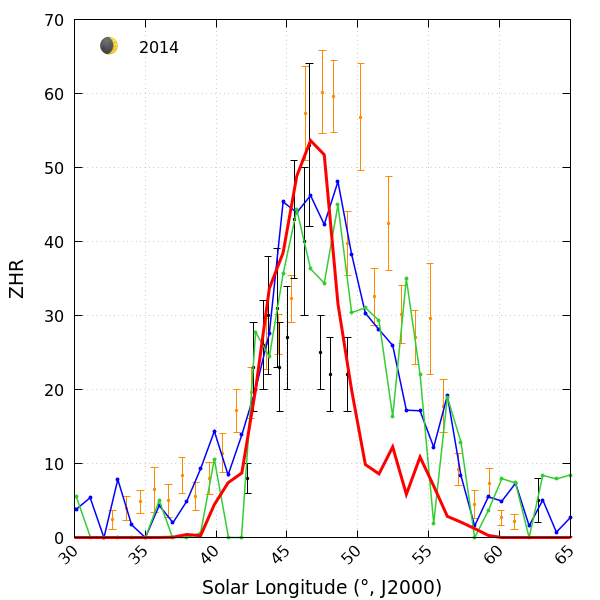}
      \includegraphics[width=.24\textwidth]{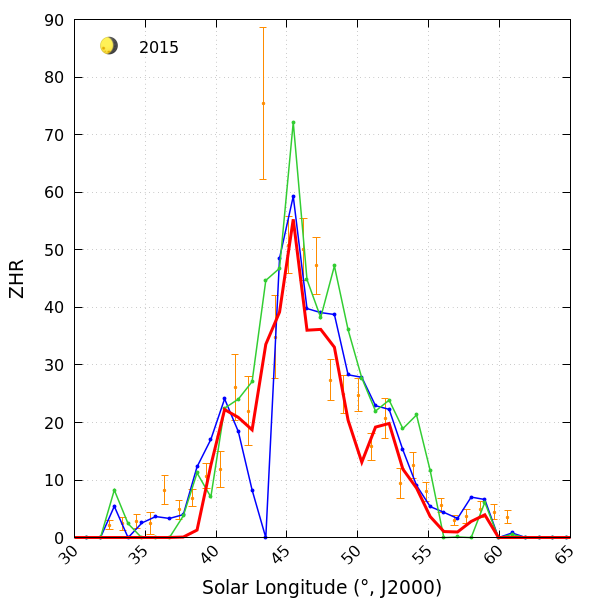}\\
      \includegraphics[width=.24\textwidth]{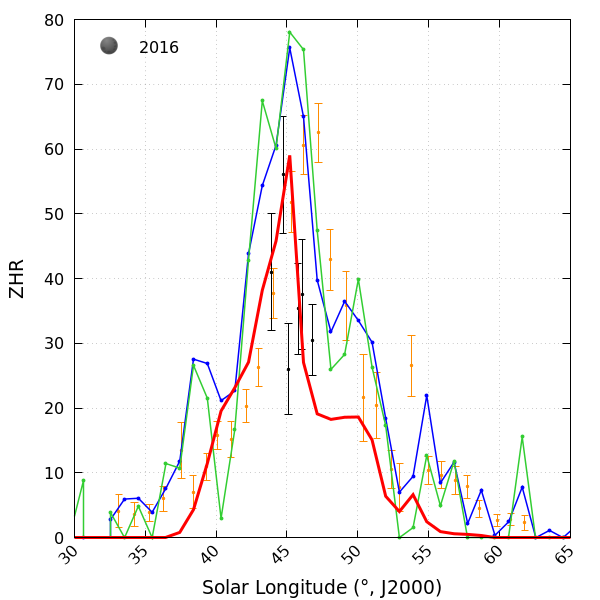}
      \includegraphics[width=.24\textwidth]{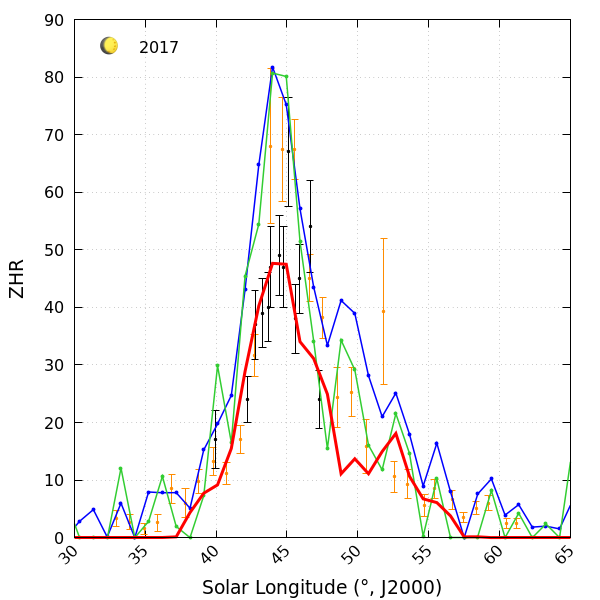}
      \includegraphics[width=.24\textwidth]{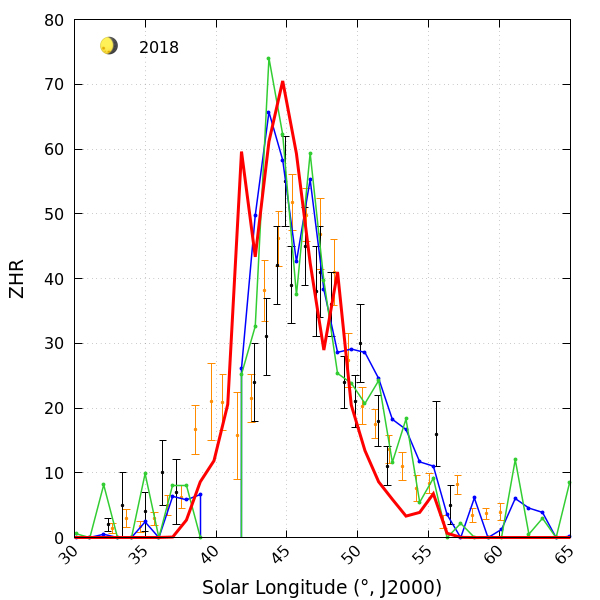}
      \includegraphics[width=.24\textwidth]{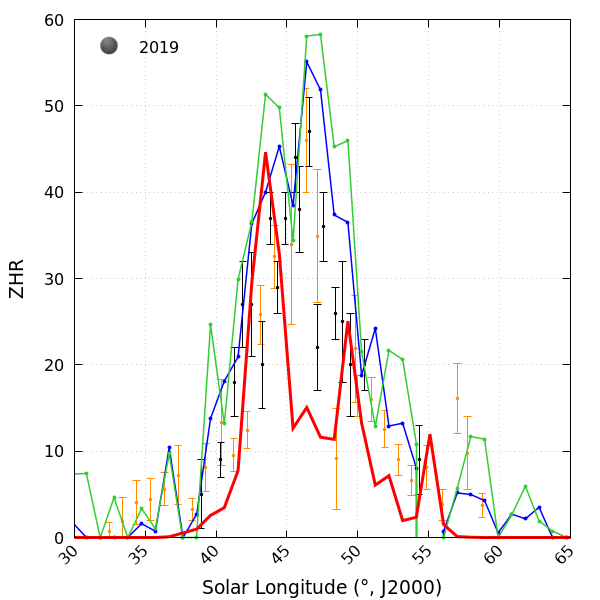}\\
      \includegraphics[width=.24\textwidth]{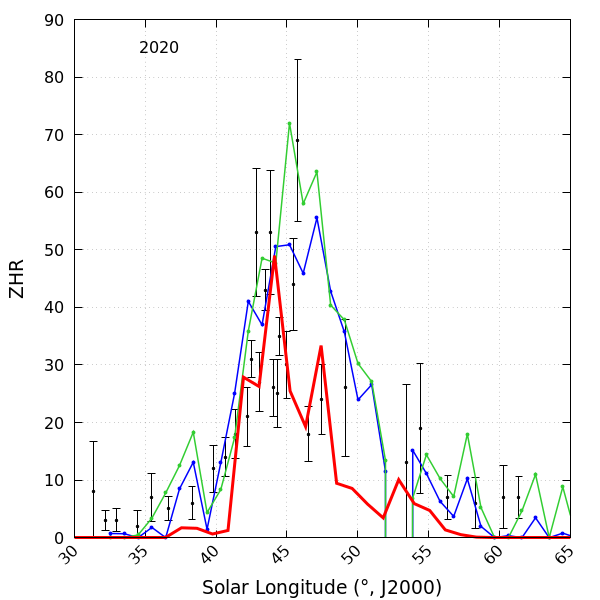}
      \caption{Activity profiles of the $\eta$-Aquariids between 2002 and 2020, without the outbursts of 2004 and 2013. Outburst years are detailed in Appendix \ref{appendix:outbursts}.}
      \label{fig:new_etas}
  \end{figure*}
  
       \begin{figure*}[!ht]
    \centering
      \includegraphics[width=.24\textwidth]{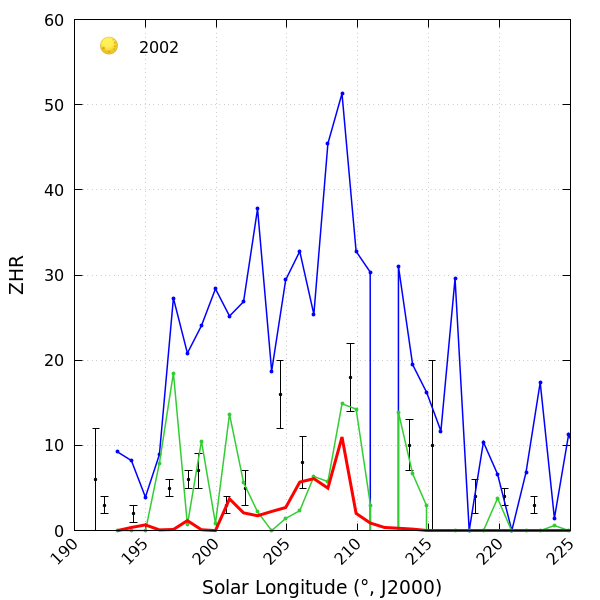}
      \includegraphics[width=.24\textwidth]{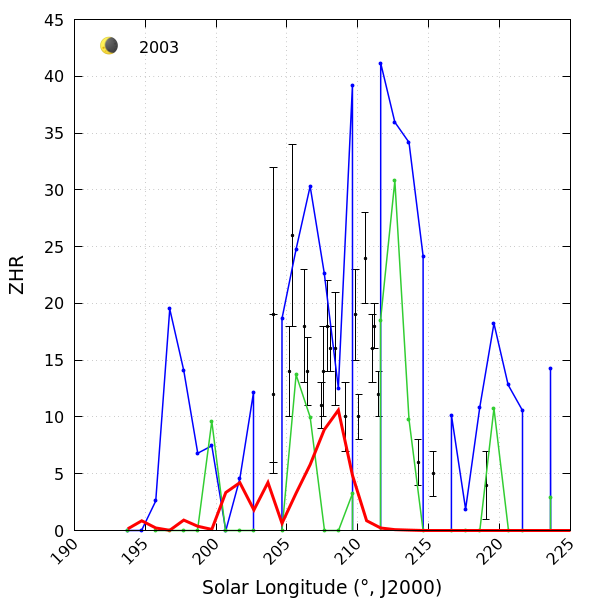}
      \includegraphics[width=.24\textwidth]{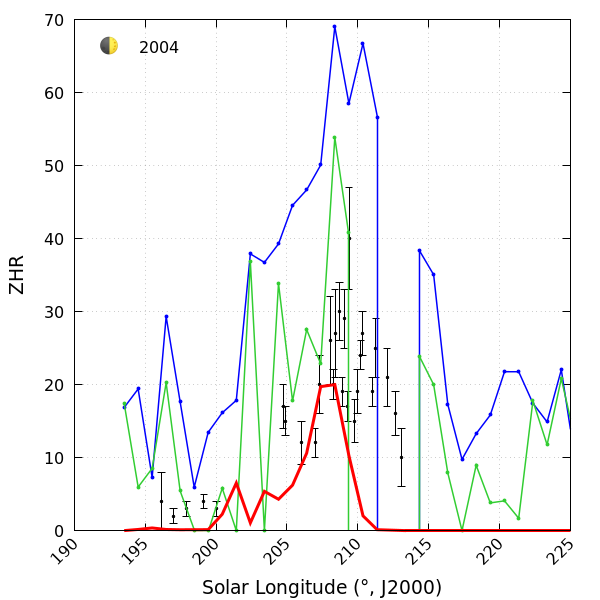}
      \includegraphics[width=.24\textwidth]{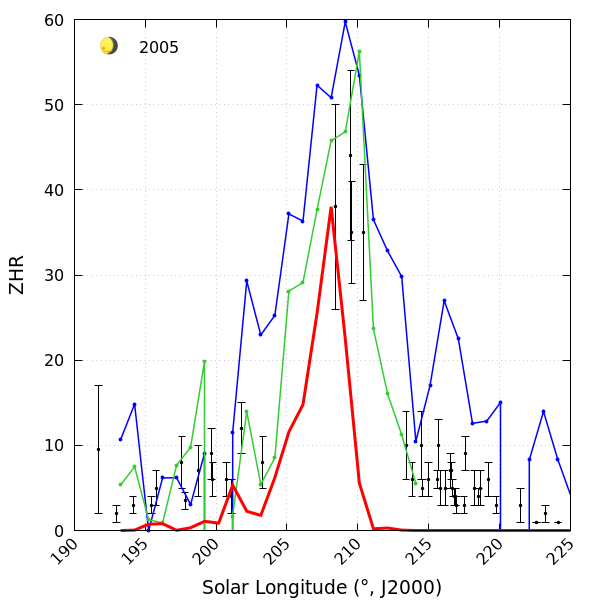}\\
      \includegraphics[width=.24\textwidth]{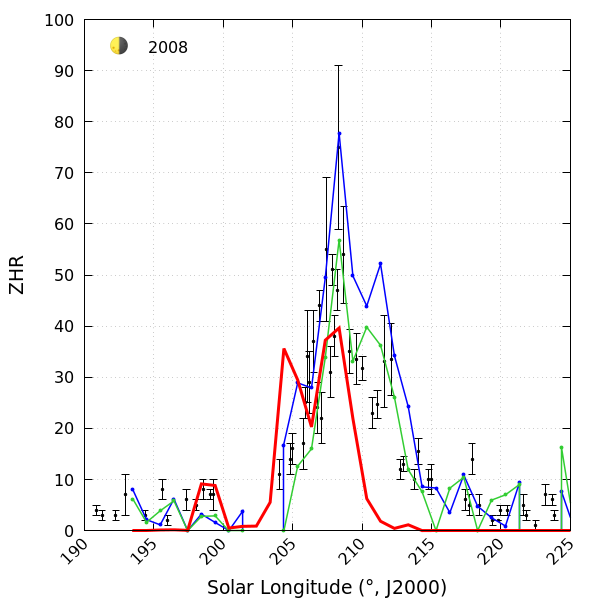}
      \includegraphics[width=.24\textwidth]{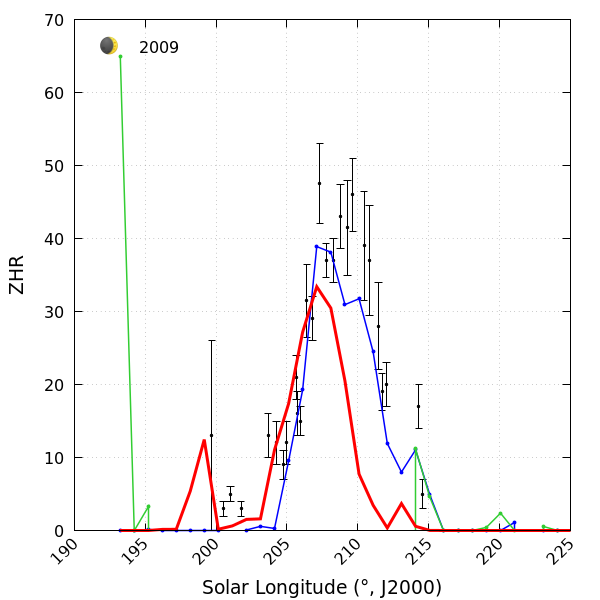}
      \includegraphics[width=.24\textwidth]{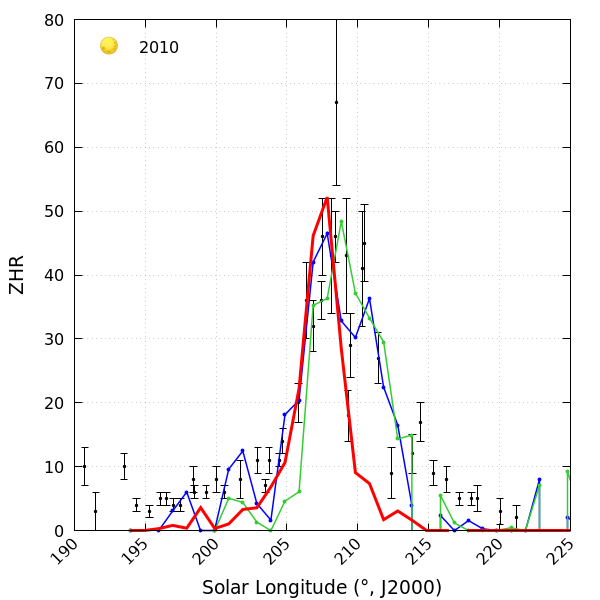}
      \includegraphics[width=.24\textwidth]{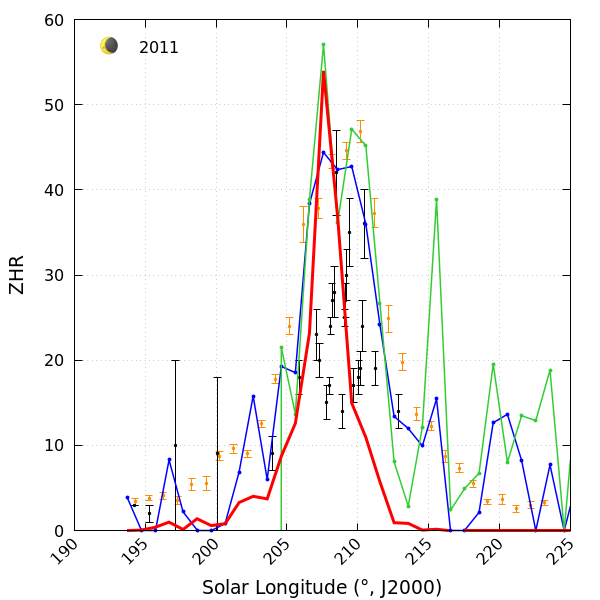}\\
      \includegraphics[width=.24\textwidth]{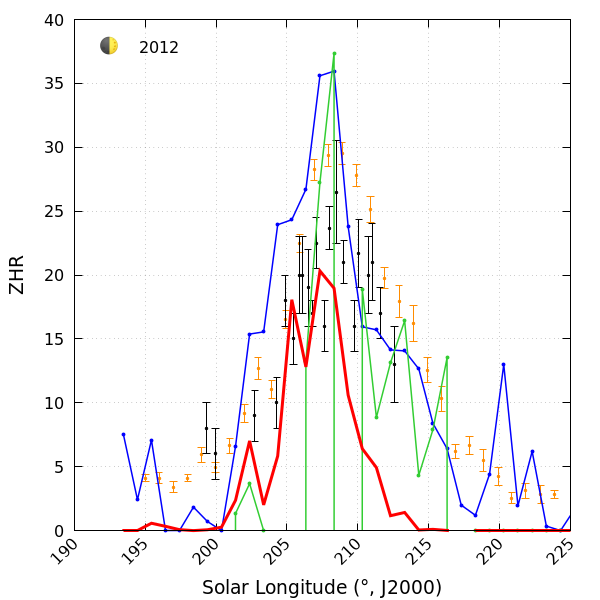}
      \includegraphics[width=.24\textwidth]{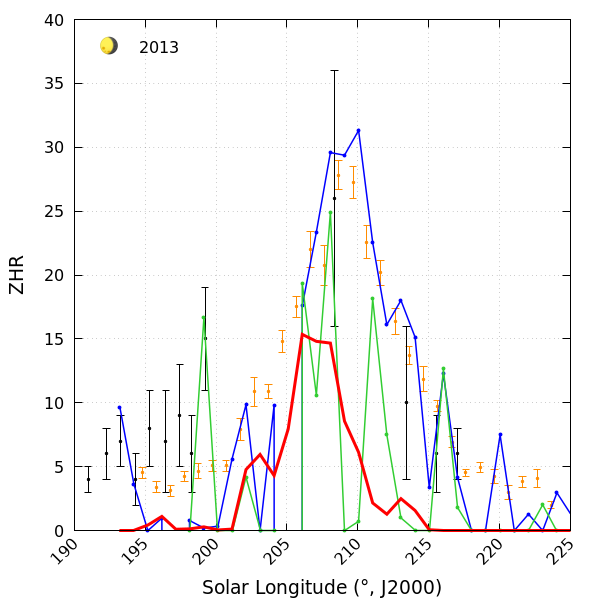}
      \includegraphics[width=.24\textwidth]{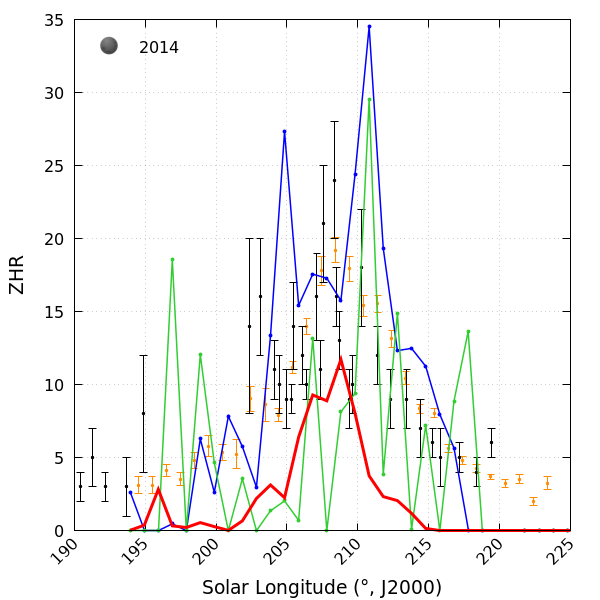}
      \includegraphics[width=.24\textwidth]{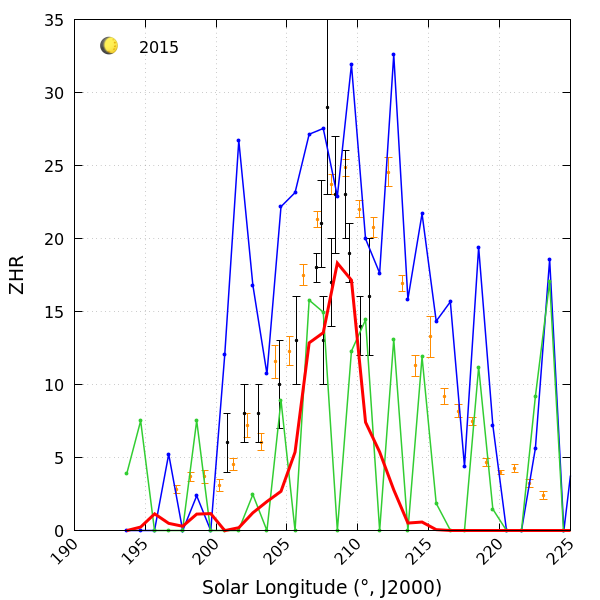}\\
      \includegraphics[width=.24\textwidth]{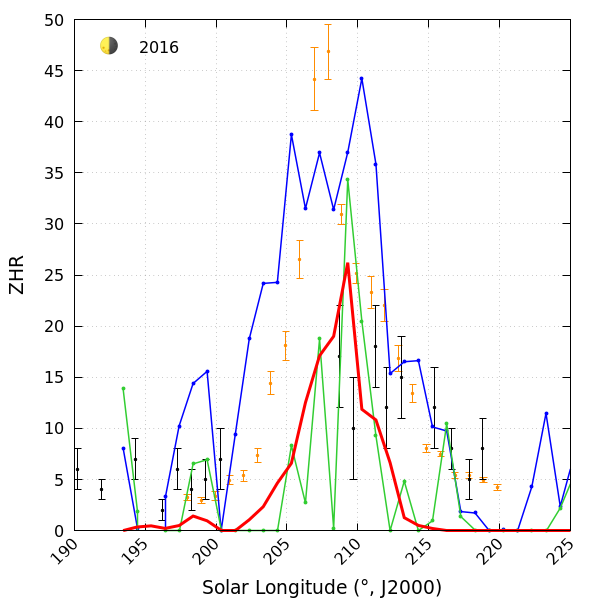}
      \includegraphics[width=.24\textwidth]{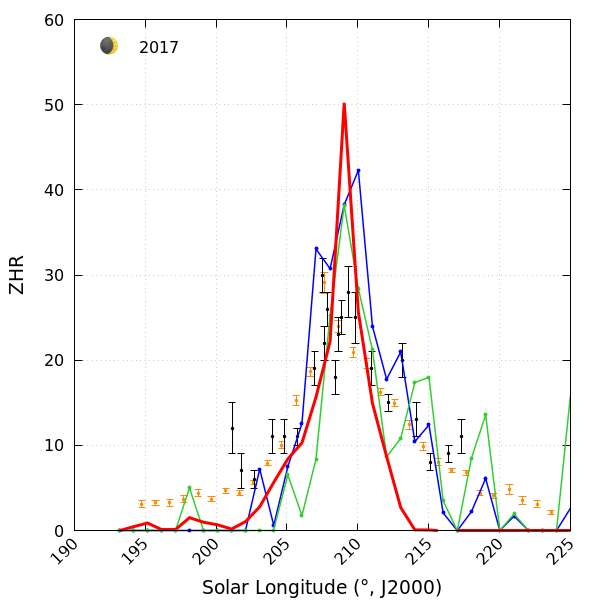}
      \includegraphics[width=.24\textwidth]{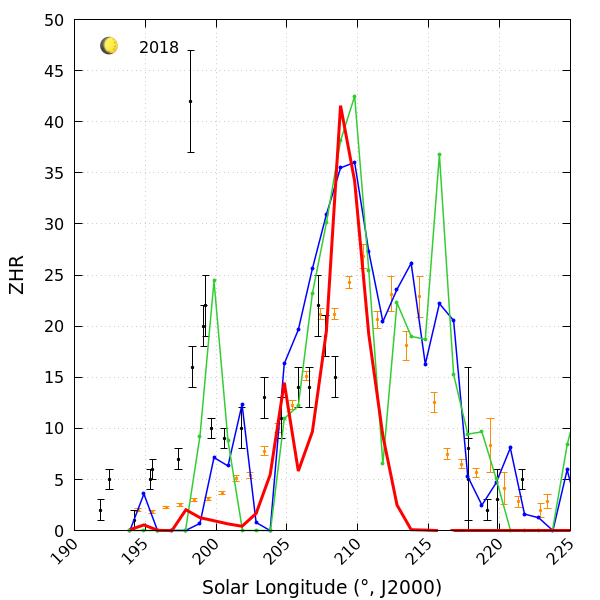}
      \includegraphics[width=.24\textwidth]{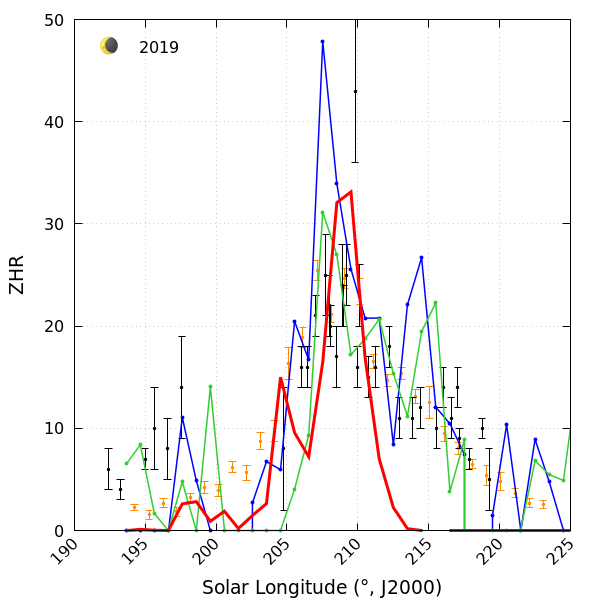}
      \caption{Activity profiles of the Orionids between 2002 and 2019, without the outbursts of 2006 and 2007. Outburst years are detailed in Appendix \ref{appendix:outbursts}.}
      \label{fig:new_oris}
  \end{figure*}
  
  \clearpage
 \section{Past Halleyid Outburst Characteristics}\label{appendix:outbursts}
    \begin{figure*}[!ht]
     \centering
     \includegraphics[width=0.29\textwidth]{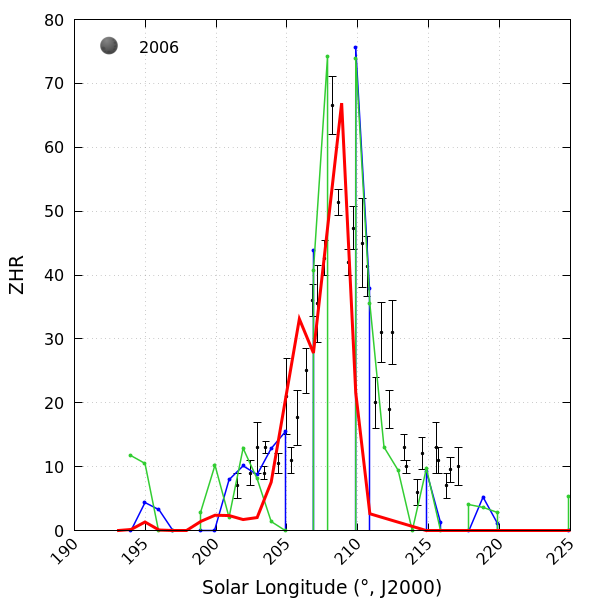}
     \includegraphics[width=0.287\textwidth]{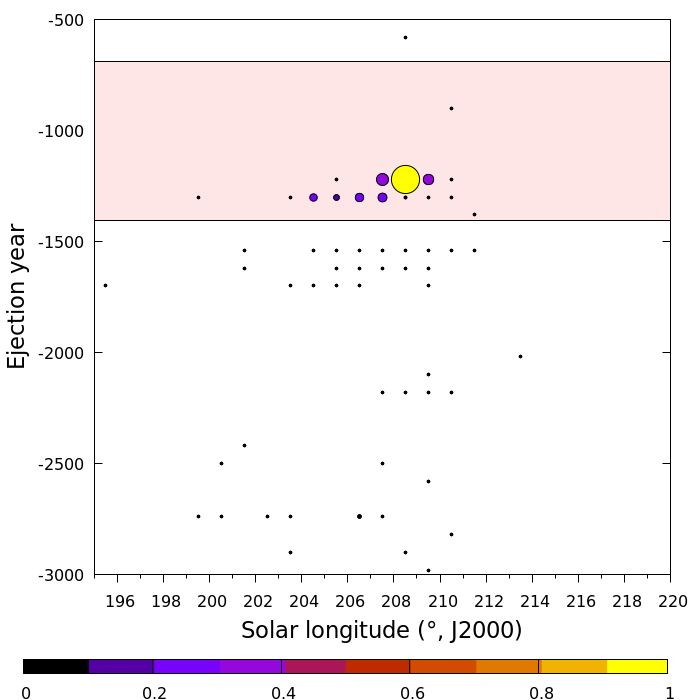}
     \includegraphics[width=0.29\textwidth]{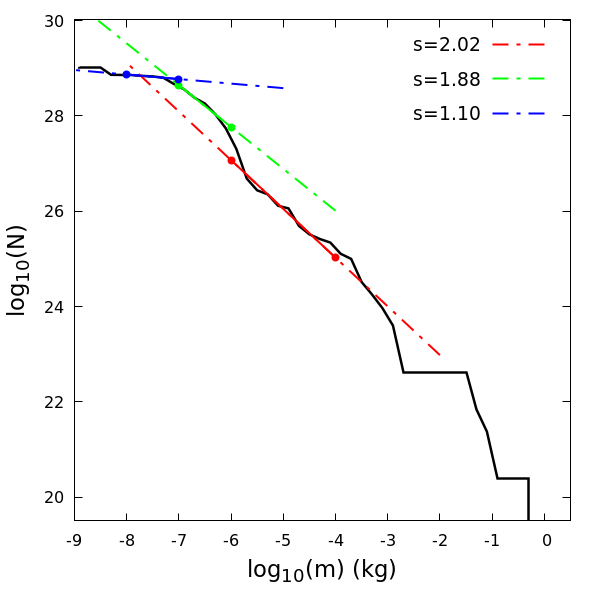}\\
     \includegraphics[width=0.29\textwidth]{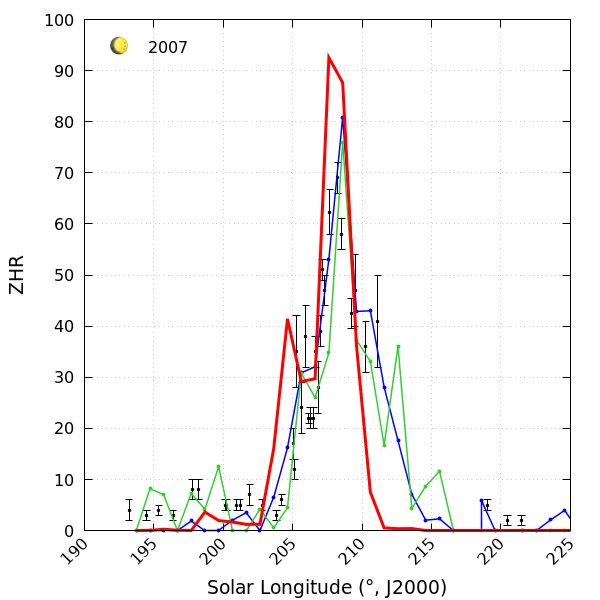}
     \includegraphics[width=0.287\textwidth]{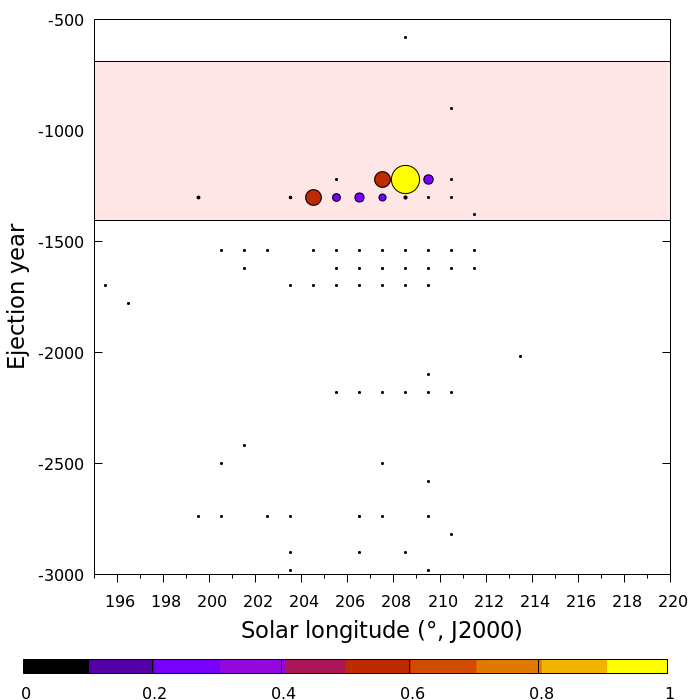}
     \includegraphics[width=0.29\textwidth]{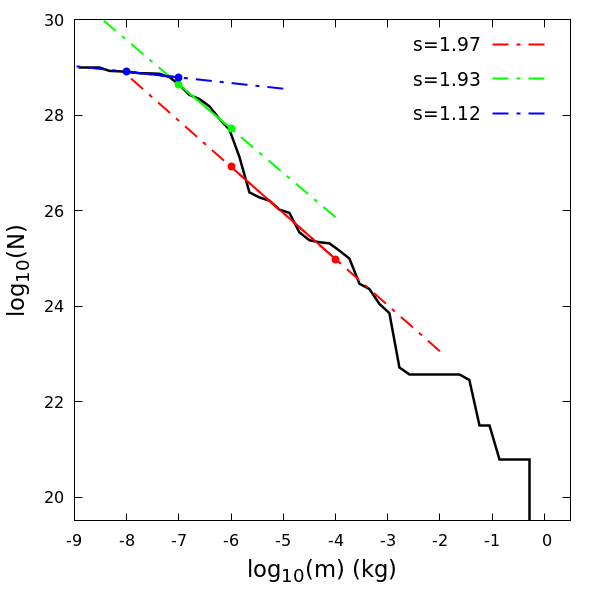}\\
     \includegraphics[width=0.29\textwidth]{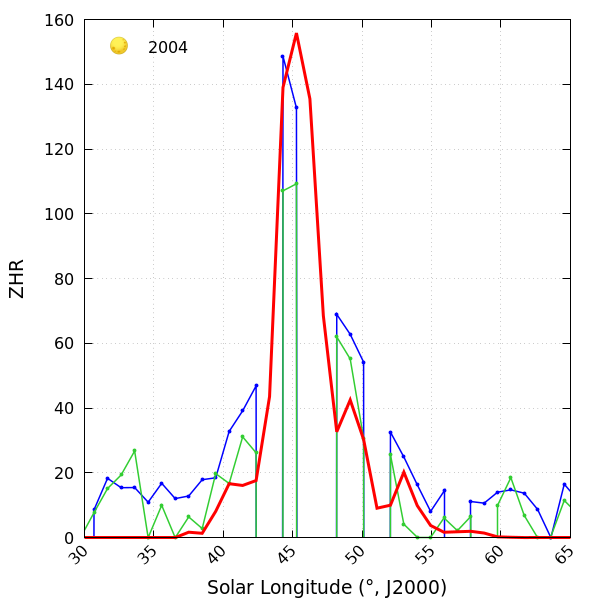}
     \includegraphics[width=0.287\textwidth]{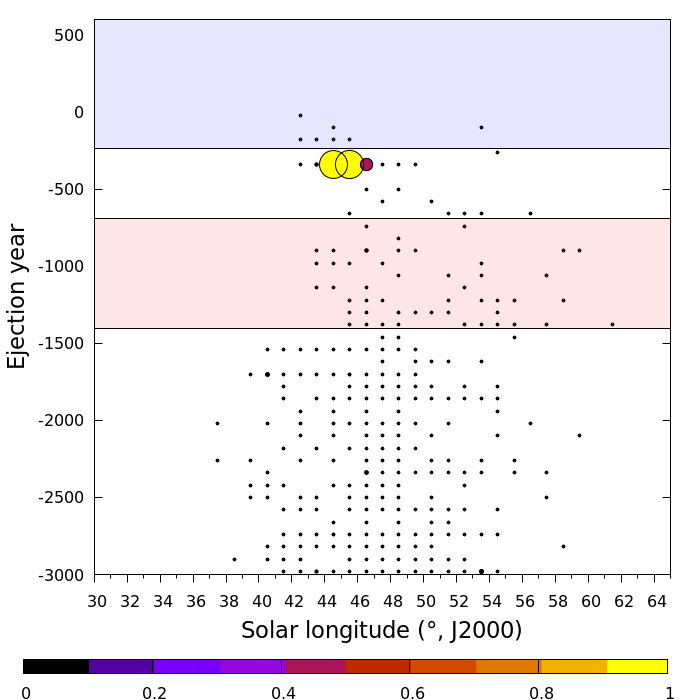}
     \includegraphics[width=0.29\textwidth]{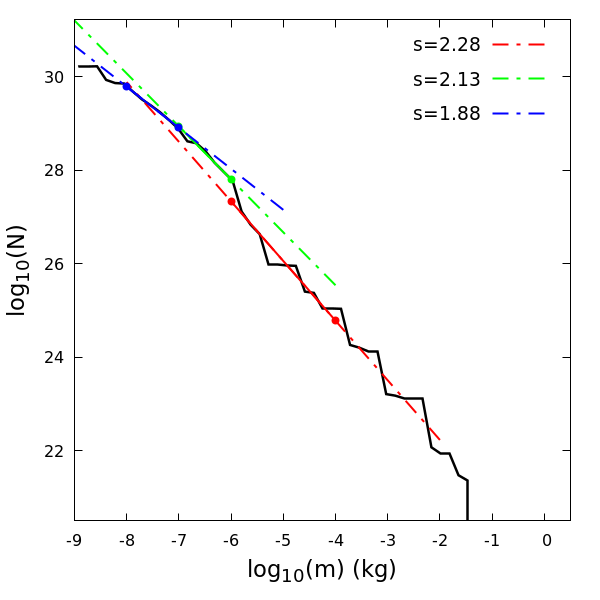}\\
     \includegraphics[width=0.29\textwidth]{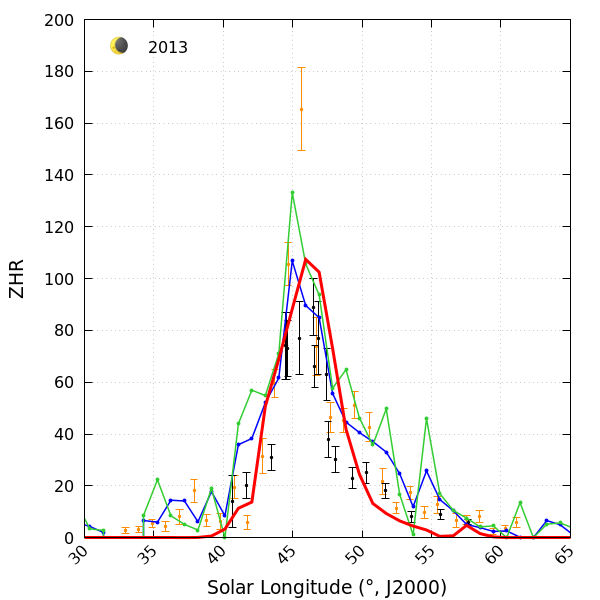}
     \includegraphics[width=0.287\textwidth]{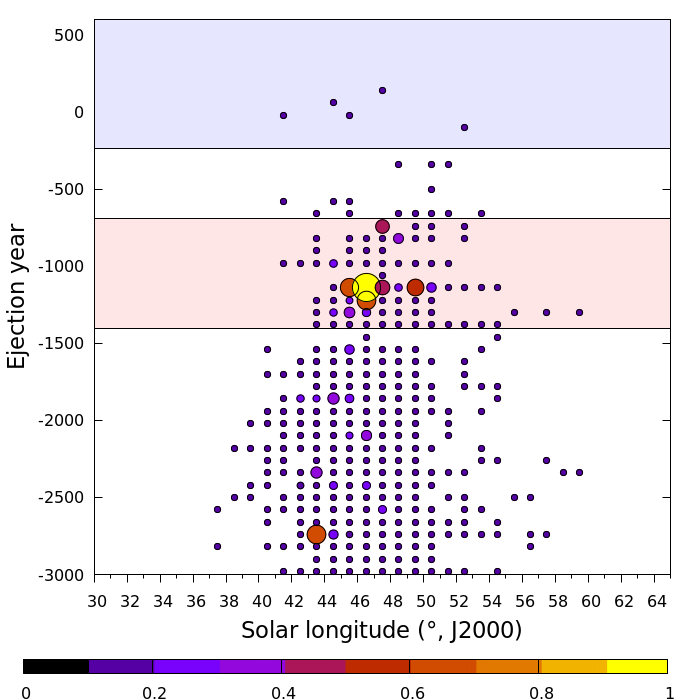}
     \includegraphics[width=0.29\textwidth]{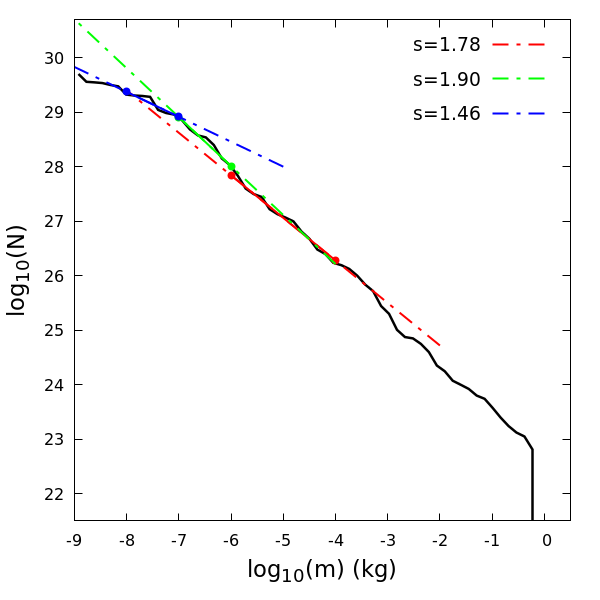}
       \caption{From left to right: Simulated (red line) and measured ZHR profiles, relative contribution of simulated trails and cumulative mass distribution. The trails relative contribution is indicated by the circles size (small to large) and color (black: low significance to yellow: highest contribution). The red band in the middle panel indicates the times when 1P/Halley was in the 1:6 resonance, blue the 2:13 as discussed in Section~\ref{section:resonances}}.
   \end{figure*}

\end{appendix}
\end{document}